\documentclass[3p]{elsarticle}
\usepackage{multirow}
\usepackage{lineno,hyperref}
\modulolinenumbers[5]
\usepackage{amssymb,amsmath}
\newcommand{\vect}[1]{ \boldsymbol{#1}} 
\usepackage{graphicx}
\usepackage{adjustbox}
\usepackage{tabularx}

\hypersetup{colorlinks,
	citecolor=blue,
	filecolor=black,
	linkcolor=blue,
	urlcolor=blue,
}

\usepackage[utf8]{inputenc}
\usepackage{listings}
\usepackage{color} 
\usepackage{subfigure}

\definecolor{codegreen}{rgb}{0,0.6,0}
\definecolor{codegray}{rgb}{0.5,0.5,0.5}
\definecolor{codepurple}{rgb}{0.58,0,0.82}
\definecolor{backcolour}{rgb}{0.95,0.95,0.92}
\usepackage[framemethod=TikZ]{mdframed}
\mdfsetup{%
	middlelinewidth=2pt,
	roundcorner=20pt}
\lstdefinestyle{mystyle}{
    backgroundcolor=\color{backcolour},   
    commentstyle=\color{codegreen},
    keywordstyle=\color{magenta},
    numberstyle=\tiny\color{codegray},
    stringstyle=\color{codepurple},
    basicstyle=\footnotesize,
    breakatwhitespace=false,         
    breaklines=true,                 
    captionpos=b,                    
    keepspaces=true,                 
    numbers=left,                    
    numbersep=5pt,                  
    showspaces=false,                
    showstringspaces=false,
    showtabs=false,                  
    tabsize=2
}

\journal{}

\begin{document}
\begin{frontmatter}

\title{Wave induced thrust on a submerged hydrofoil: pitch stiffness effects}

\author[a1]{Jingru Xing}
\author[a2]{Dimitris Stagonas \corref{cof1}}
\author[a1]{Phil Hart}
\author[a3]{Chengchun Zhang}
\author[a4]{Jianhui Yang}
\cortext[cof1]{Corresponding author: Dr. D. Stagonas, stagonas.dimitris@ucy.ac.cy}
\cortext[cof2]{Corresponding author: Dr. L. Yang,  liang.yang@cranfield.ac.uk}
\author[a1]{Liang Yang\corref{cof2}}

\address[a1]{Division of Energy and Sustainability, Cranfield University, UK}
\address[a2]{Department of Civil and Environmental Engineering, University of Cyprus, Cyprus}
\address[a3]{Key Laboratory of Engineering Bionics Ministry of Education, Jilin University, China}
\address[a4]{TOTAL E\&P UK Limited, UK}

\begin{abstract}
Submerged flapping foils can convert wave energy directly into thrust, which could be potentially utilised for green marine propulsion. This study analyses the wave-induced flapping hydrofoil propulsion using an in-house developed, new computational fluid dynamics (CFD) framework. The numerical model was initially validated against a few benchmarked problems and then used for the numerical investigation of wave-induced flapping hydrofoil propulsion. The transition between drag and thrust can be observed from the vortex flow pattern. The pitch stiffness and other physical parameter were non-dimensionalised for the first time. The optimal wave conditions and the optimal pitch stiffness are given for the future green marine system design. 
\end{abstract}
\end{frontmatter}


\begin{itemize}
    \item First high-fidelity CFD study of wave devouring propulsion system with passive flapping hydrofoil movement induced by waves.
    \item The numerical simulation results agree well with referenced experimental benchmark for different wave length
    \item Optimal pitch stiffness are investigated for peak thrust generation efficiency with non-dimensionlisation.
\end{itemize}

\section{Introduction}
\label{Introduction}


Thrust generation from a foil flapping in a fluid is often studied topic with various applications. Knoller \cite{knoller1909gesetzedes} and Betz \cite{betz1912beitrag} first explained the mechanism of thrust production by a heaving foil in a uniform flow. The so-called `Knoller-Betz' effect was verified experimentally by Katzmayr \cite{katzmayr1922effect}, who demonstrated that a foil can generate thrust in an oscillating flow and recommended that thrust generation may also be possible for a foil oscillating in a uniform flow. McKinney and DeLaurier \cite{mckinney1981wingmill} presented an analytical and experimental investigation wind energy extraction by a windmill with flapping airfoils to extract wind energy, while Wu et al. \cite{wu1972extraction} studied energy generation through the  water wave-foil interaction and reported that energy can only be extracted for waves have with a vertical velocity component to both the mean free stream and the span of the flapping hydrofoil.

Even before these studies, the observation of birds and fishes with foil shaped body parts has inspired a vast number of works focused on understanding the mechanism of motion induced thrust generation. In a seminal work about the modes of aquatic animal propulsion, Lighthill \cite{lighthill1975mathematical} reported a mechanism leading to the formation of jet-like wake and thrust production. The jet-like wake of Lighthill is the well known reverse Von Karman vortex street, proposed by Von Karman and Burgers \cite{von1935general}. Following Lighthill \cite{lighthill1975mathematical}, this presence of this mechanism was numerically and experimentally confirmed by several researchers like, Koochesfahani et al. \cite{koochesfahani1989vortical}, Jones et al. \cite{jones1996wake}, Buchholz et al. \cite{buchholz2008wake}, Godoy-Diana et al. \cite{godoy2009model}, Lagopoulos et al. \cite{lagopoulos2019universal} and Ma et al.  \cite{ma2021study}. 

Building on the concept of motion induced thrust generation and the work of Wu et al. \cite{wu1972extraction}, Isshiki \cite{isshiki1982theory} theoretically proved the possibility of wave devouring propulsion (WDP) through the use of a submerged, passive flapping hydrofoil subject to the action of waves. Later, Isshiki et al. \cite{isshiki1983theory, isshiki1984theory} conducted a series of the experiments and further theoretical discussions considering the WDP of a hydrofoil placed in waves with shallow draft and having its heave and pitch motion restricted by springs. For a wide range of wave lengths, the authors reported that the foil's advance speed varied non-linearly with the heave and pitch amplitude, with a maximum velocity occurring only under certain conditions. Grue et al. \cite{grue1988propulsion}, developed a mathematical formulation considering the vortex distribution along the foil's and the wake's centreline. Despite the simplifications in their mathematical approach Grue et al. \cite{grue1988propulsion} results confirmed that the foil can capture a large amount of the incoming wave energy and apply it for self-propulsion. More recently, Kumar et al. \cite{kumar2019thrust} employed a wave flume and conducted experiments on the generation of thrust by a surge restricted flapping foil system. For these experiments the foil was rigidly fixed to an elastic flat plate, which was in turn exposed to regular waves. The results presented by Kumar et al. \cite{kumar2019thrust} showed an increase in thrust generation with the plate area and an empirical equation was also proposed. 

In addition to the aforementioned works, considerable effort and computational time has been devoted in simulation foil-flow interaction using Computational Fluid Dynamics (CFD). Pedro et al. \cite{pedro2003numerical} investigated the mechanism of thrust produced by a motion restricted (either pitch or heave) hydrofoil in a  viscous flow with relatively low Reynolds number $Re=1000$. The fluid-structure interaction was simulated by combining  CFD with the arbitrary Lagrangian-Eulerian (ALE) method. The authors demonstrated that when the phase angle between heave and pitch decreased (from 115$^\circ$ to 30$^\circ$), the average thrust coefficient also decreased, while drag was generated for phase angles less than 60$^\circ$, and the propulsive efficiency reached a maximum of 64\% for Strouhal number of 0.35. Later, Xie et al. \cite{xie2014investigation} investigated the energy extraction performance of a flapping foil with an elliptical shape using the commercial ANSYS CFX. It was shown that the power extracted from the incoming flow was primarily the result of heave, while the contribution pitch was rather limited. With a marine propulsion application in mind, De Silva et al. \cite{de2012numerical} also used a commercial code to investigate the active oscillation a foil in a wavy flow. Simulations were conducted in the FLUENT solver employing a dynamic mesh and the Volume of Fraction (VOF) method for capturing the air-water interface. The problem was parameterised for eight design parameters seeking to identify the optimum value per parameter in order to maximise the thrust producing performance of the oscillating hydrofoil. De Silva et al. \cite{de2012numerical} reported maximum foil propulsion for a phase angle of -90$^\circ$ and increasing foil efficiency with the wave amplitude (e.g. from 0.5 m to 3 m). 

Filippas et al. \cite{filippas2014hydrodynamic} used the unsteady boundary element method (BEM) to also analyse flapping hydrofoils under the action of propagating waves and compared their results with experimental measurements. The authors emphasised that free surface effects cannot be neglected, especially for hydrofoils with a relatively small draft (low submersion level). These findings were aligned with the observations and the results of Isshiki \cite{isshiki1982theory} and Grue et al. \cite{grue1988propulsion}. Even more recently Lagopoulos et al. \cite{lagopoulos2019universal} applied a boundary data immersion method (BDIM) to study the drag-to-thrust wake transition for hydrofoils submerged in uniform flows. $St_T$, a new version of Strouhal's number using the foil kinematics was proposed for predicting the wake conditions. In particular, it was proven that for $St_T$ greater than 1, a reverse Von Karman vortex street was generated and thrust was produced. Later on, Zurman-Nasution et al. \cite{zurman2020influence} reported on the limitations of using 2D simulations to study 3D flapping foil motion through BDIM and suggested 0.15 $\leq St_A \leq$ 0.45 as a valid range for such simulations.

 For practical applications, a so called wave devouring propulsion system (WDPS) was considered by Jakobsen \cite{jakobsen1981foilpropeller} for ships. The authors experimented with full-size models but in a controlled environment and showed that if the energy of the waves is sufficient, then the WDPS produced ship speed could reach or even exceed the speed produced by the ship's conventional propulsion system. Terao and Sunahara \cite{terao2012application} tested in a wave tank different WDPS types and reported that the wave drifting force on a floating structure can be reduced due to the WDPS operation. Bøckmann et al. \cite{bockmann2016model} conducted physical model tests and time-domain simulations of a tanker with two hydrofoils (referred to in their work as wavefoils) placed at the bow of the ship's hull and recommended that the implementation of the wavefoils had lowered the resistance of the ship by 9-17$\%$ in typical sea conditions. Driven by the previous works, Moreira et al. \cite{moreira2020dual} presented a 2D parametric study of a dual flapping hydrofoils system on a ship, focusing on the evaluation of the ship's propulsion as well as the wave energy harvesting performance of foils. The results reported showed that a larger horizontal spacing between the hydrofoils improved both the structure's energy harvesting performance and its hydrodynamic coefficients. The energy harvesting potential of flapping hydrofoils was also explored by Lahooti et al. \cite{lahooti2019multi} who concluded that a bluff body placed upstream of a flapping hydrofoil would be 30\% more efficient at energy harvesting for Reynolds number, Re=1000. In the same year, Yang et al. \cite{yang2019systematic} presented results on the performance optimisation of a propulsion system employing six tandem hydrofoils on the wave glider; it was showed that (in 2D) a single hydrofoil delivers its optimum (propulsion) performance when  its pivot axis is set on its leading edge and a torsion spring is applied at the pivot axis. Several years of research have led to commercialisation attempts. To the author's best knowledge, the first successful voyage of the WDPS ship Mermaid II was reported in 2008 and since then a number of wave-propelled concepts have become commercially available, as for example, the Wave Glider from Liquid Robotics \cite{hine2009wave}, the Autonomous Surface Vehicles (ASV) from AutoNaut \cite{johnston2017marine}, and the, equipped with retractable bow foils, M/F Teistin by Wavefoil \cite{yrke2019full}.

Overall, hydrofoils have the potential to harness energy from the flow field and through their flapping motion convert it to thrust. In practical applications involving waves, the mode of motion of the hydrofoil is often controlled through the use of springs. For example, an infinite heave/pitch spring stiffness will impose a pure heave or pitch only motion, while relaxing the heave and pitch spring stiffness will lead to a coupled motion. For each mode, the dynamic separation at the foil has been considerably explored and the generation of more complex wake patterns has been associated with the pure (heave or pitch) motions, while the coupled motion has been shown to lead to a less intense (when compared with that of any of the pure motions) flow separation at the hydrofoil. For a coupled motion, the thrust coefficient was found to decrease as the phase angle between heave and pitch decreases, while a 90$^\circ$ phase difference is considered optimum for propulsion efficiency, with drag being generated at phase angles smaller than 60$^\circ$. Propulsion jets occur when the thrust induced foil velocity exceeds the ambient flow velocity, while the drag-to-thrust wake transition is driven by the foil kinematics shown to include the oscillating amplitude and frequency, and the cycle-averaged swept trajectory. The pitch / heave spring stiffness is a measure of direct control over the foil kinematics and thus of the drag-to-thrust wake transition. Nonetheless, to the best of the author's knowledge, the stiffness impact on the generation of thrust has seldom been explored and explained. The present work focuses on the pitch stiffness effects but the methodology and the methods used can be directly applied for exploring the heave stiffness as well.

Specifically, a relatively recently developed CFD model is used to simulate the performance of a hydrofoil subject to the action of propagating regular waves. The model is initially validated against experiments with hydrofoils in uniform flow and regular waves and then it is used to explore and explain the influence of pitch stiffness on thrust generation. In particular, the experimental case with the highest foil propagation velocity is considered first. The test is reproduced in the numerical domain using the physical stiffness for the numerical spring and then it is repeated for a range of stiffness. Soon after the procedure is repeated for shorter and longer waves thereby considering wave steepness within the range of 0.029 $<$ H/L $<$ 0.048. Compared with the majority of previous numerical studies where the motion is forced a hydrofoil exposed to a flow field, the present work benefits from the consideration of wave (flow field) induced, and thus passive, flapping hydrofoil motion. This is made possible through the use of the, so called, 'one-fluid' CFD approach, which is described first in the remainder. Then, theoretical information on the hydrofoil geometry, the modes of foil motion and the numerical reproduction of the ambient flow conditions is provided, before introducing the important dimensionless parameters used. The results of the model convergence and validation tests are then presented and discussed, followed by the main results about the influence of pitch stiffness on the thrust generation and on the hydrofoil's propagation velocity, and by the work's conclusions. 

\section{Material and Method}
\label{Material and Method}
\subsection{Numerical solver}
\label{Numerical solver}
An 'in-house' developed code employing the 'one-fluid' formulation was recently proposed by Yang et al. \cite{yang2015immersed, yang2018one, yang2018unified,chen2021numerical} as an addition to the family of immersed methods. In marked contrast with the traditional representation of structural dynamics in CFD through a set of Newton-Euler equations, the 'one-fluid' framework uses Lagrangian multipliers to constraint the motions and ensure rigidity. In other words, the objects simulated within the `one-fluid' numerical domain  are treated as fluids (hence the name 'one-fluid'). The simulation is performed on a fixed Eulerian Cartesian grid with uniform mesh size. The Level Set Method (LSM) is used to capture the interface between the different phases and prevent the development of indeterminate regions among the fields between, e.g., two phases. The comparative advantages of the formulation include the use of the fluid solver for resolving the fluid - solid (structure) dynamics, as well as the use of Heaviside/LSM for handling topological changes at the fluid-structure interface removing the requirement for updating the mesh or for re-meshing. Both the former and the latter increase the computational efficiency of simulations. The governing equations of the approach follow.

\subsubsection{`One-fluid’ formulation - Governing equations}
\label{`One-fluid’ formulation - Governing equations}
\begin{figure}[ht!]
    \centering
    \includegraphics[width=0.7\textwidth]{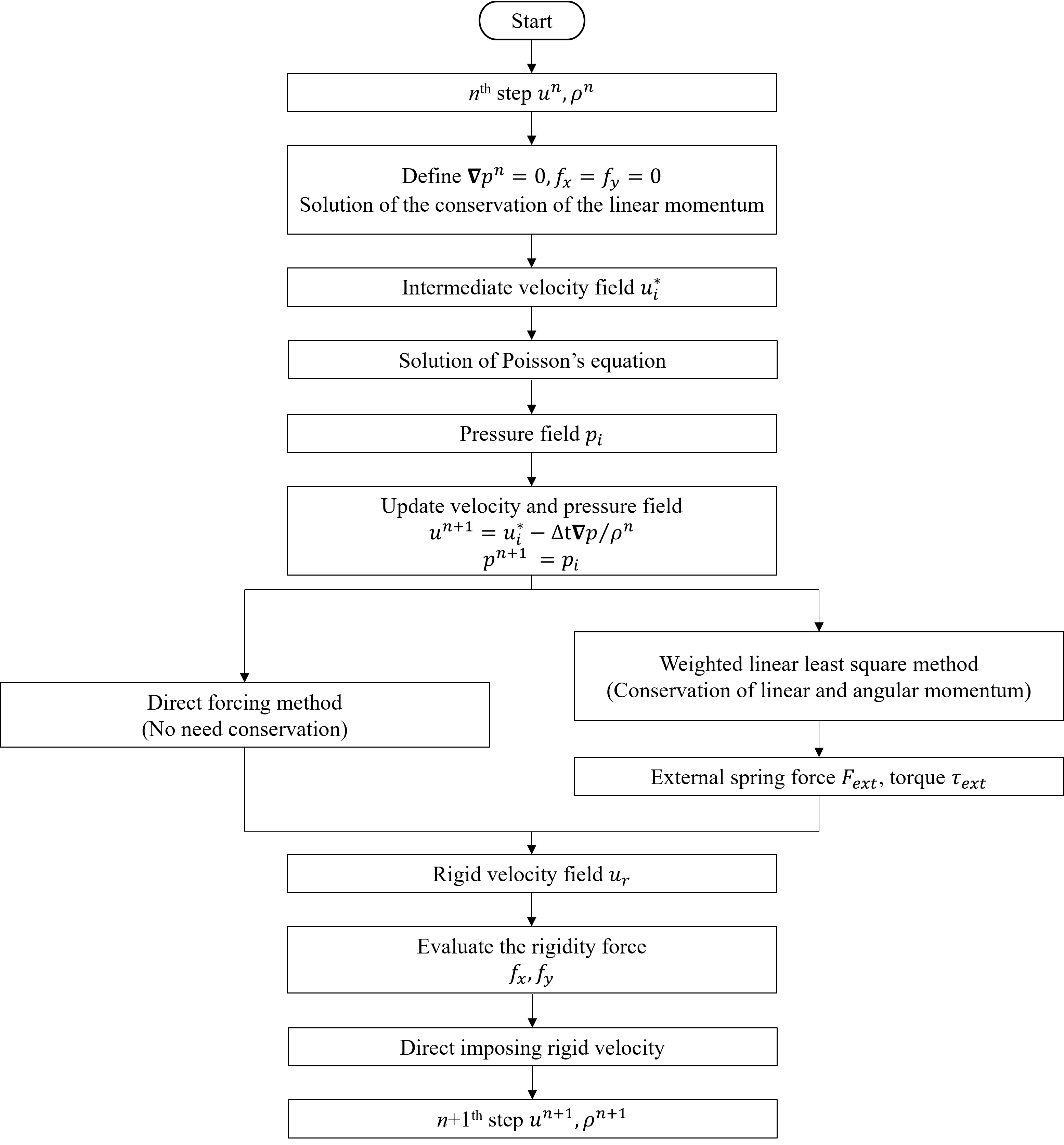}
    \caption{The flow chart of the `one-fluid' framework.}
    \label{fig:onefluid-fc}
\end{figure}

The principle flow chart of the 'one-fluid' algorithm employed in this study is presented in Fig.\ref{fig:onefluid-fc}.

The incompressible Navier–Stokes equation following  the exclusion of surface tension and the inclusion of the viscous stress force $\vect{f}$ at the right of the equation, reads Eq.\ref{eq:n-s-1:3}.
\begin{subequations}
	\label{eq:n-s-1}
	\begin{align}
		\rho \left[ {\partial \vect{u}\over \partial t} + ({\vect \nabla} \vect{u})\vect{u} \right] & =-{\vect \nabla} p+  \vect{f} +  \rho \vect{g} \label{eq:n-s-1:1}  \\
		{\vect \nabla} \cdot \vect{u} & =  0 \label{eq:n-s-1:2} \\ 
		\vect{f} & =
		\begin{cases} 
		\vect{\nabla} \cdot 2\mu_a\vect{\bf{D}}(\vect{u}) & \text{in }  \Omega_a\\
		\vect{\nabla} \cdot 2\mu_w\vect{\bf{D}}(\vect{u}) & \text{in }  \Omega_w\\
		\rho \left[ \left(\frac{\partial \mathrm{P} (\vect{u})}{\partial t}\right)+(\vect{\nabla} \mathrm{P} (\vect{u})) \mathrm{P}(\vect{u})\right]+\vect{\nabla} p-\rho \vect{g} & \text{in }  \Omega_r\\
		\end{cases} \label{eq:n-s-1:3}
	\end{align}
\end{subequations}
where $\rho=\rho_{a} H_{a}+\rho_{w} H_{w}+\rho_{r} H_{r}$ is the density field, $\vect{u}$ is the velocity vector field, $p$ is the pressure, $\vect{f}$ is the force on each phase $\Omega$, $\vect{g}$ is the gravitational acceleration, $\mu$ is the dynamic viscosity and $\vect{\bf{D}}(\boldsymbol{u})=\frac{1}{2}\left(\nabla \boldsymbol{u}+(\nabla \boldsymbol{u})^{T}\right)$ is the strain rate tensor. The subscripts indicate the different phases, i.e., $a$ for air, $w$ for water, and $r$ for rigid body.

As already mentioned, the LSM is used for capturing the interface between the fluid phases in the numerical domain, while the characteristic function $\chi$ of any region $\Omega_\alpha$ can be uniformly expressed as
\begin{equation}
	\label{eq:characteristic}
	\chi=H_{\alpha}(\phi) \quad \alpha=a, w, r
\end{equation}
where
\begin{equation}
	\label{eq:heaviside_def}
	H_\alpha (\vect{x}) = \left\{
	\begin{array}{l l}
		1 & \quad \text{if}  \quad   \vect{x} \in \Omega_\alpha    \\
		0 & \quad \text{if}  \quad   \vect{x} \notin \Omega_\alpha 
	\end{array} \right. .
\end{equation}
is a Heaviside function for a phase $\alpha$. $\vect{x}=(x,y)$ is the coordinate of any point in the domain. $H_\alpha$ is introduced to identify different phase region. Once the Heaviside function has been constructed, the corresponding density and viscosity field can be assigned. For more information refer to Yang's study \cite{yang2018one}. 

\subsubsection{Numerical and discretisation scheme}
\label{Numerical and discretisation scheme}
All the simulations are performed with a fixed Eulerian Cartesian grid and are numerically discretised using a grid-based immersed method. The Cartesian staggered Finite Volume scheme (Marker-and-Cell grid) is adopted for the spatial discretisation of the study domain. The convective terms of the Navier-Stokes equations are discretised with a second order Total Variation Diminishing (TVD) Runge-Kutta scheme in the conservative framework. The second-order accurate Runge-Kutta scheme processes the time treatment of the computational momentum and level set equations. Adaptive time steps are applied to the calculation of the Courant-Friedrichs-Lewy (CFL) number \cite{yang2018one}, which for all the simulations presented in this paper is kept equal to $0.5$. 

\subsubsection{Boundary conditions}
\label{Boundary conditions}
Dirichlet type, the common inlet boundary condition for free surface flow simulations, are considered to be the boundary condition. Dirichlet boundary condition directly specifies the values of related physical quantities at the boundary. The flow profiles are based on experimental inlet settings. The velocity of the boundary inlet is given directly. The uniform and regular waves used in this study will be explained in detail in Section \ref{Validation with experimental results}.

\subsection{Hydrofoil geometry, modes of motion and spring stiffness}
\label{Hydrofoil geometry, modes of motion and spring stiffness}
\begin{figure}[ht!]
    \centering
    \includegraphics[width=0.4\textwidth]{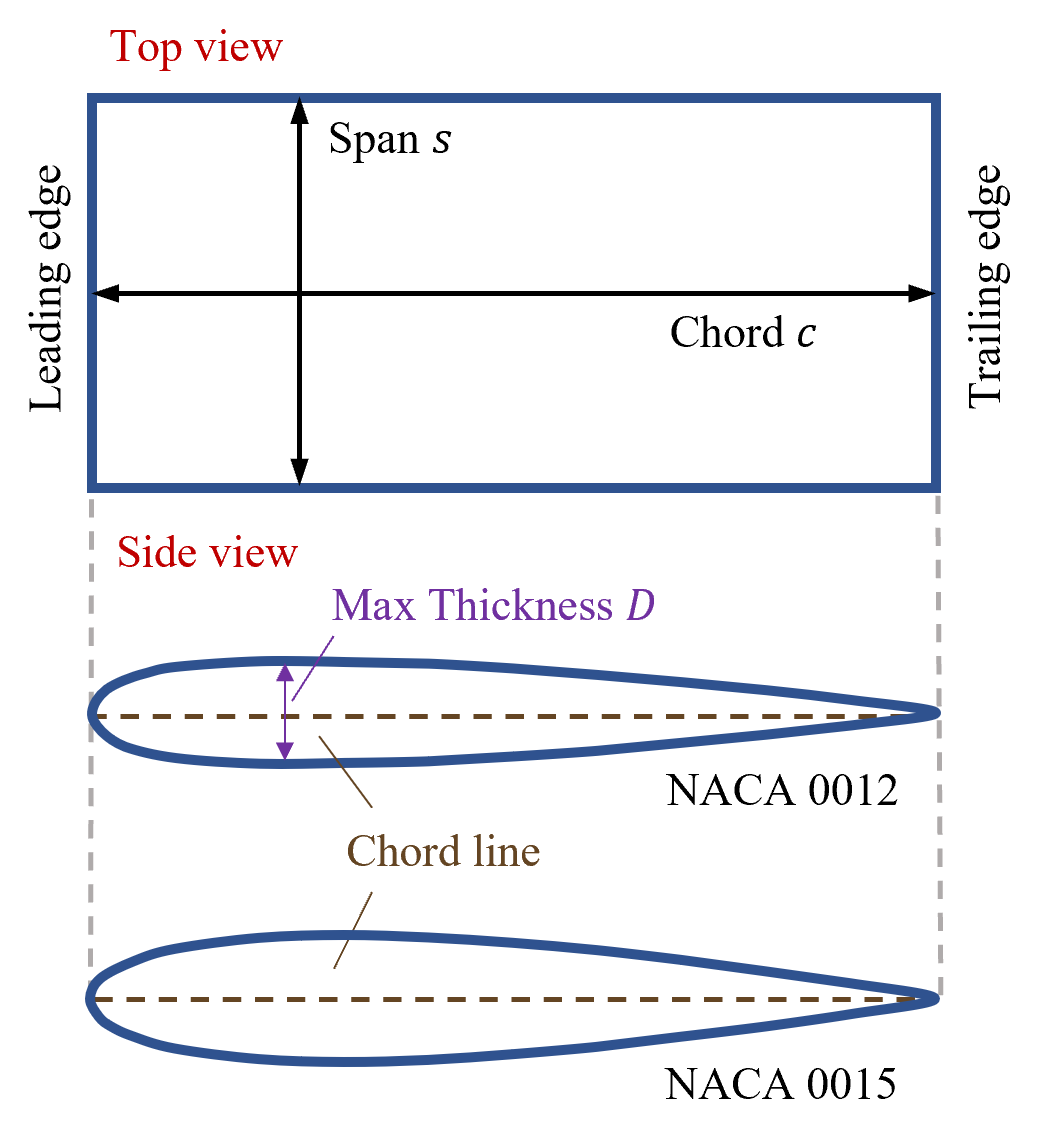}
    \caption{The characteristic of hydrofoil.}
    \label{fig:hy_chara}
\end{figure}

As it is reported later in this paper, and for validation purposes the experimental works of Read et al. \cite{read2003forces} and  Isshiki et al.  \cite{isshiki1983theory, isshiki1984theory} are simulated with the numerical model described above. As such, the work and the results presented consider two different NACA type foils, namely NACA0012 and NACA0015. Each symmetrical foil is further characterised by the chord length $c$ --- defined as the straight line connecting the leading to the trailing edge, which for symmetric foils coincides with the mean camber line ---, and the foil maximum thickness $D$ often expressed as a percentage of $c$, as illustrated in Fig.\ref{fig:hy_chara}. For 3D applications, as for example in the experiments, the breadth or span $s$ of the foil is also considered. For this 2D study, it is treated as a normalised unit, i.e. the value of span $s$ defaults to 1 and will not be repeated hereafter.

The hydrofoil in our simulations performs either enforced (active) oscillations around a stationary equilibrium position or it is subject to flow induced (passive) flapping motions. The latter motions is the combined outcome of simultaneous heave and pitch oscillations, see Fig.\ref{fig:hy_motions}. Forced oscillations take the form of, pure heave motion, where the the hydrofoil translates in the vertical direction or pure pitch, where the hydrofoil rotates around a pivot point $P$, see also Fig.\ref{fig:hy_motions}. In accordance with the experiments of Read et al. \cite{read2003forces} $P$ is set at the chord and at a distance of $c/3$ from the leading edge of the foil, while pure heave and pitch motions are prescribed in the numerical domain through Eq.\ref{eq:heavepitch_tra:1} and \ref{eq:heavepitch_tra:2}, and the flapping foil motion is the combined output of the heave and pitch superposition, Fig.\ref{fig:hy_motions}:
\begin{figure}[ht!]
    \centering
    \includegraphics[width=0.7\textwidth]{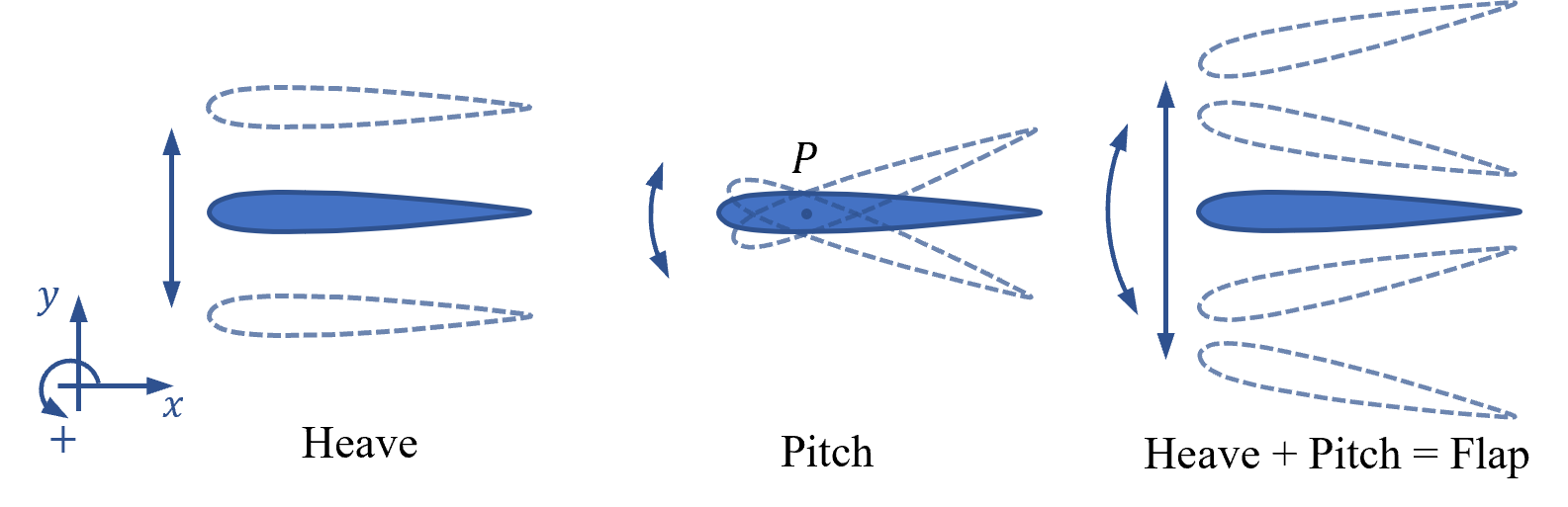}
    \caption{The motions of hydrofoil.}
    \label{fig:hy_motions}
\end{figure}

\begin{subequations}
\label{eq:heavepitch_tra}
	\begin{align}
    h(t)=h_{0} \sin (\omega t) \label{eq:heavepitch_tra:1}\\
    \theta(t)=\theta_{0} \sin (\omega t+\psi) \label{eq:heavepitch_tra:2}
	\end{align}
\end{subequations}
where, $h(t)$ is the displacement of the hydrofoil in the y direction or in other words the instantaneous vertical position of the foil axis. $\theta(t)$ is the instantaneous angle formed between the oncoming flow velocity ($U_{\infty}$) and the chord, also referred to as the angle of rotation. $h_{0}$ and $\theta_{0}$ is the heave and pitch amplitude. $\omega$ is the angular frequency (rad/sec), $2{\pi}f$, where $f$ is the oscillating frequency, and $\psi$ is the phase difference between the pitch and heave motion. When it comes to propulsion efficiency, Platzer et al. \cite{platzer2008flapping} suggested $\psi=90^o$ as the optimal phase difference between pitch and heave. 

\begin{figure}[ht!]
    \centering
    \begin{tabular}{cc}
        \includegraphics[width=0.35\textwidth]{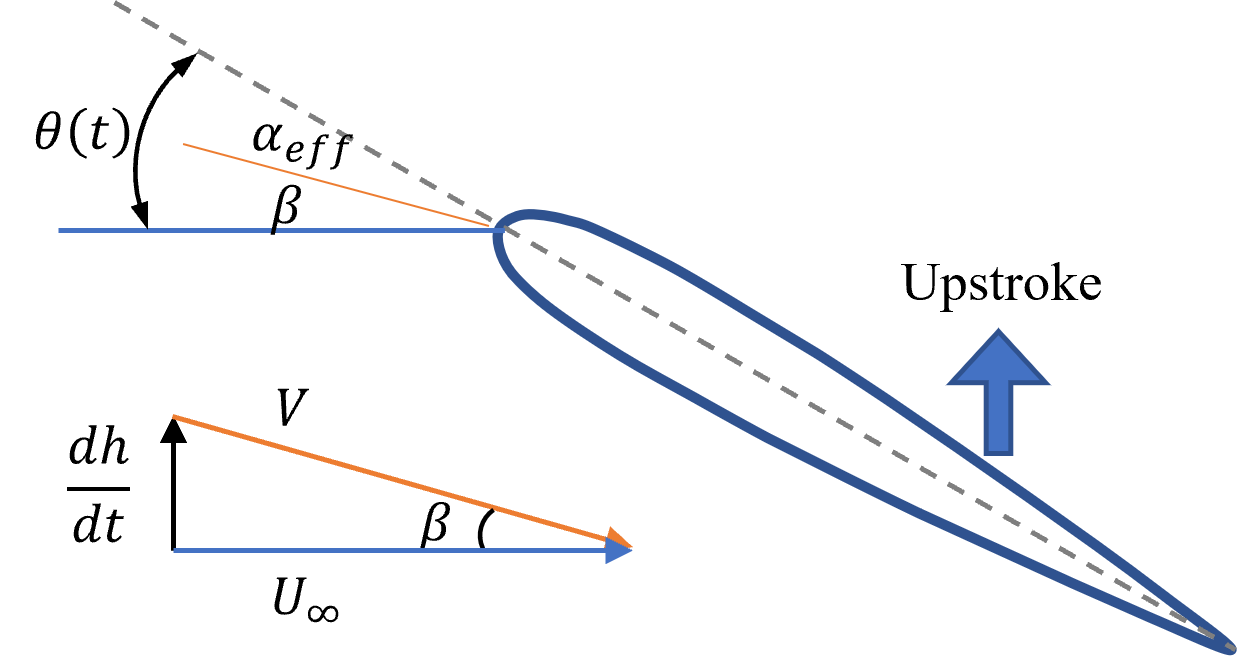} &
        \includegraphics[width=0.35\textwidth]{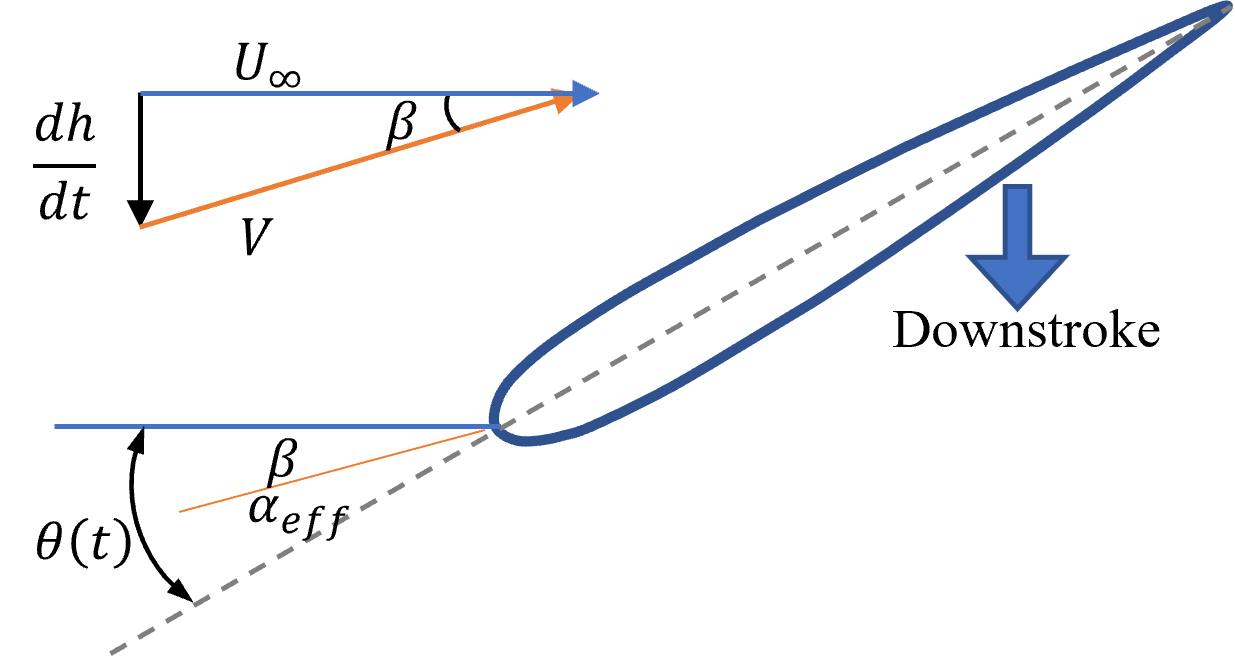} \\
    (a) Upstroke & (b) Downstroke
    \end{tabular} 
    \caption{The illustration for the effective AOA $\alpha_{eff}$ of flapping hydrofoil in uniform flow ($\frac{dh}{dt}$ is the heave velocity, $V$ is the velocity of the stream relative to the hydrofoil.).}
    \label{fig:aoa_eff}
\end{figure}

For foils subject to an oncoming flow, the resulting angle of attack (AoA) is related to the heave velocity and pitch angle as illustrated in Fig.\ref{fig:aoa_eff} and expressed from Eq.\ref{eq:eff_aoa} \cite{read2003forces, wu2020review}.
\begin{equation}
	\label{eq:eff_aoa}
	\alpha_{eff}(t)=\theta(t)-\arctan \left(\frac{\dot h(t)}{U_{\infty}}\right)
\end{equation}
where $\alpha_{eff}(t)$ and $\theta(t)$ is the instantaneous effective AOA and pitch angle and $\dot h(t)$=$\frac{dh}{dt}$ is the heave velocity.

\begin{figure}[ht!]
    \centering
    \begin{tabular}{ccc}
        \includegraphics[width=0.3\textwidth]{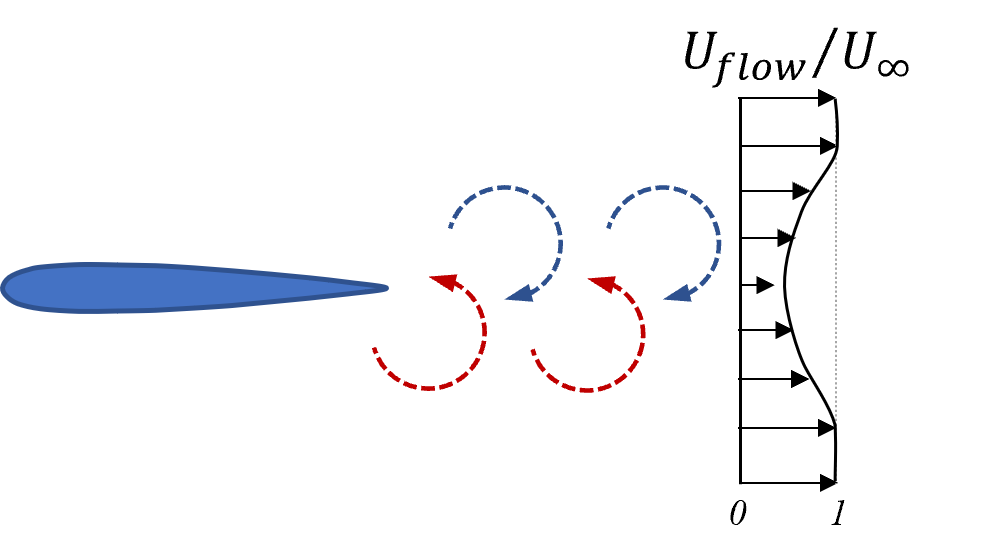} &
        \includegraphics[width=0.3\textwidth]{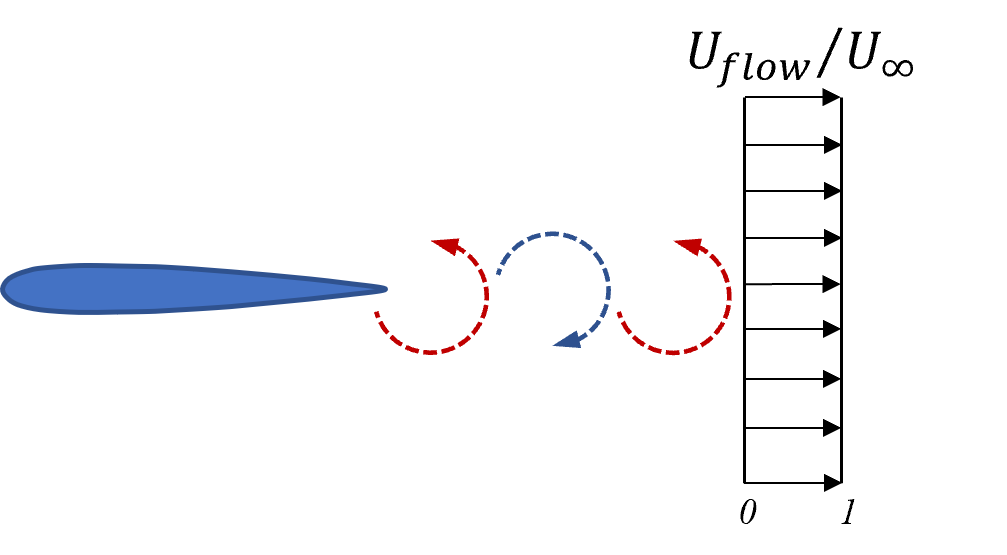} &
        \includegraphics[width=0.3\textwidth]{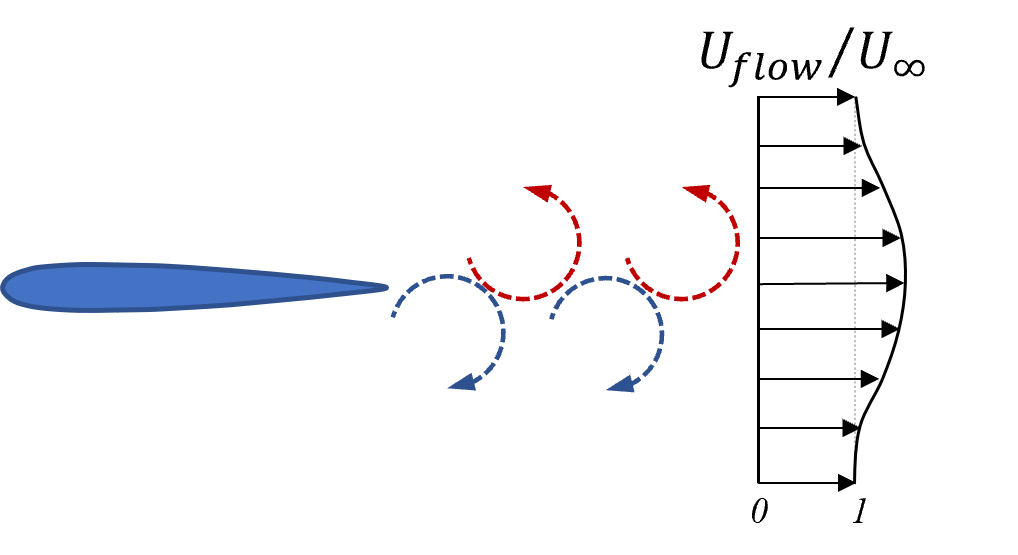} \\
    (a) vK street & (b) Neutral regime & (c) RvK street
    \end{tabular} 
    \caption{The illustration for the drag-to-thrust wake transition of flapping hydrofoil.}
    \label{fig:wake_tran}
\end{figure}

Foil undulations imposed by such harmonic motions are known to generate wake patterns leading to the generation of either drag or thrust. Specifically, when a Bénard von Kármán vortex street (vK)  \cite{von1935general} develops, the velocity of the flow in the wake is smaller than the ambient / oncoming flow velocity and the foil is introduced into a drag dominated regime. When the vortexes in the wake at the foil's trailing edge align, then the velocity of the flow in the wake is uniform with the ambient / oncoming flow velocity and the foil experiences a neutral regime. While, upon the development of a reversed Bénard von Kármán vortex street (RvK) the foil experiences thrust since the velocity of the flow in the wake becomes larger then the ambient / oncoming flow velocity. For flapping foils in uniform flows, these drag-to-thrust wake transition stages have been extensively studied and the foil's transition towards thrust has been shown to occur with a delay compared with the drag-to-thrust wake transition \cite{lagopoulos2019universal, streitlien1998thrust, ramamurti2001computational, bohl2009mtv}. The three fundamental wake patterns are illustrated in Fig.\ref{fig:wake_tran}.

\begin{figure}[ht!]
    \centering
    \includegraphics[width=1.0\textwidth]{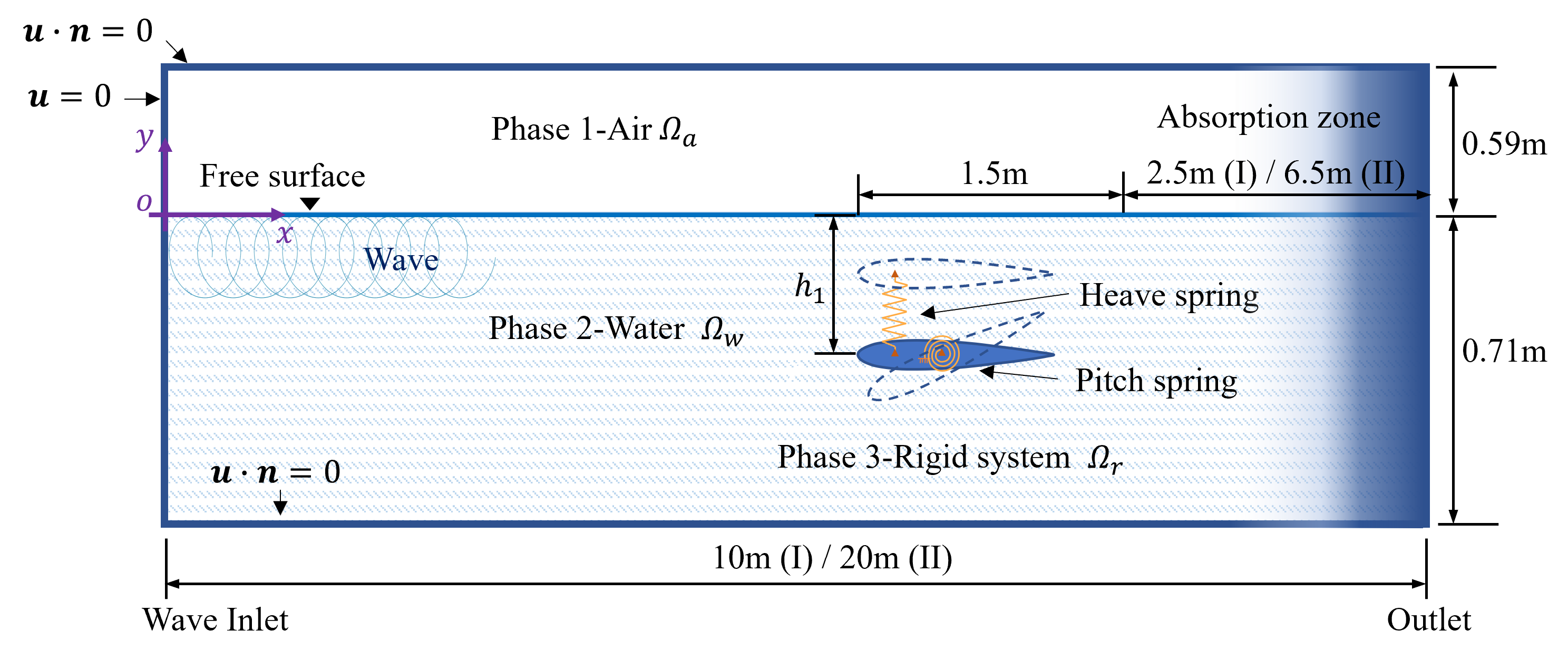}
    \caption{The schematic diagram of the passive hydrofoil simulation.}
    \label{fig:sch_wavefoil}
\end{figure}
The existence of drag and thrust regimes is also confirmed by the effective AoA equation (Eq.\ref{eq:eff_aoa}), which was shown to have two solutions over a large range of frequencies and phase differences between heave and pitch;  one solution corresponding to drag production, and a second corresponding to thrust generation, Read et al. \cite{read2003forces}. This double solution in the AoA equation is associated with the vertical (up or down) pitch of the foil with respect to the instantaneous oncoming flow. As such, for foils controlled by springs it stands to reason that drag-to-thrust and vice versa will depend on the spring stiffness, which controls the rate of change of the pitch angle and affects the pitching amplitude as well. In our simulations the spring effect is expressed in the form of an instantaneous force and moment applied on the foil given by Eq.\ref{eq:hp_spring} and illustrated in Fig.\ref{fig:sch_wavefoil}.
\begin{subequations}
	\label{eq:hp_spring}
	\begin{align}
	    F_h(t)=k_h \Delta y \label{eq:hp_spring:1}  \\
		M_p(t)=k_p \Delta \theta \label{eq:hp_spring:2} 
	\end{align}
\end{subequations} 
where, $F_h(t)$ is the heave force and $M_p(t)$ is the pitch moment. $\Delta y$ and $\Delta \theta$ is the vertical and rotational deviation of the hydrofoil from its initial position at time $t$, respectively, while $k_h$ and $k_p$ is the heave and pitch stiffness. This definition enables the direct comparison with the experiments of Isshiki et al. \cite{isshiki1983theory, isshiki1984theory} for which $k_h$ is set to 10388 N/m and $k_p$ to 129.24 N$\cdot$m/rad. The pitch stiffness effect is then explored by keeping $k_h$ constant and varying $k_p$ for different oncoming flow conditions. We return to this in Section \ref{The effect of the pitch stiffness}.   

In such computations, buoyancy is expected to influence the foil kinematics. This unwanted influence is removed from the simulations by setting the density of the hydrofoil to be the same as the density of the fluid, $\rho=\rho_{hydrofoil}=\rho_{fluid}$.

\subsection{Numerical flow conditions}
\label{Numerical flow conditions}
Convergence tests and an initial validation of the computations are performed with a hydrofoil subject to a uniform flow. Excluding the solids, the flow velocity $U_{flow}$ is the same throughout the domain and is described by Eq.\ref{eq:uni_flow}.
\begin{subequations}
	\label{eq:uni_flow}
	\begin{align}
    u (x,y,t) &= U_{flow} x \\
    v (x,y,t) &= 0
	\end{align}
\end{subequations}
The current flows steadily along the $x$ positive direction with zero velocity in the $y$ direction. Across the domain, the current enters from the left boundary and exits from the right boundary, with the same velocity at both boundaries to ensure the conservation of mass. Then the hydrofoil is exposed to the action of regular waves. For these simulations, wave generation and absorption is set on the left and right hand side of the numerical domain, see also Fig.\ref{fig:sch_wavefoil}. The flow conditions, in the form of horizontal and vertical velocity, at the inlet are calculated using linear wave theory and specifically: 
\begin{subequations}
	\label{eq:linear-wave}
	\begin{align}
    u (x,y,t) &= \frac{\pi H}{T}\frac{\cosh [k(y+d)]}{\sinh (k d)}\cos{(k x-\omega_w t)}\\
    v (x,y,t) &= \frac{\pi H}{T}\frac{\sinh [k(y+d)]}{\sinh (k d)}\sin{(k x-\omega_w t)}
	\end{align}
\end{subequations}
where, $u$ and $v$ are the instantaneous horizontal and vertical particle velocities at $(x,y)$ location, $H$ is the wave height, $T$ is the wave period, $d$ is the water depth, $k=\frac{2 \pi}{\lambda}$ is the wave number,  $\omega_w=\frac{2 \pi}{T}$ is the angular wave frequency, and $\lambda$ the wavelength calculated with the linear dispersion equation, Eq.\ref{eq:wave-length}. Using linear wave theory, the wave conditions in the numerical domain were tuned to match the wave measured in the experiments of Isshiki et al. \cite{isshiki1984theory}. It is noted, that the use of linear wave theory for generating non-linear waves is a limitation of our study, which can lead to the generation of spurious long wave components. The application of a more precise wave generation procedure goes beyond the scope of the present work and the spurious effects partly mitigated through the use an absorption zone that the end of the numerical fume.    
\begin{equation}
	\label{eq:wave-length}
	\lambda=\frac{g T^2}{2 \pi} \tanh{(k d)}
\end{equation}

The length of the wave absorption zone at the outlet (right hand side boundary) varies in accordance with the incoming wave length and it is set to 2.5 m for $\lambda < 3.75 $m and 6.5 m for $\lambda < 7.70$ m. At the absorption zone wave reflection is prevented through the application of the relaxation method proposed by Bihs et al. \cite{bihs2016new}, Eq.\ref{eq:wave_absorp}.
\begin{subequations}
	\label{eq:wave_absorp}
	\begin{align}
		\Gamma (\tilde{x}) & =1-\frac{e^{\left(\tilde{x}^{3.5}\right)}-1}{e-1}  \quad \text {for} \quad \widetilde{x} \in[0,1] \label{eq:wave_absorp:1}  \\
		u_r& =\Gamma(\tilde{x}) u_c \label{eq:wave_absorp:2} \\ 
		v_r& =\Gamma(\tilde{x}) v_c \label{eq:wave_absorp:3}
	\end{align}
\end{subequations}
where, $\Gamma (\tilde{x})$ is the relaxation function and $\tilde{x}$ is a scale factor, i.e. the distance from the wave inlet boundary to the absorption zone over the entire domain length. $u_r$ and $v_r$ are the relaxed velocity components in the absorption zone while $u_c$ and $v_c$ are the original computational ones without relaxation.

\subsection{Main and dimensionless parameters}
\label{Main and dimensionless parameters}
Any hydrofoil subject to the interaction with a viscous force experiences a drag and a lift force, e.g. Fig.\ref{fig:hy_force}. As described previously, the occurrence of a RvK vortex street is associated with drag-to-thrust wake transition and once the wake velocity exceeds that of the ambient flow thrust jets are produced and the foil experiences a thrust force --- or in other words a horizontal force acting in the opposite to the drag force direction, from the trailing to the leading edge ---, see e.g. Lagopoulos et al. \cite{lagopoulos2019universal}. As accustomed lift and drag forces are expressed through the dimensionless lift and drag coefficients ($C_{l}$) and ($C_{d}$), given by the force acting on the foil normalised with the dynamic pressure of the oncoming flow at the foil's platform area, Eq.\ref{eq:clcd-1}:
\begin{figure}[ht!]
    \centering
    \includegraphics[width=0.6\textwidth]{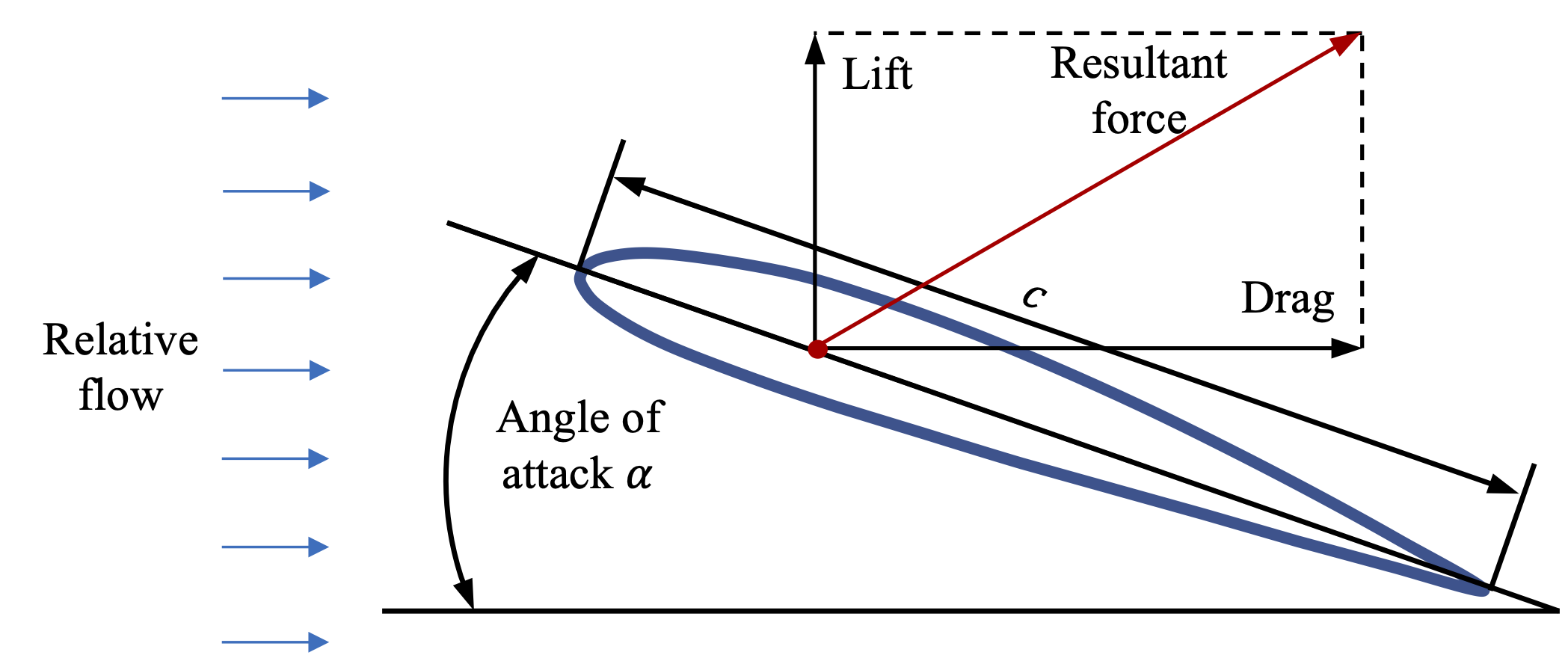}
    \caption{The forces on the foil facing flows.}
    \label{fig:hy_force}
\end{figure}
\begin{subequations}
	\label{eq:clcd-1}
	\begin{align}
		C_{l}=\frac{F_L}{\frac{1}{2} \rho U_{\infty}^{2} c} \label{eq:clcd-1:1}  \\
		C_{d}=\frac{F_D}{\frac{1}{2} \rho  U_{\infty}^{2} c} \label{eq:clcd-1:2} 
	\end{align}
\end{subequations}
where, $F_L$ and $F_D$ is the lift and drag force and $c$ is the foil chord. $F_T=-F_D$ and $C_{t}=-C_{d}$ are the thrust force and coefficient. In addition, Eq.\ref{eq:clcd-1} refers to instantaneous coefficient values and the time-averaged, e.g. over one wave period $T$, forces and coefficients ($\overline{F_D}$, $\overline{F_L}$ and $\overline{C}$) are calculated as:
\begin{equation}
	\label{eq:ave_c}
	\overline{C}=\frac{1}{T} \int_{0}^{T} C(t) \mathrm{d} t
\end{equation}
\begin{equation}
	\label{eq:ave_fx}
	\overline{F_x}=\frac{1}{T} \int_{0}^{T} D(t) \mathrm{d} t
\end{equation}
\begin{equation}
	\label{eq:ave_fy}
	\overline{F_y}=\frac{1}{T} \int_{0}^{T} L(t) \mathrm{d} t
\end{equation}
where, $\overline{C}$ can be any of $\overline{C}_{l}$, $\overline{C}_{d}$ and $\overline{C}_{t}$.

The interaction between a moving foil and the oncoming flow is described through the non-dimensional Strouhal and Reynolds numbers. The chord based $Re$ is calculated from Eq.\ref{eq:re}, while several expressions for the Strouhal number have been proposed. Typically, the peak-to-peak oscillating amplitude, $A$, of the foil's trailing edge and it's oscillating frequency, $f$, are normalised with the characteristic flow velocity, $U_{\infty}$, for estimating $St_A$, Eq.\ref{eq:st} \cite{triantafyllou1991wake}. A thickness-based Strouhal number $Sr_{D}=fD/U_{\infty}$ and a chord-length-based Strouhal number $Sr_{c}=fc/U_{\infty}$ have also been proposed by Godoy-Diana et al. \cite{godoy2009model} and Cleaver et al. \cite{cleaver2012bifurcating}, but $ St_A$ is often preferred in order to better characterise the wake pattern, i.e. RvK street, with a single factor. 
Read et al. \cite{read2003forces} reported optimum thrust generation for 0.2 $<St_A<$ 0.4.      
\begin{subequations}
	\label{eq:rest-1}
	\begin{align}
		Re&=\frac{\rho c U_{\infty}}{\mu} \label{eq:re}  \\
		St_A&=\frac{f A}{U_{\infty}} \label{eq:st}
	\end{align}
\end{subequations}
where $\rho$ is the density of the surrounding fluid, $U_{\infty}$ is the characteristic flow speed, $\mu$ is the dynamic viscosity of the fluid, $f$ is the oscillating frequency and $A$ is the oscillating amplitude, which is often taken as the width of the wake. 
For forced harmonic oscillating motions, 2$h_0$ is an estimate for the width of the foil wake, thus $A$ can be approximated with 2$h_0$, and since $f$=$\frac{\omega}{2\pi}$ Eq.\ref{eq:st} is also written as Eq.\ref{eq:st-2}.
\begin{equation}
	\label{eq:st-2}
	St_A=\frac{\omega h_0}{\pi U_{\infty}}
\end{equation}

\begin{figure}[ht!]
    \centering
    \includegraphics[width=0.5\textwidth]{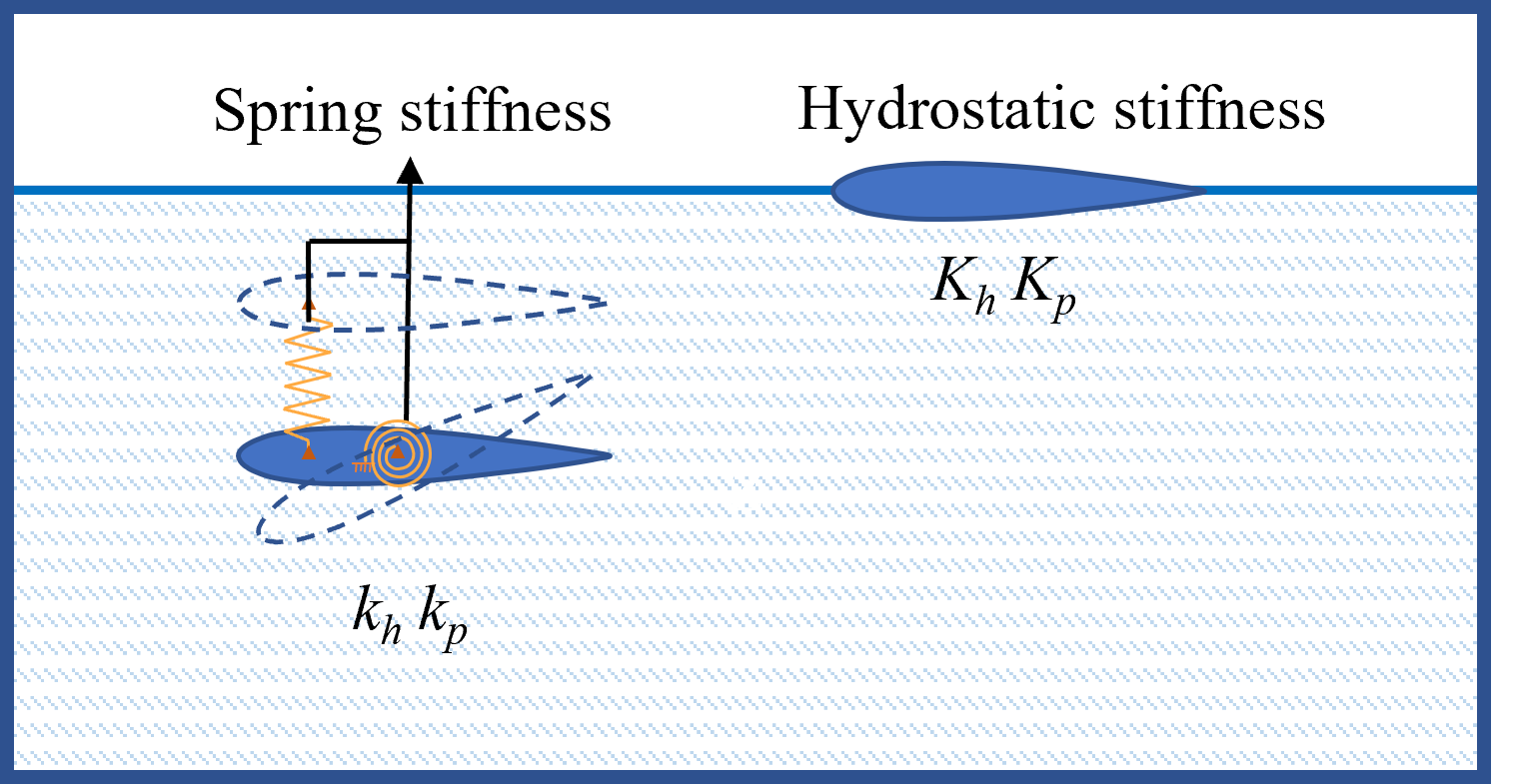}
    \caption{The heave and pitch spring stiffness (left) and the hydrostatic stiffness of the foil itself (right).}
    \label{fig:hp_hydrosp}
\end{figure}

The non-dimensional properties of the spring stiffness can be explored by the study of the hydrostatic stiffness of foil itself, see Fig.\ref{fig:hp_hydrosp}. The hydrostatic stiffness of a rigid object in still water is related to its size. A 3D NACA0015 foil model is constructed for the calculation of the hydrostatic stiffness. The span of this 3D model is considered as 1 m as it is the 3D conversion of 2D foil model. The hydrostatic stiffness of the foil in heave $K_h$ is 3794.43 N/m while it in pitch $K_p$ is 47.52 N$\cdot$m/rad, calculated by ANSYS AQWA. It can be noticed that both of the non-dimensional stiffness $k_h/K_h$ and $k_p/K_p$ in experiments are around 2.7. Therefore, only the effect of pitch stiffness is chosen as an example for discussion in the Section \ref{The effect of the pitch stiffness}.

\section{Convergence tests}
\label{Convergence tests}
\begin{figure}[ht!]
    \centering
    \begin{tabular}{cc}
        \includegraphics[width=0.4\textwidth]{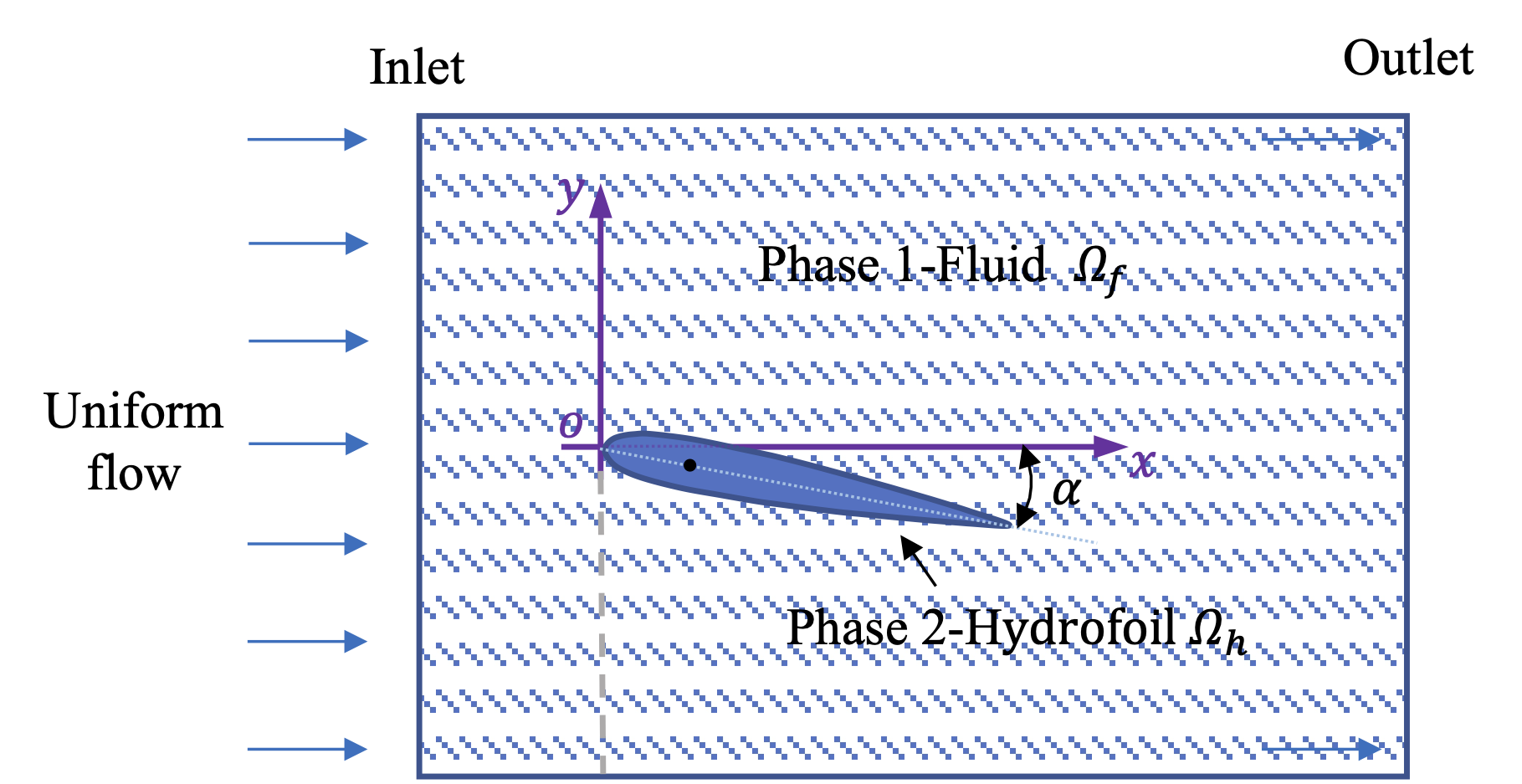} &
        \includegraphics[width=0.4\textwidth]{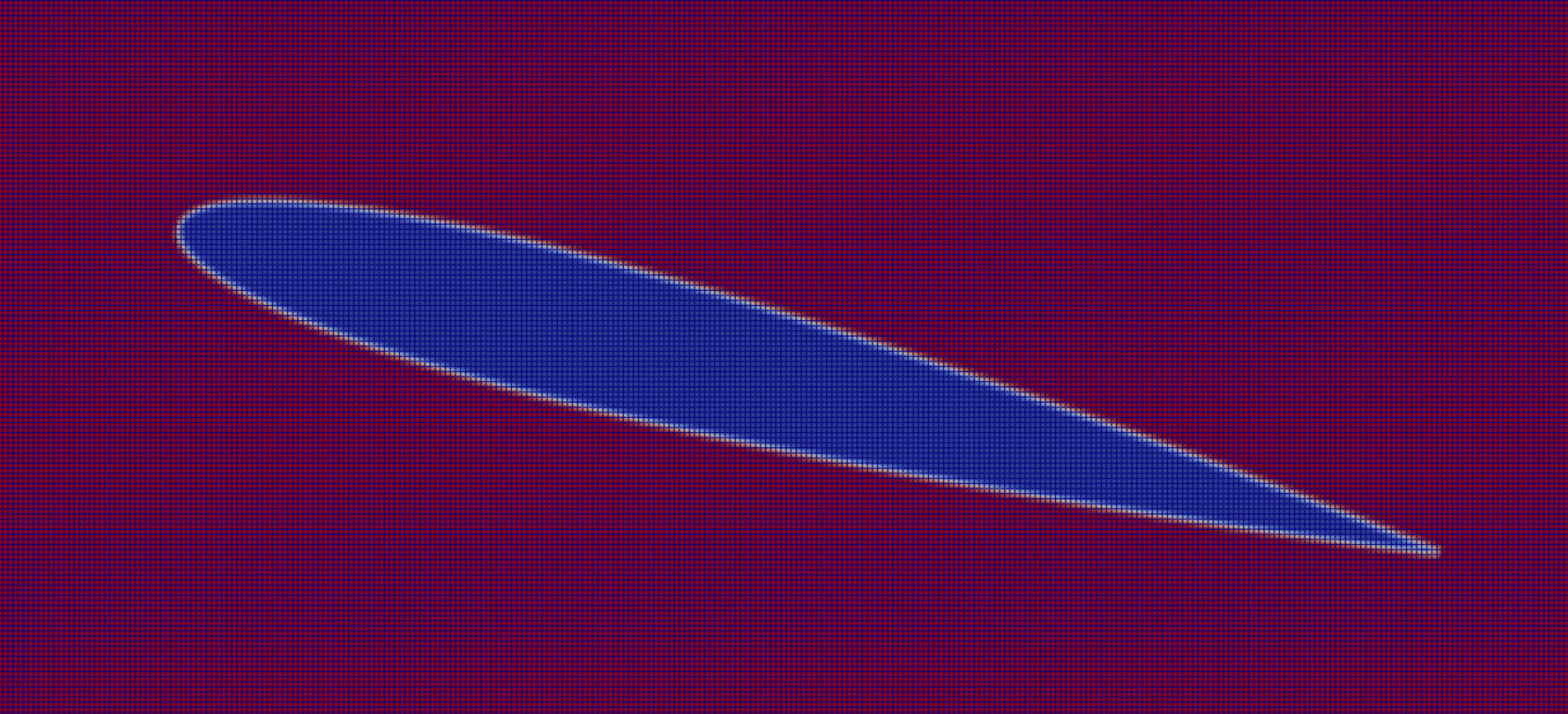} \\
    (a) schematic diagram & (b) structured grid ($\Delta x =  \frac{1}{256}$ m)
    \end{tabular}
    \caption{A schematic diagram (left) and partial structured grid (right) of the simulation for the fixed hydrofoil in uniform flow.}
    \label{fig:sch_fix}
\end{figure}

The convergence of the solution for the the drag and lift coefficients is examined here considering the case of a NACA0012 foil subject to the uniform flow, Fig.\ref{fig:sch_fix}. The dimensions of the numerical tank is 7 m$\times$7 m, and the hydrofoil is fixed at the middle of the domain and at a distance of 1m from the inlet (left hand side boundary). As shown in Fig.\ref{fig:sch_fix} the origin of the coordinates system is set at fore of the hydrofoil, directed positively from the leading (LE) to the trailing (TE) edge and from the fore point and upwards. The density of the hydrofoil is set equal to the water density $\rho$=1 $\mathrm{kg} / \mathrm{m}^{3}$, and the dynamic viscosity $\mu$ is 10$^{-3}$ $ \mathrm{kg} / \mathrm{m} \cdot \mathrm{sec}$. The velocity of the oncoming uniform flow is $U_{\infty}$=1 m/sec, the AoA $\alpha$ is set to -15$^\circ$ (positive at counterclockwise rotation), and for simplicity the hydrofoil chord length is taken as $c=$1 m. Therefore, the Reynolds number is calculated to $Re=1000$ and withing the range of naturally occurring ﬂapping foils 100$<Re<$10000 \cite{godoy2009model}.

\begin{figure}[ht!]
    \centering
    \includegraphics[width=0.7\textwidth]{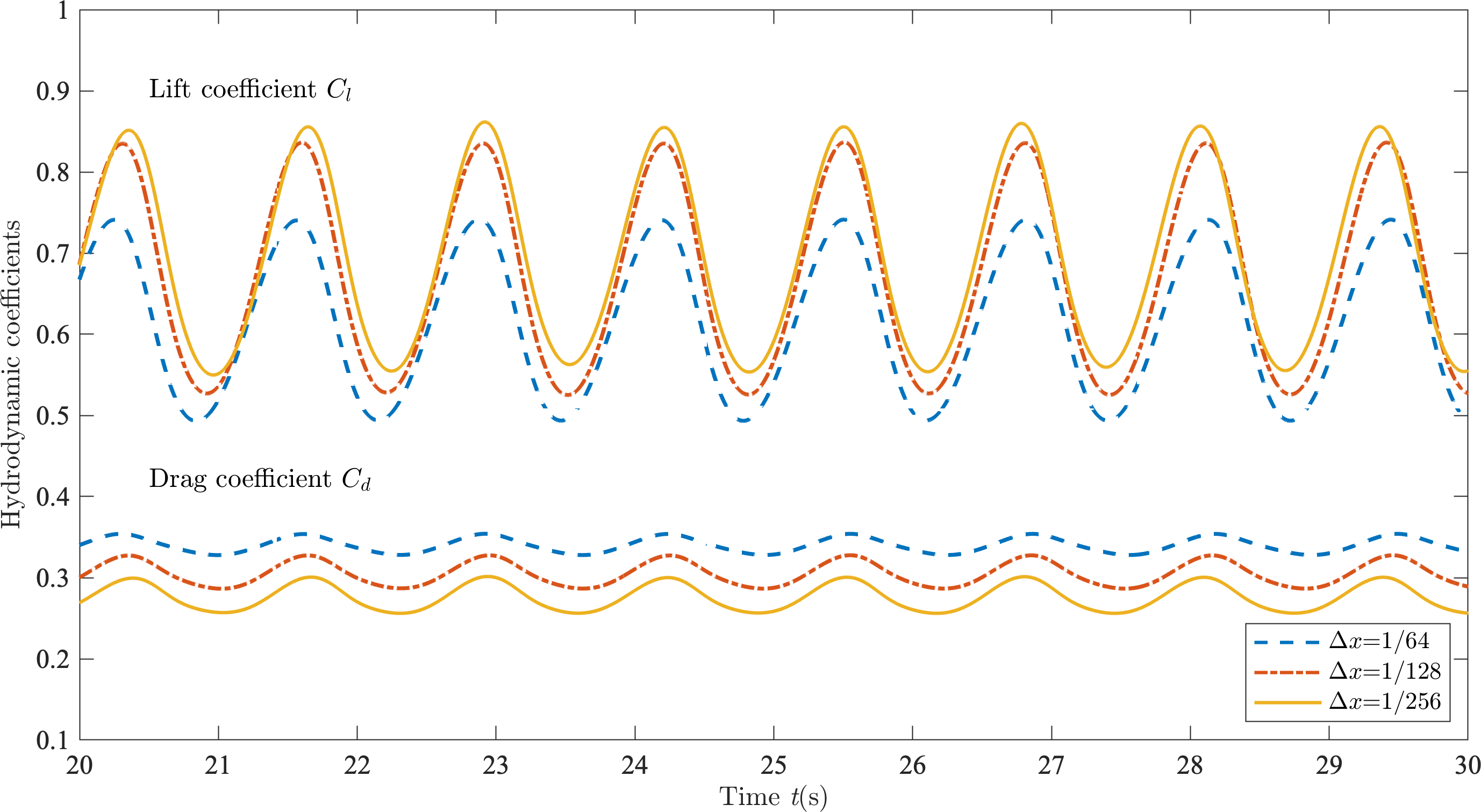}
    \caption{Validation of convergence of fixed hydrofoil facing uniform flow simulations.} 
    \label{fig:con_clcd}
\end{figure}

Convergence tests are conducted for three different mesh sizes, Table.\ref{tab:wd_grids} and the results for both $C_{l}$ and $C_{d}$ are presented in Fig.\ref{fig:con_clcd}. As $\Delta x$ decreases and the mesh is refined by a factor of two per test run- the differences between the runs reduce. As soon as the wake becomes fully developed (within the time period prior to the 20\textsuperscript{th} sec shown in Fig.\ref{fig:sch_fix}, the solution for $C_{l}$ and $C_{d}$ is seen to converge towards a harmonic sinusoidal curve of constant amplitude and period. In particular, from smaller $\Delta x$ the sinusoids become steeper since a finer mesh entails less numerical viscosity and thus less numerical damping. As expected, the hydrofoil experiences a considerably higher and more strongly variable lift than drag force, while the constantly positive $C_{d}$ indicates that the hydrofoil experiences the effect of a constant, albeit periodic drag force no thrust is generated.
\begin{table}[]
\centering
\begin{tabular}{ccc}
\hline
 &  Mesh size $\Delta x$ (m) & Numbers of cells \\ \hline
 1 & 0.0156 & $448 \times 448$ \\
 2 & 0.0078 & $896 \times 896$ \\  
 3 & 0.0039 & $1792 \times 1792$ \\ 
\hline
\end{tabular}
\caption{Grid resolution for numerical comparison of fixed hydrofoil model.}
\label{tab:wd_grids}
\end{table}

\section{Validation with experimental results}
\label{Validation with experimental results}
\subsection{Hydrofoil in uniform flow}
\label{Hydrofoil in uniform flow}
The numerical model is initially validated for the case of a flapping hydrofoil in uniform flow, using the experimental results of Read et al.  \cite{read2003forces}. For their experiments, Read et al. \cite{read2003forces} forced the harmonic (sinusoidal) oscillation in heave and pitch of a towed NACA0012 foil and reported on measurements of lift and thrust forces, and pitching torque for a wide range of AoA and $\psi$. The towing speed was set at 0.4 m/sec, the hydrofoil chord length were $c$=0.1 m and hence the Reynolds number of the physical tests was $Re$=4$\times$10$^{4}$. These experiments are reproduced in a 2D numerical tank with dimensions 0.5 m$\times$0.6 m with a layout similar to that of Fig.\ref{fig:sch_fix}(a). In contrast with the physical tests the numerical hydrofoil was not towed but it was set in a uniform flow with a velocity equal to the towing velocity, $U_{\infty}$=0.4 m/sec. $c$ was also 0.1 m and as previously explained the hydrofoil density was taken to be the same with the water/fluid density. The displacement in heave and pitch was forced through the application of Eq.\ref{eq:heavepitch_tra} with $h_{0}$=0.075 m and $\psi$=-90$^\circ$, thereby achieving the same trajectory for the hydrofoil as in the experiments enabling $\alpha_{eff}(t)$ to evolve symmetrically, see also Read et al. \cite{read2003forces}. The pivot point $P$ was located at the chord and at a distance of $c/3$ from the LE and the initial AoA was set to $\alpha$=-15$^\circ$. Finally, for the experiments and the numerical simulations, 0.08 $\leq St_A \leq$ 0.44 and 0.21 $\leq f \leq$ 1.17 Hz.

\begin{figure}[ht!]
    \centering
    \includegraphics[width=0.8\textwidth]{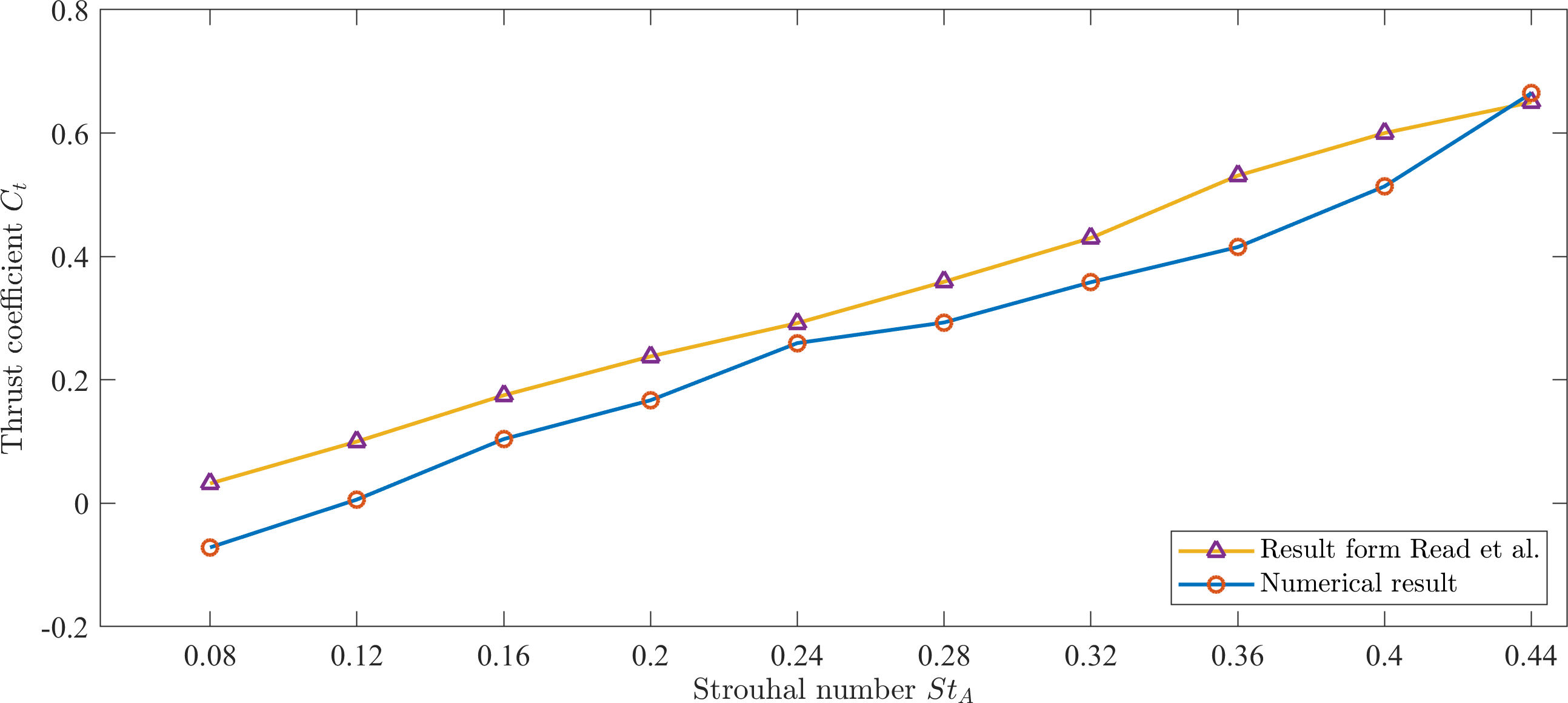}
    \caption{The comparison between the experiment (the upper line) and active hydrofoil simulation ($\Delta x = \frac{1}{1280}$ m).}
    \label{fig:case2_ct}
\end{figure}

The numerical results are compared with the experiments in Fig.\ref{fig:case2_ct}, where the thrust coefficients $C_t$ are plotted over the amplitude-based Strouhal number $St_A$ calculated using Eq.\ref{eq:st-2}. In accordance with its experimental counterpart, the numerical $C_{t}$ increases with $St_A$. However, for all $St_A$ the computed thrust coefficient is constantly smaller than the $C_{t}$ in the physical test, with the sole exception for $St_A$=0.44. This is perhaps due to the limitations of using 2D simulations as e.g. reported in Mittal and Balachandar's study \cite{mittal1995effect} where 2D simulations were shown to result in erroneous forces and wake patterns; nonetheless, exploring the effects of 3D simulations goes beyond the scope of the present work. 

\begin{figure}[!ht]
 \centering
 \begin{tabular}{ccccc}
    \multicolumn{5}{c}{\includegraphics[width=0.5\textwidth]{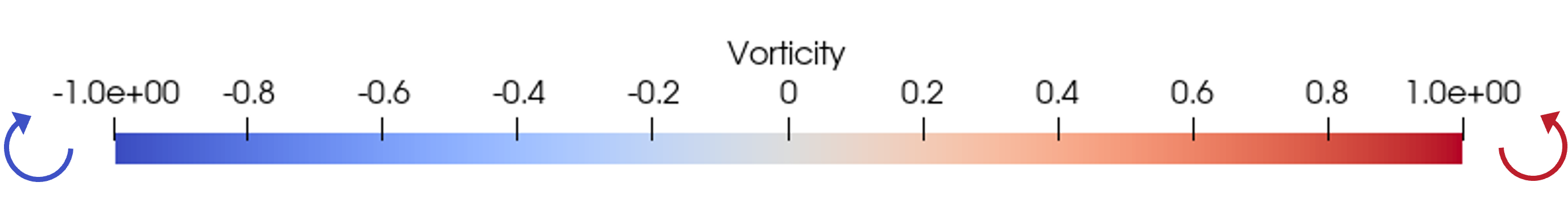}} \\
    \includegraphics[width=0.17\textwidth]{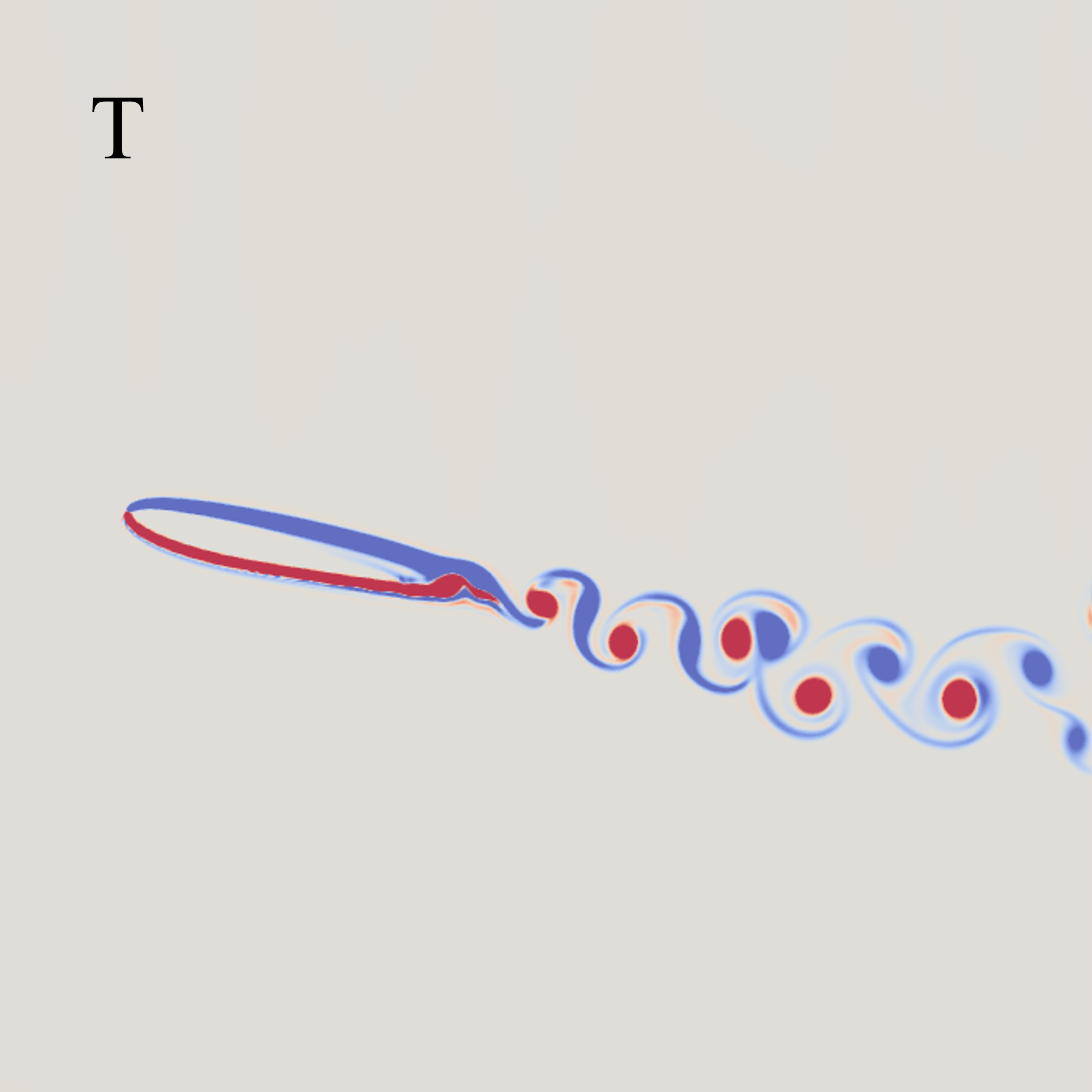} &
    \includegraphics[width=0.17\textwidth]{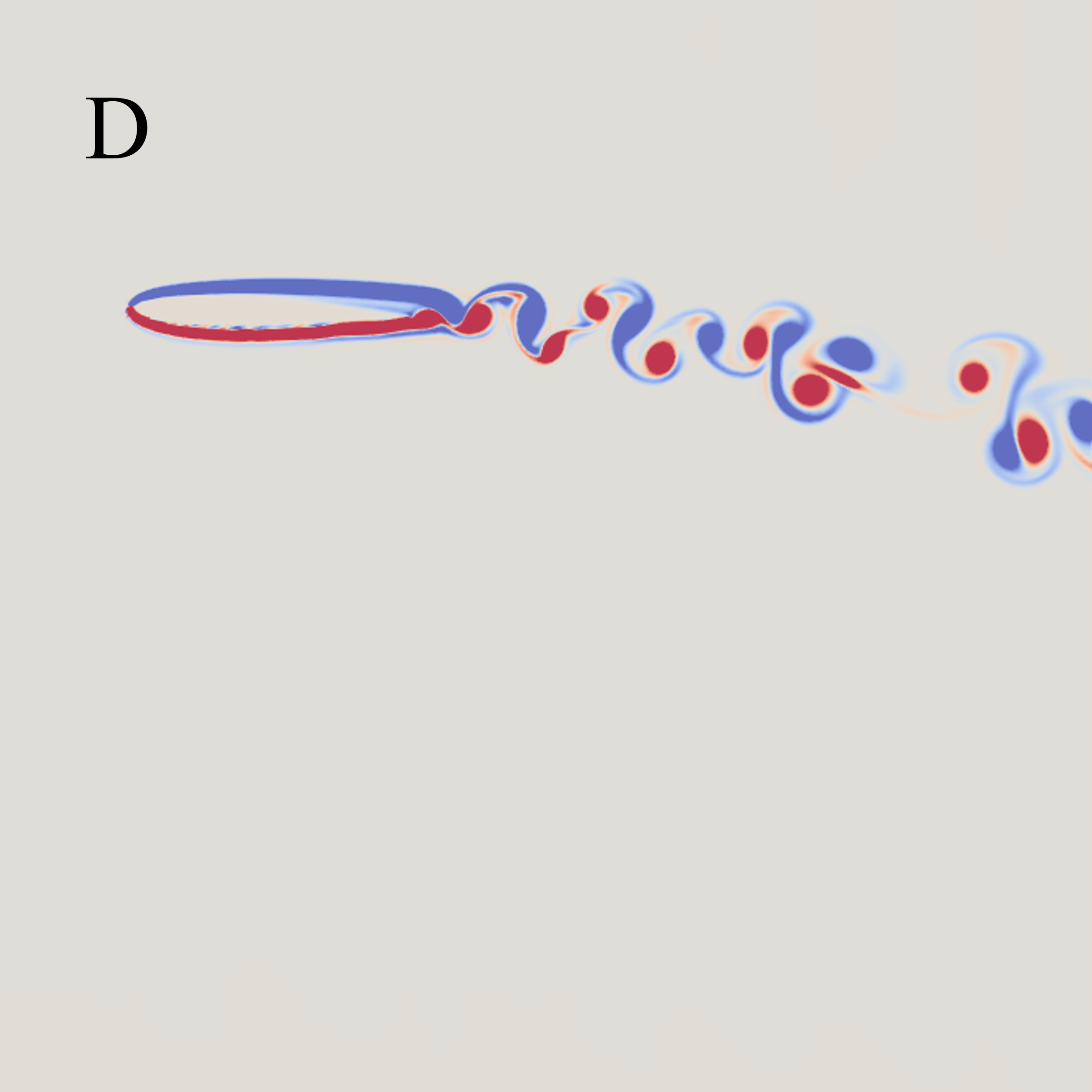} &
    \includegraphics[width=0.17\textwidth]{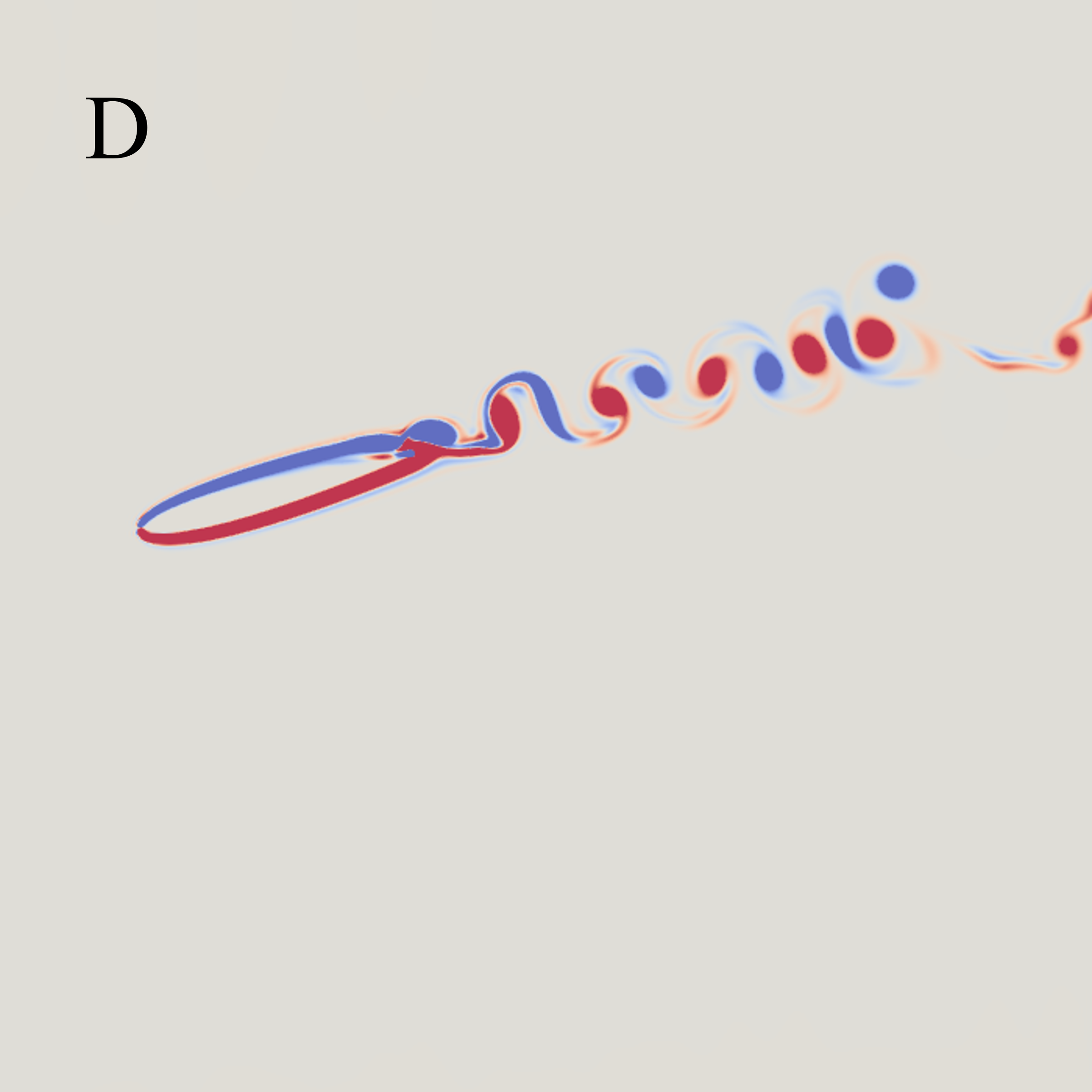} &
    \includegraphics[width=0.17\textwidth]{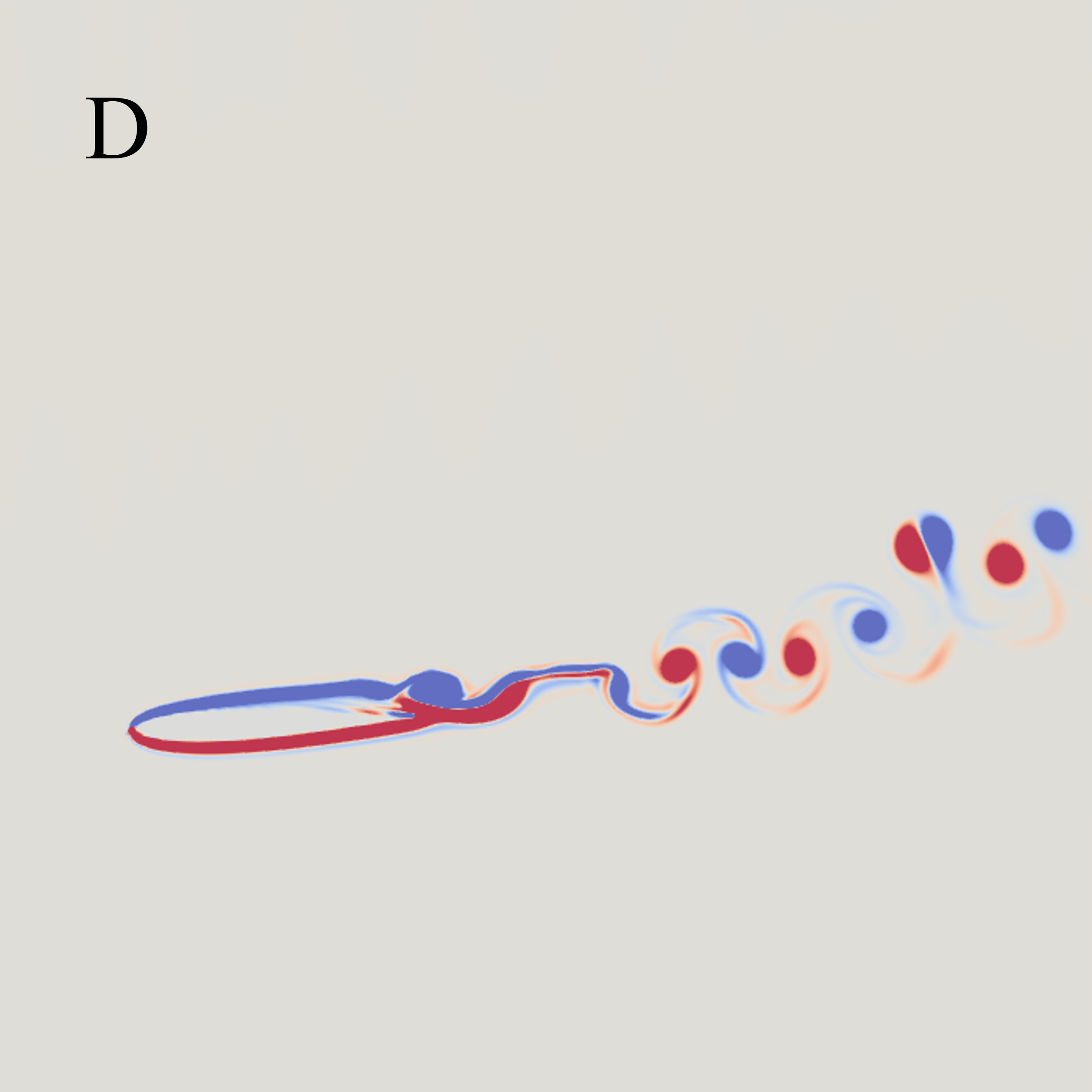} &
    \includegraphics[width=0.17\textwidth]{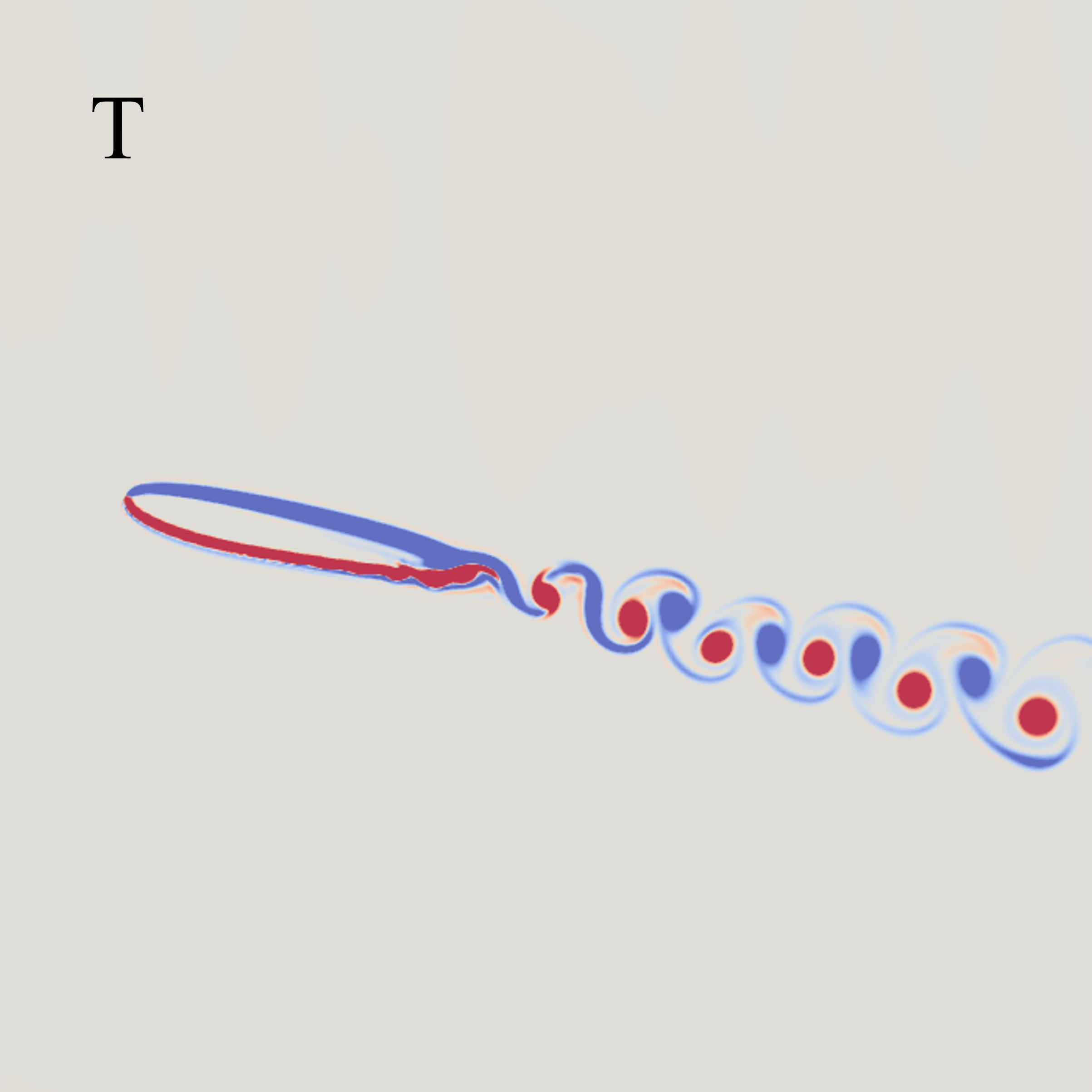} \\
    \includegraphics[width=0.17\textwidth]{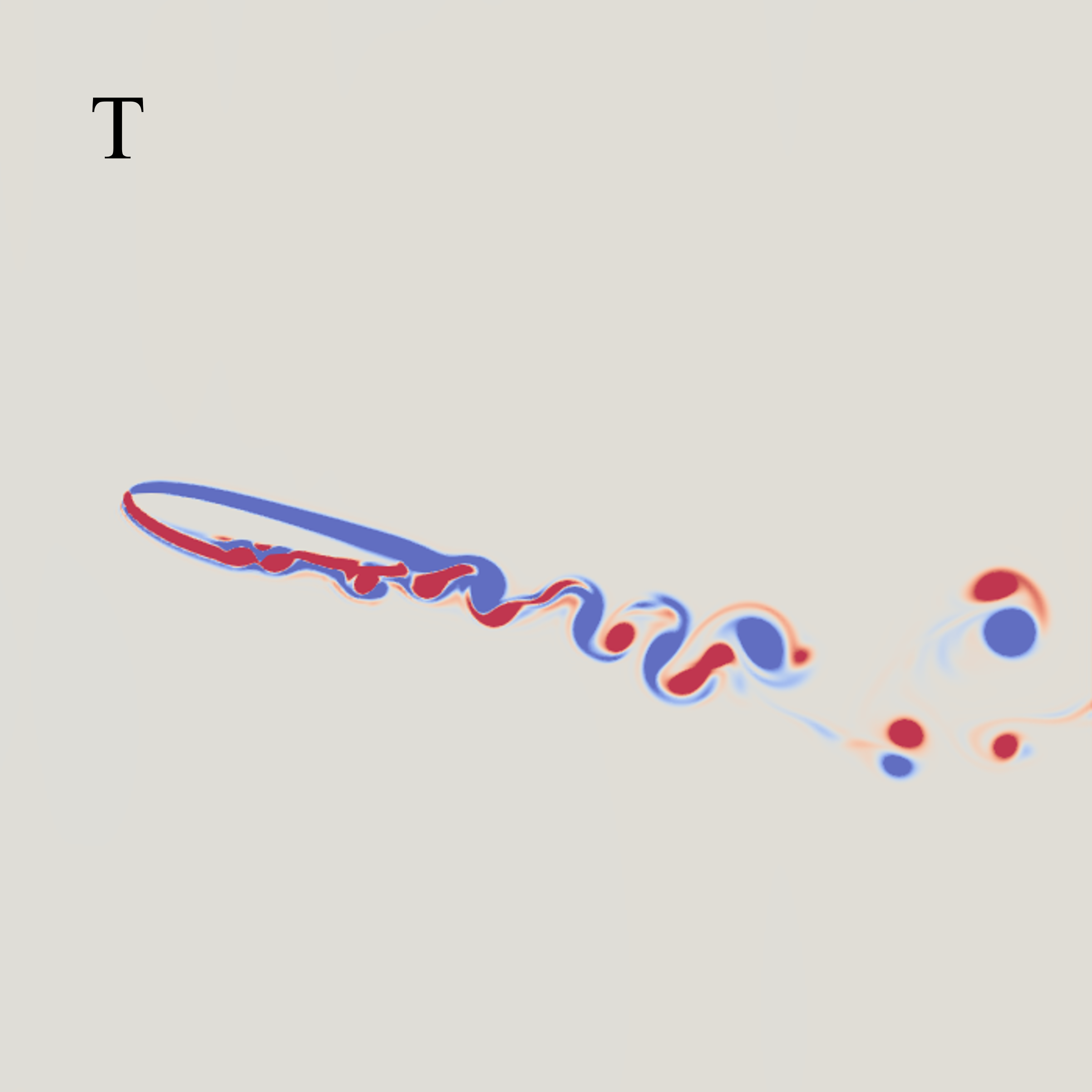} &
    \includegraphics[width=0.17\textwidth]{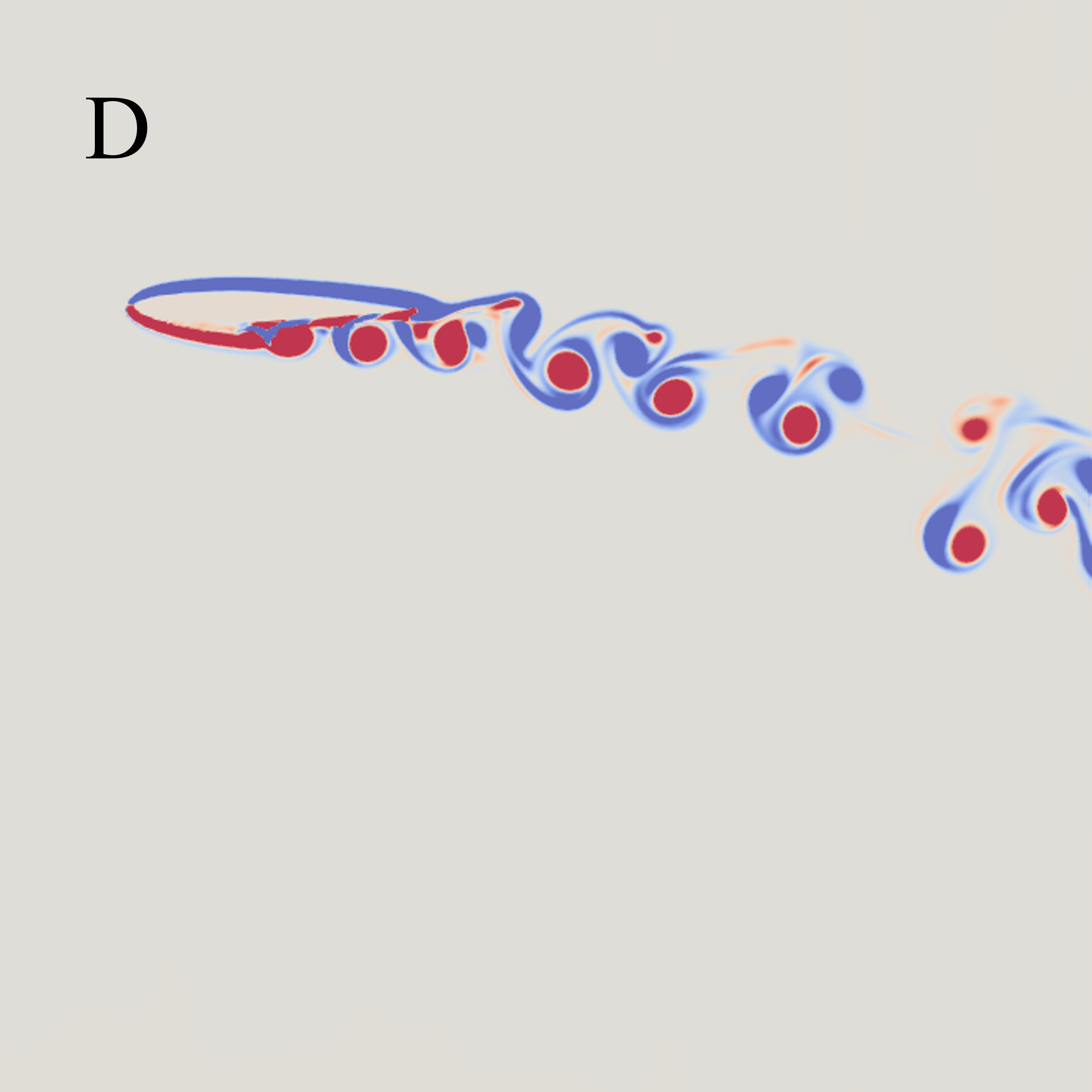} &
    \includegraphics[width=0.17\textwidth]{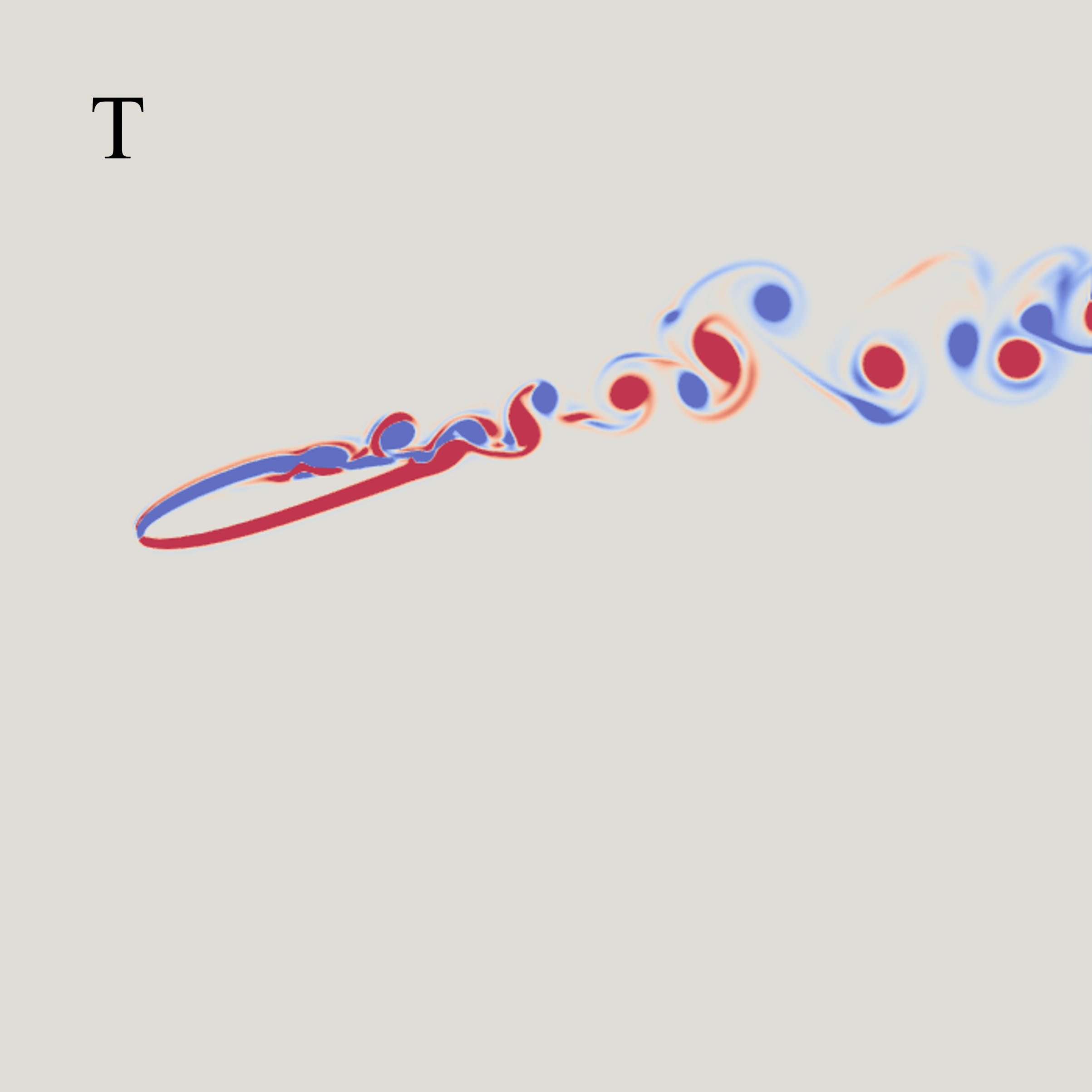} &
    \includegraphics[width=0.17\textwidth]{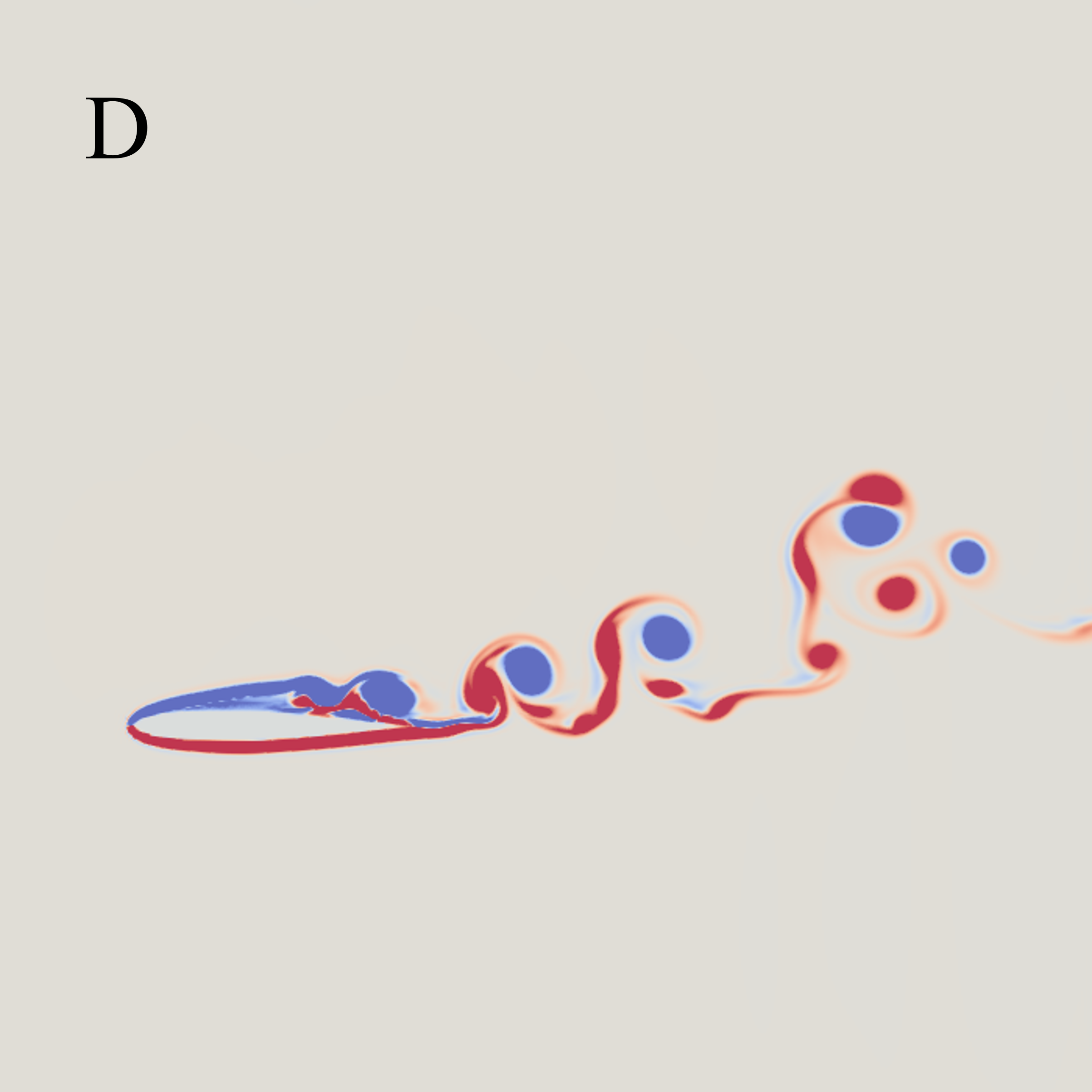} &
    \includegraphics[width=0.17\textwidth]{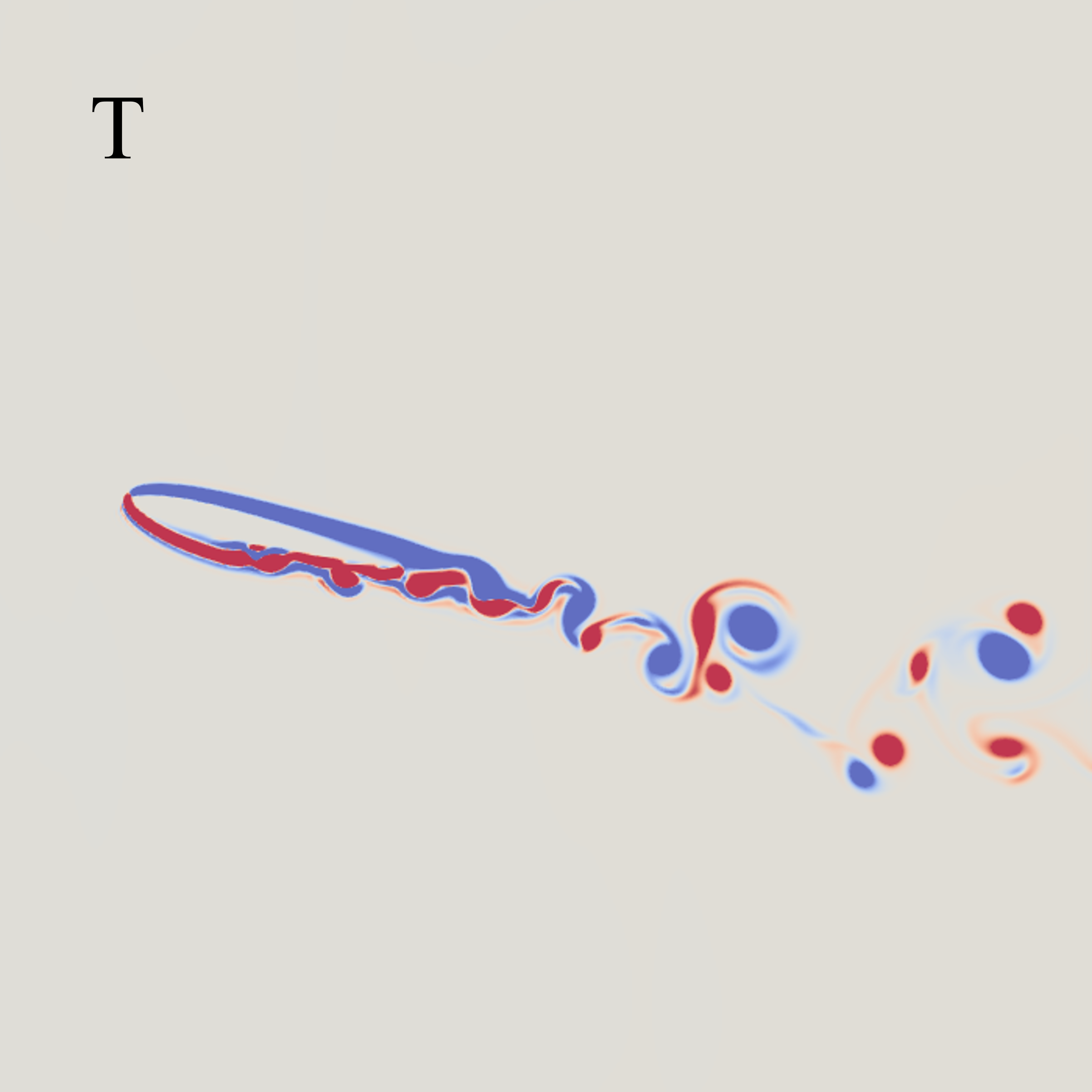} \\
    \includegraphics[width=0.17\textwidth]{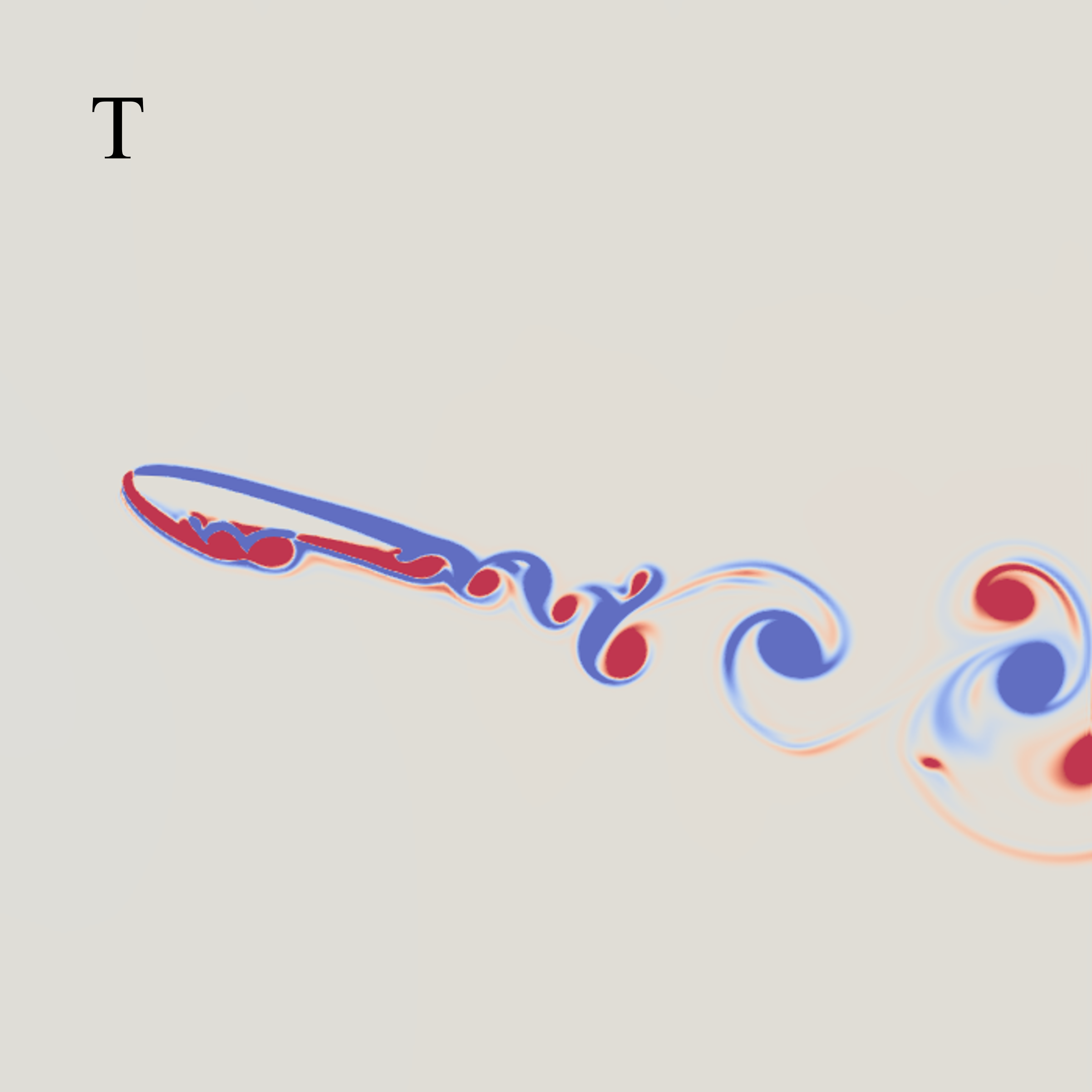} &
    \includegraphics[width=0.17\textwidth]{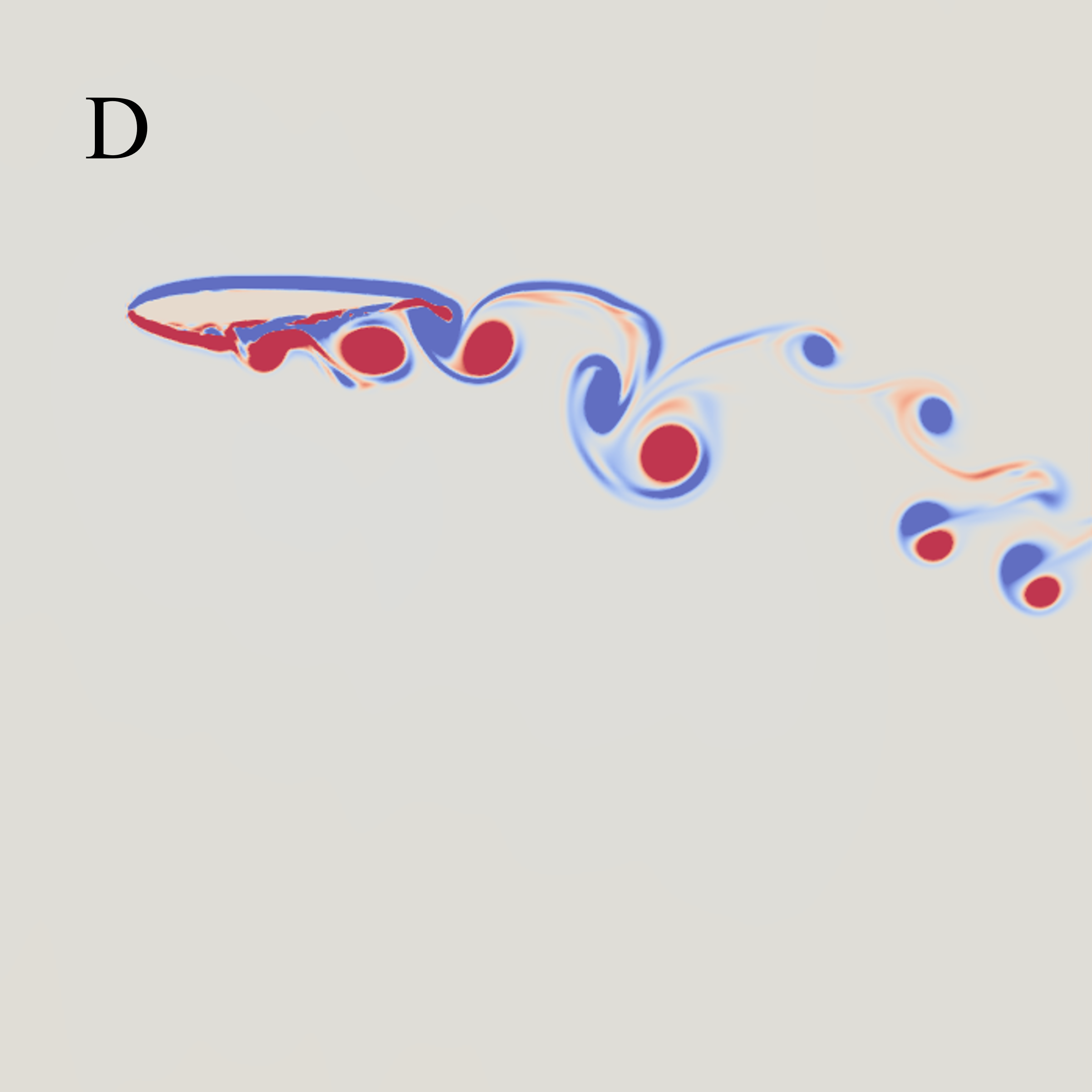} &
    \includegraphics[width=0.17\textwidth]{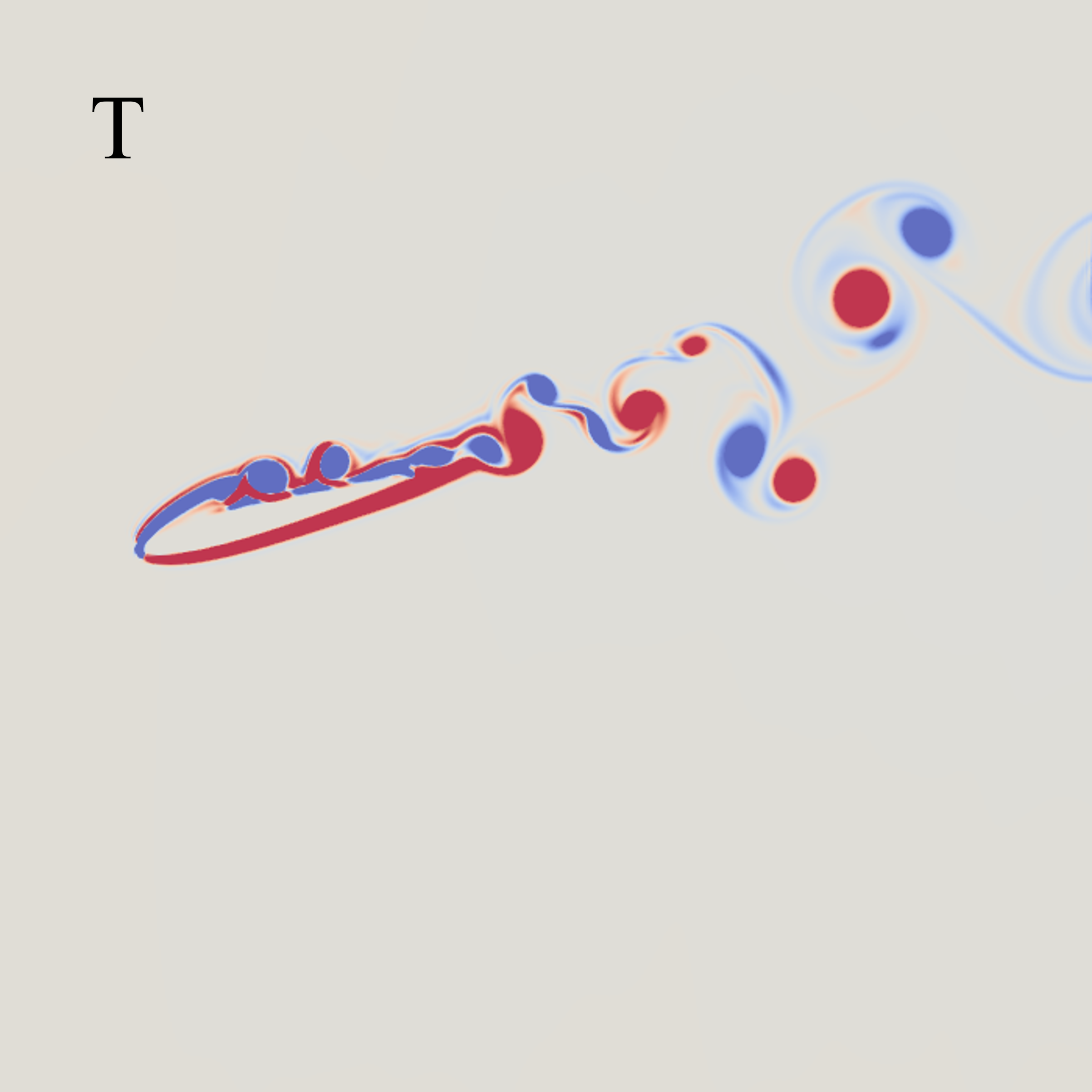} &
    \includegraphics[width=0.17\textwidth]{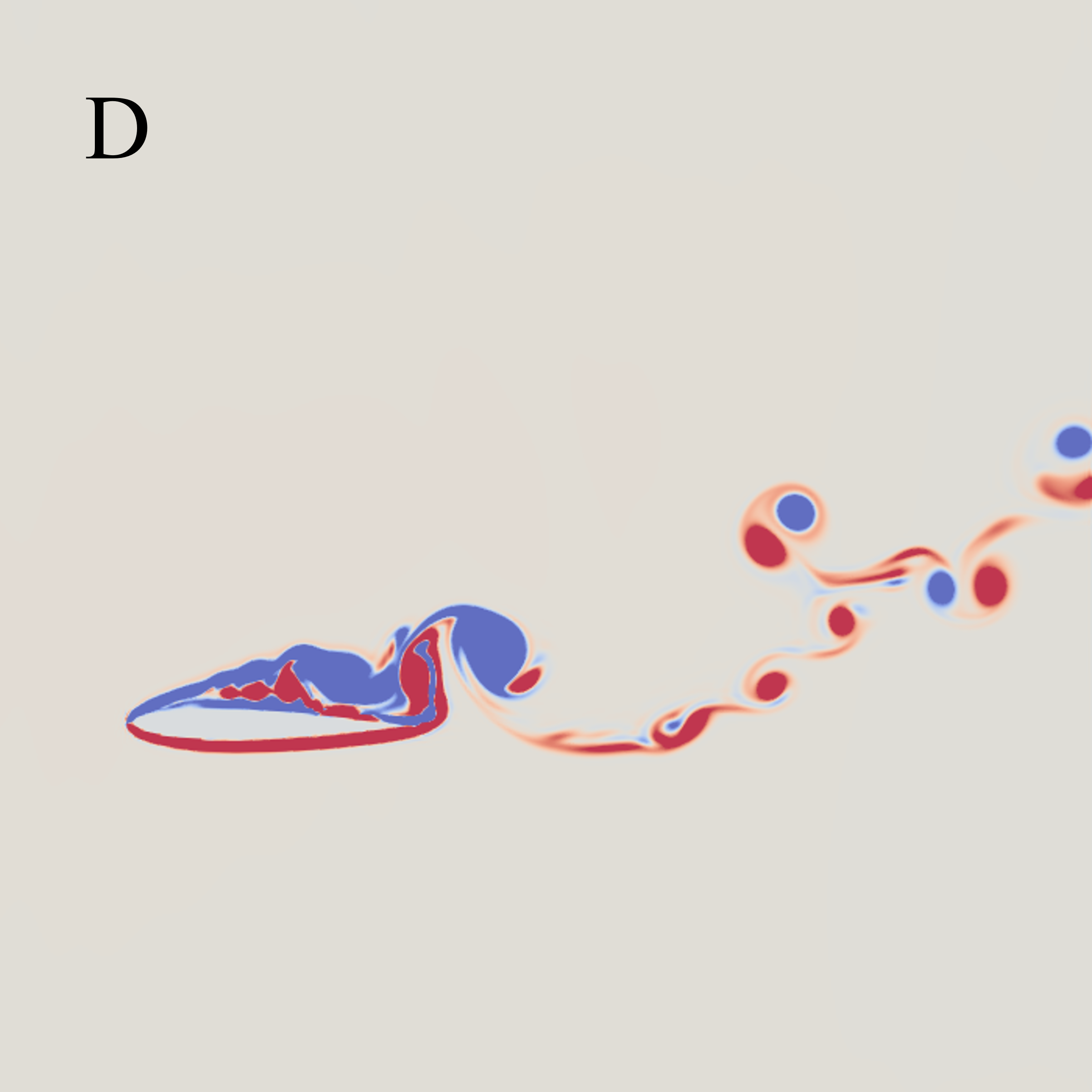} &
    \includegraphics[width=0.17\textwidth]{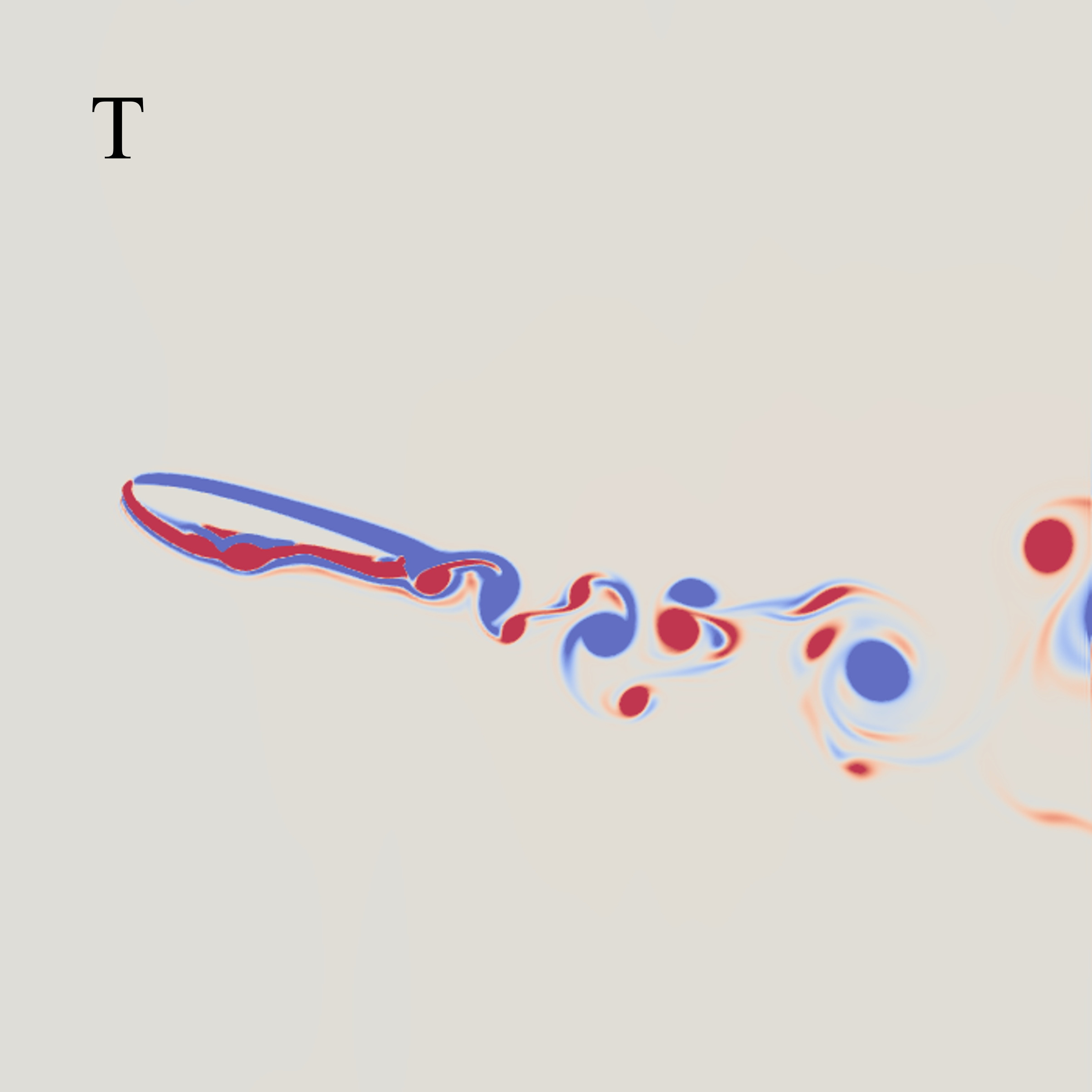} \\
    \includegraphics[width=0.17\textwidth]{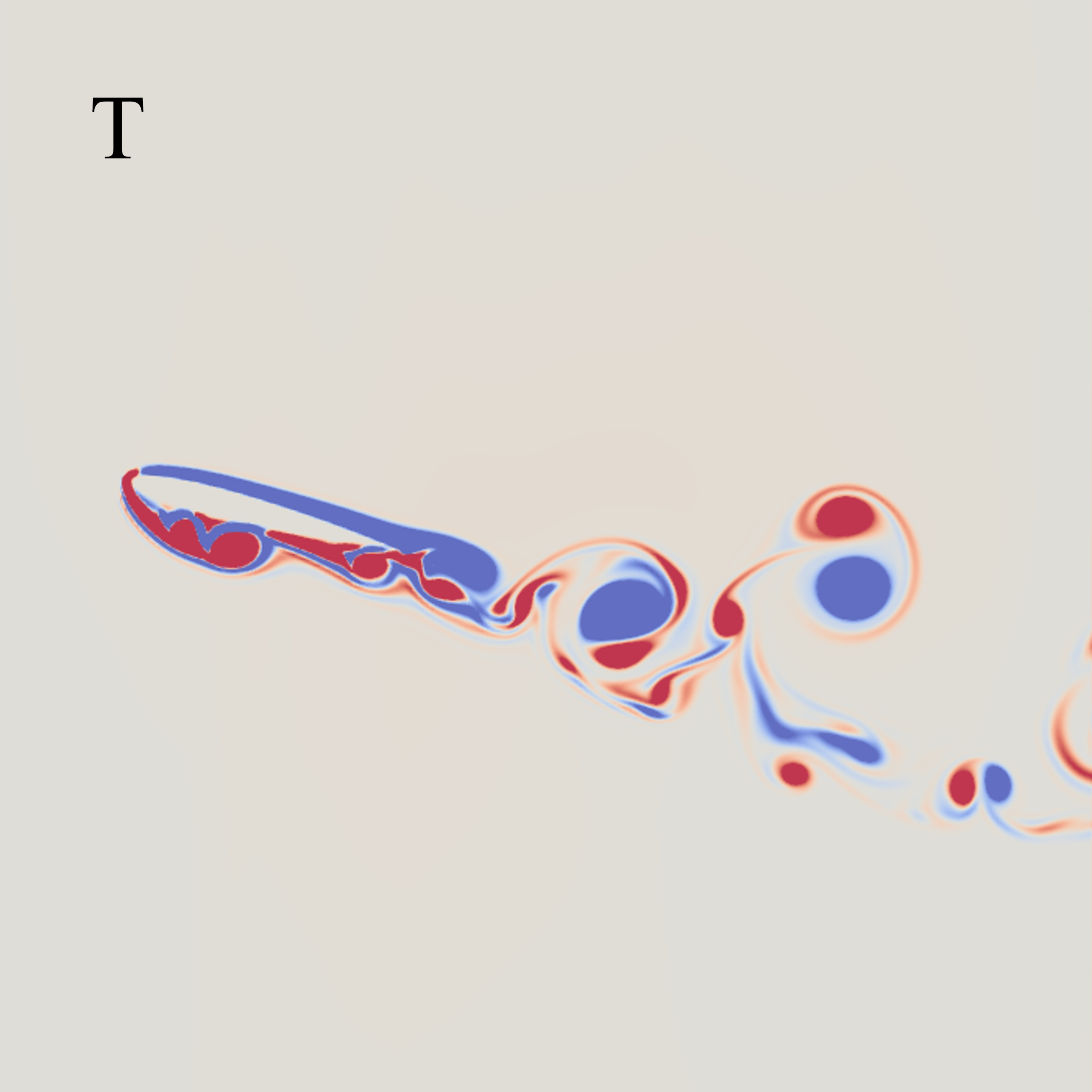} &
    \includegraphics[width=0.17\textwidth]{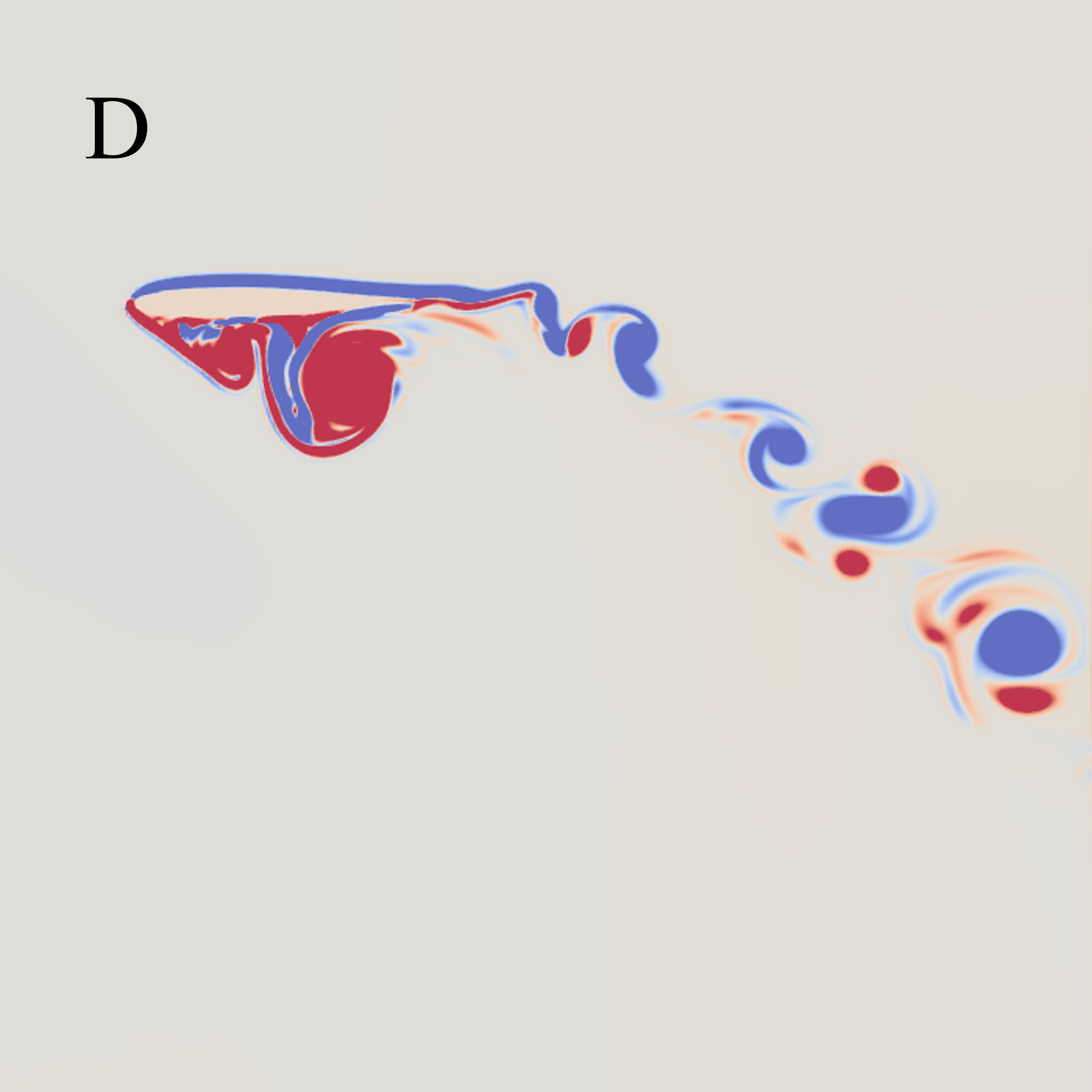} &
    \includegraphics[width=0.17\textwidth]{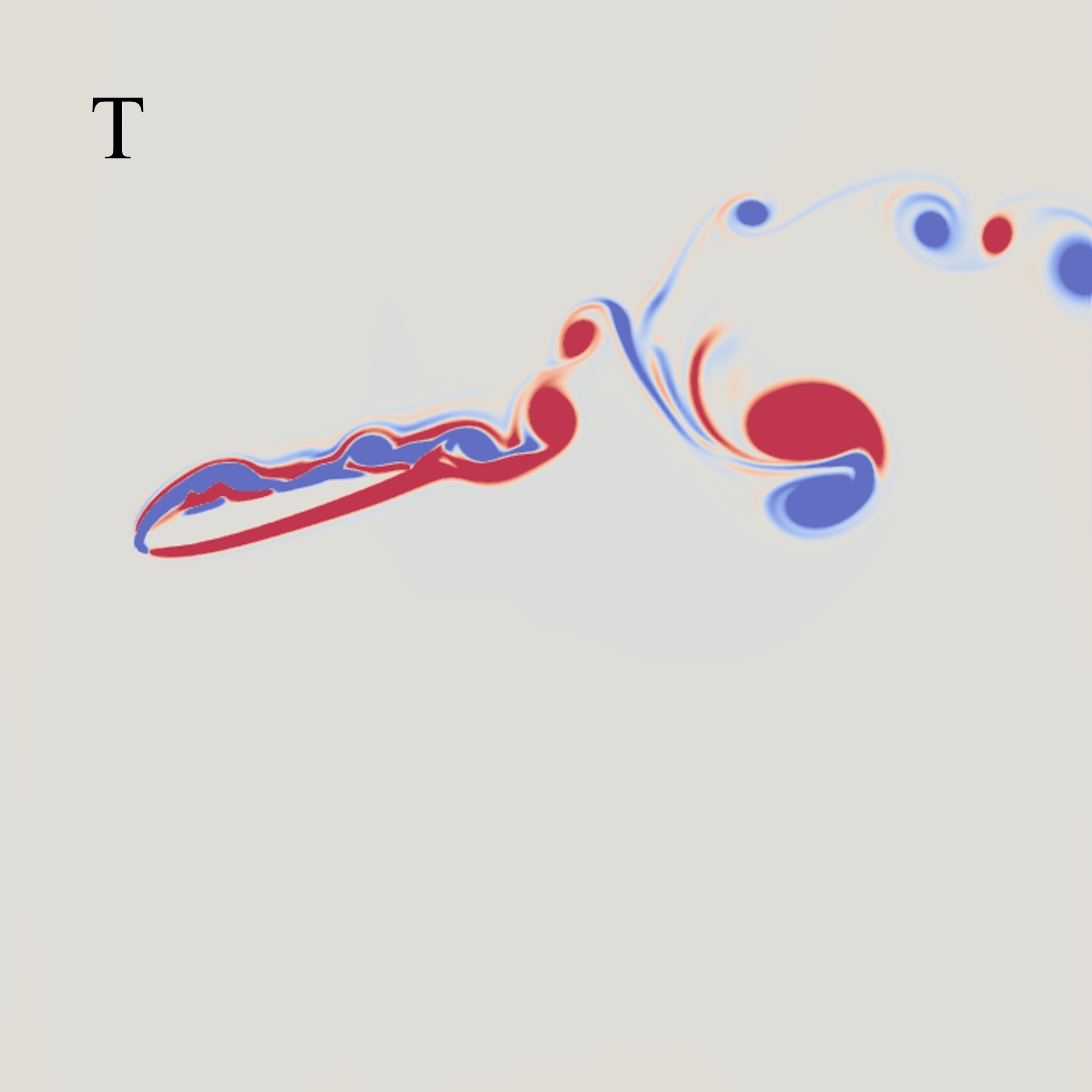} &
    \includegraphics[width=0.17\textwidth]{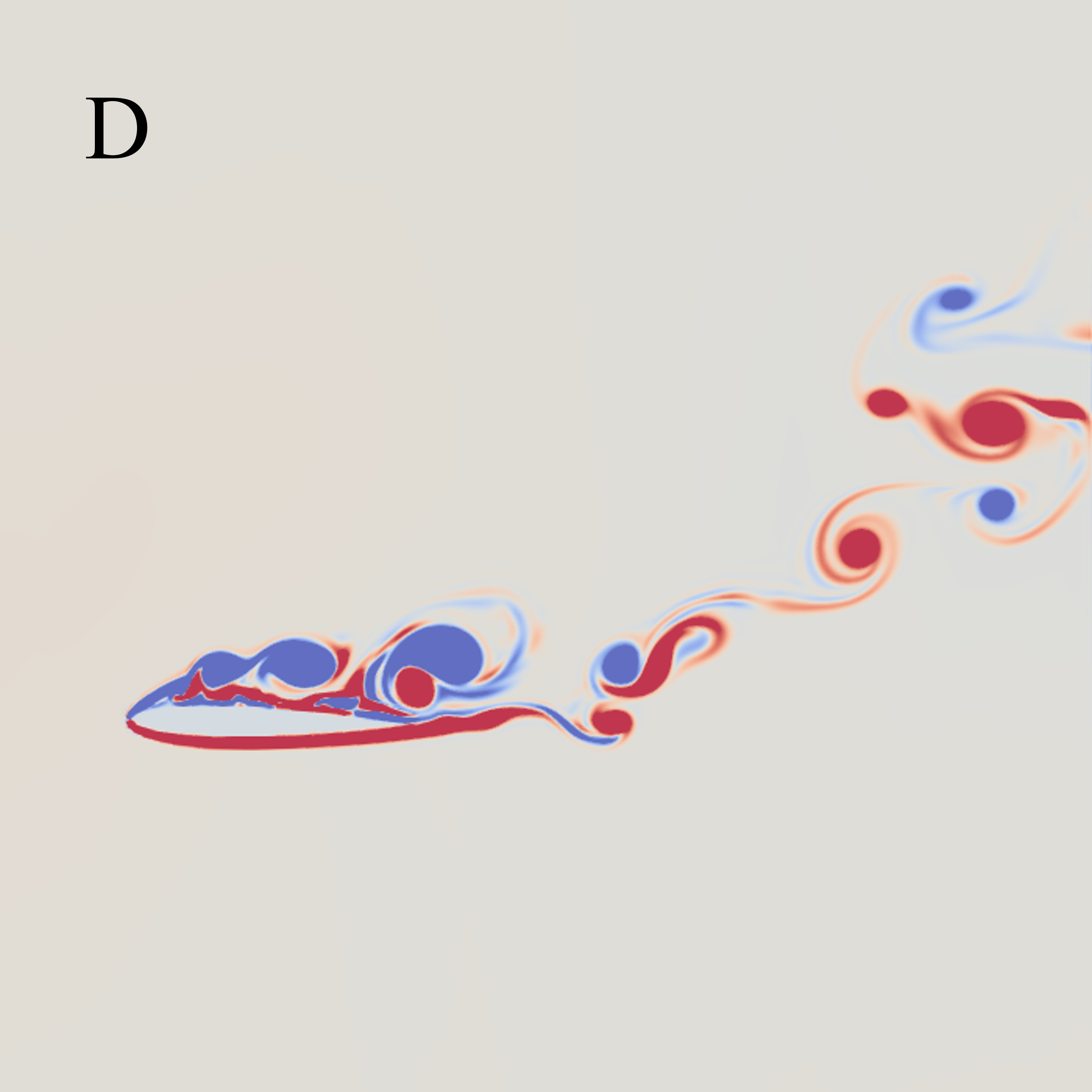} &
    \includegraphics[width=0.17\textwidth]{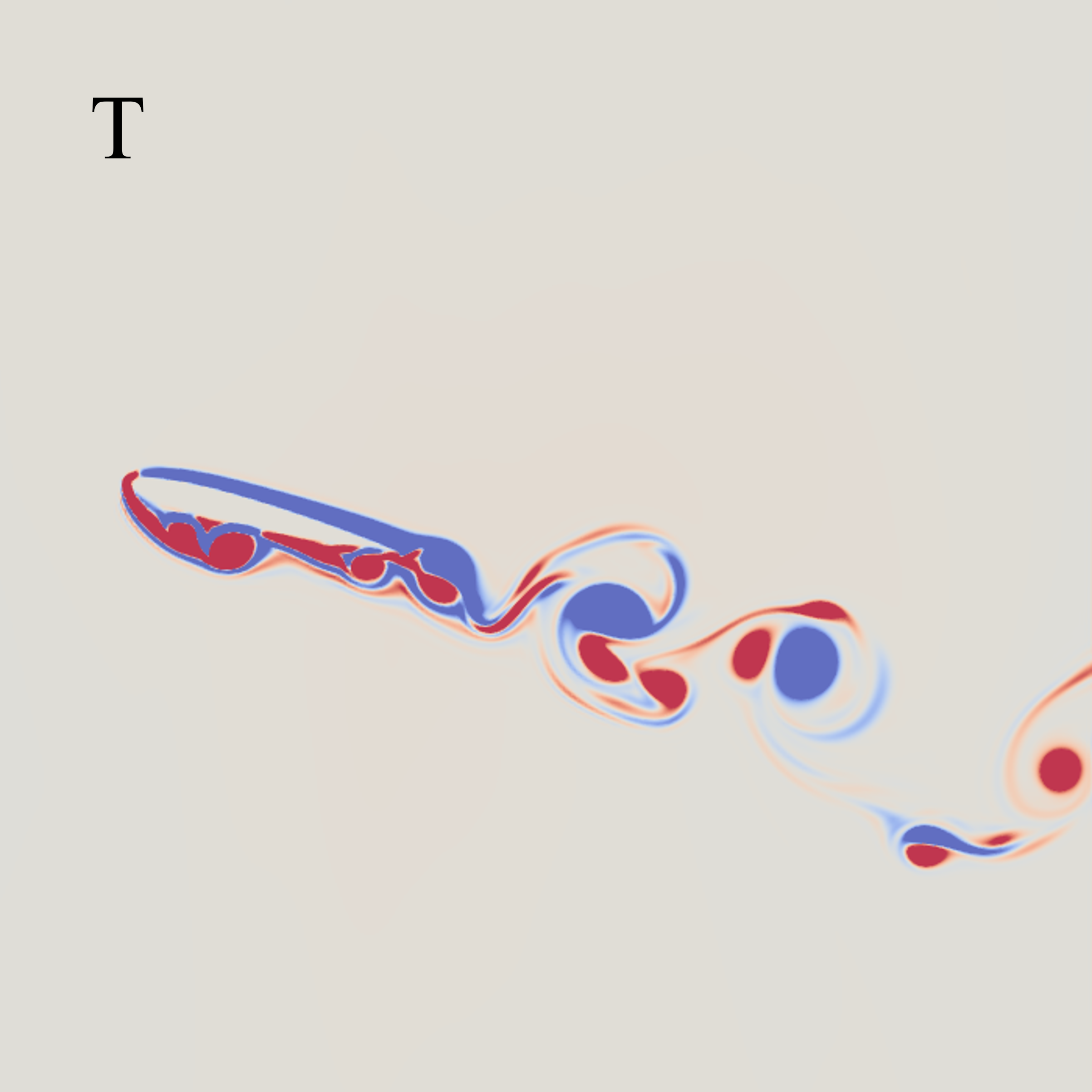} \\
    \includegraphics[width=0.17\textwidth]{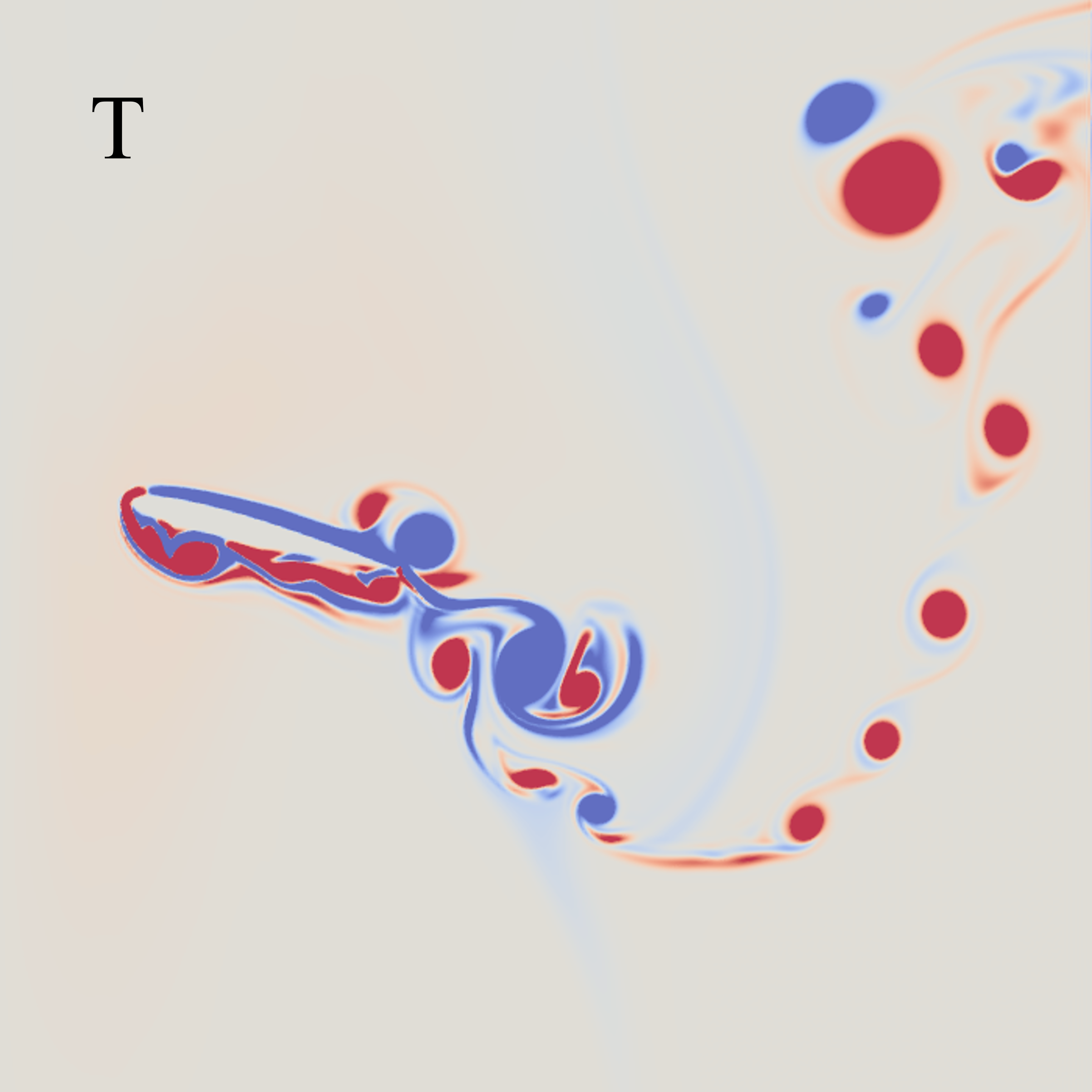} &
    \includegraphics[width=0.17\textwidth]{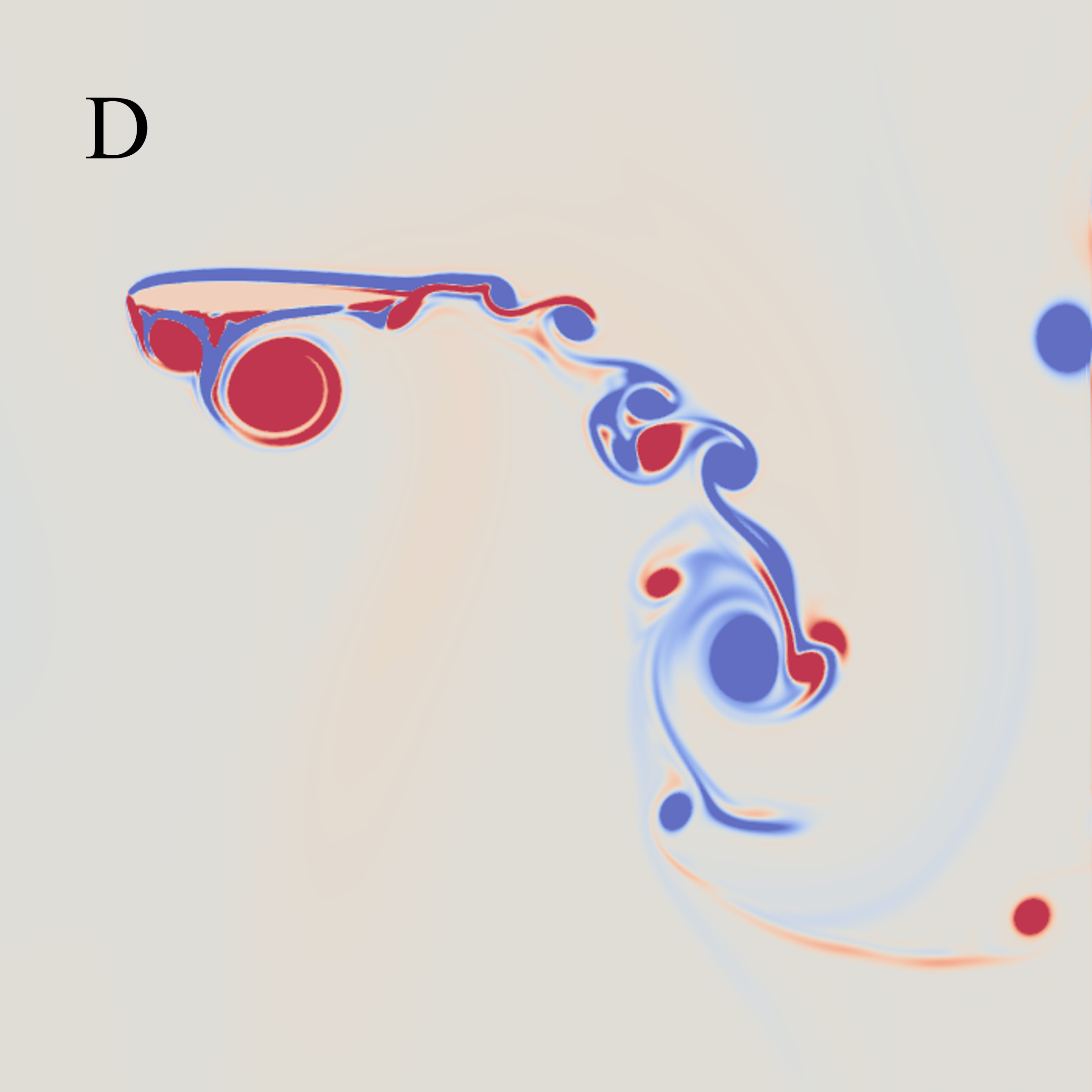} &
    \includegraphics[width=0.17\textwidth]{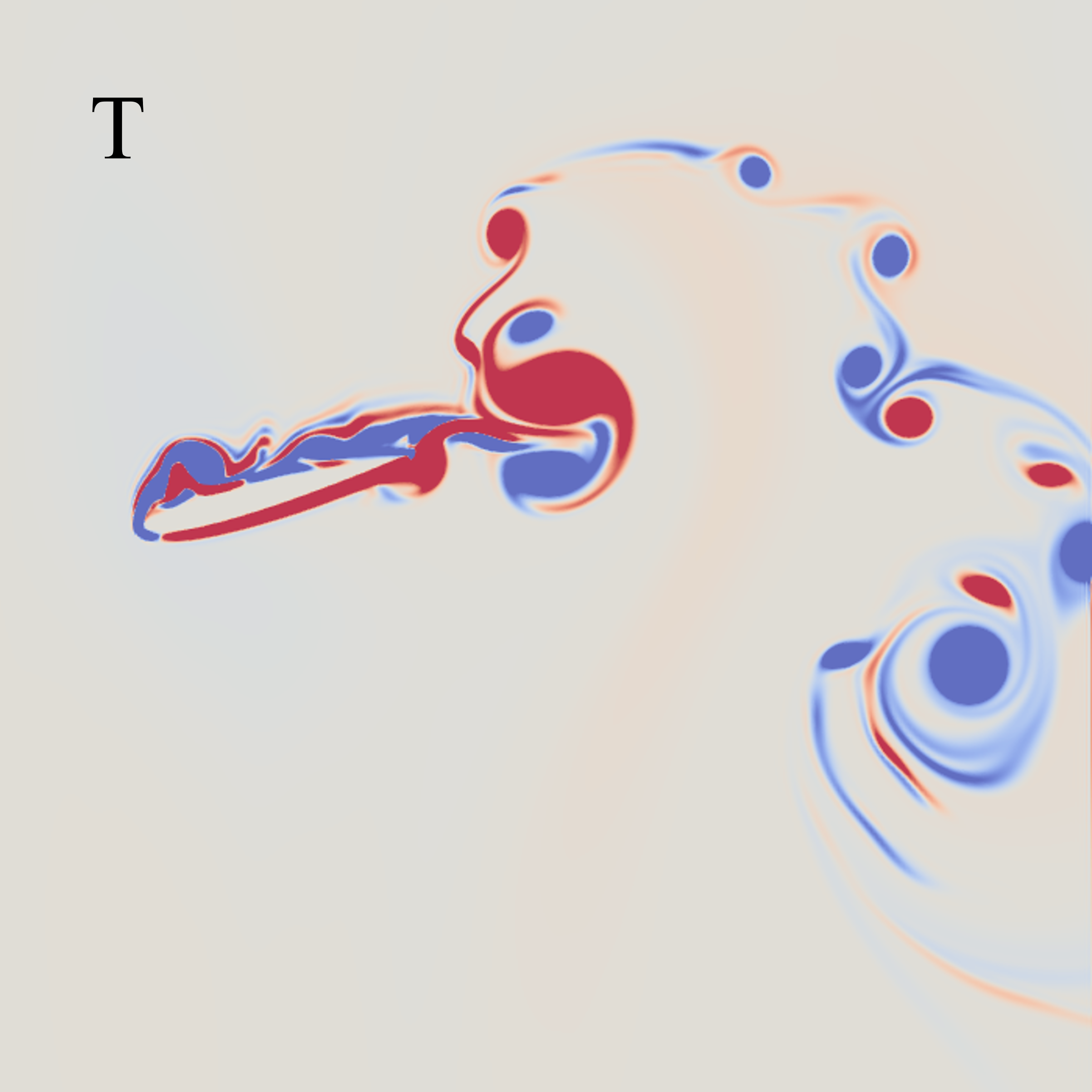} &
    \includegraphics[width=0.17\textwidth]{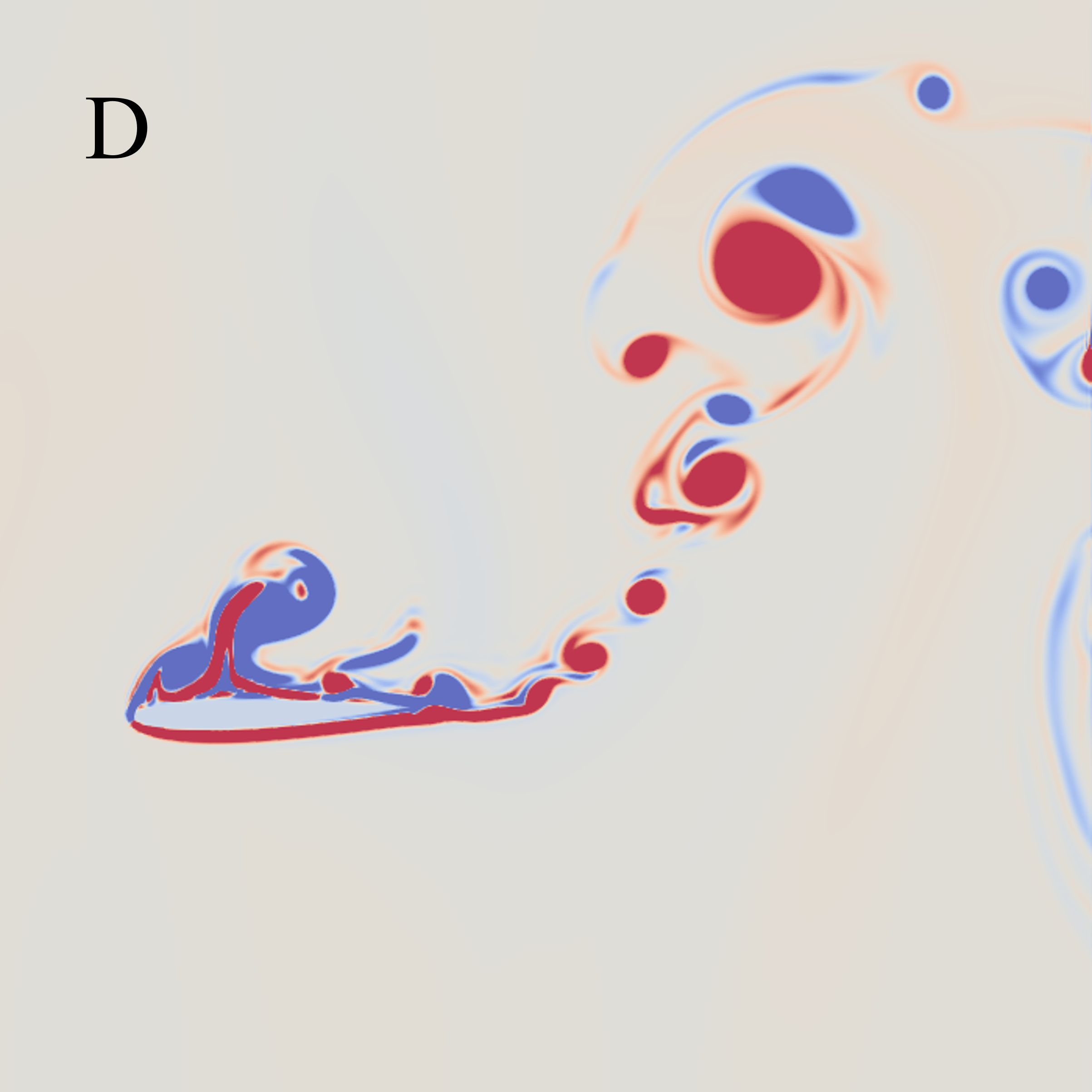} &
    \includegraphics[width=0.17\textwidth]{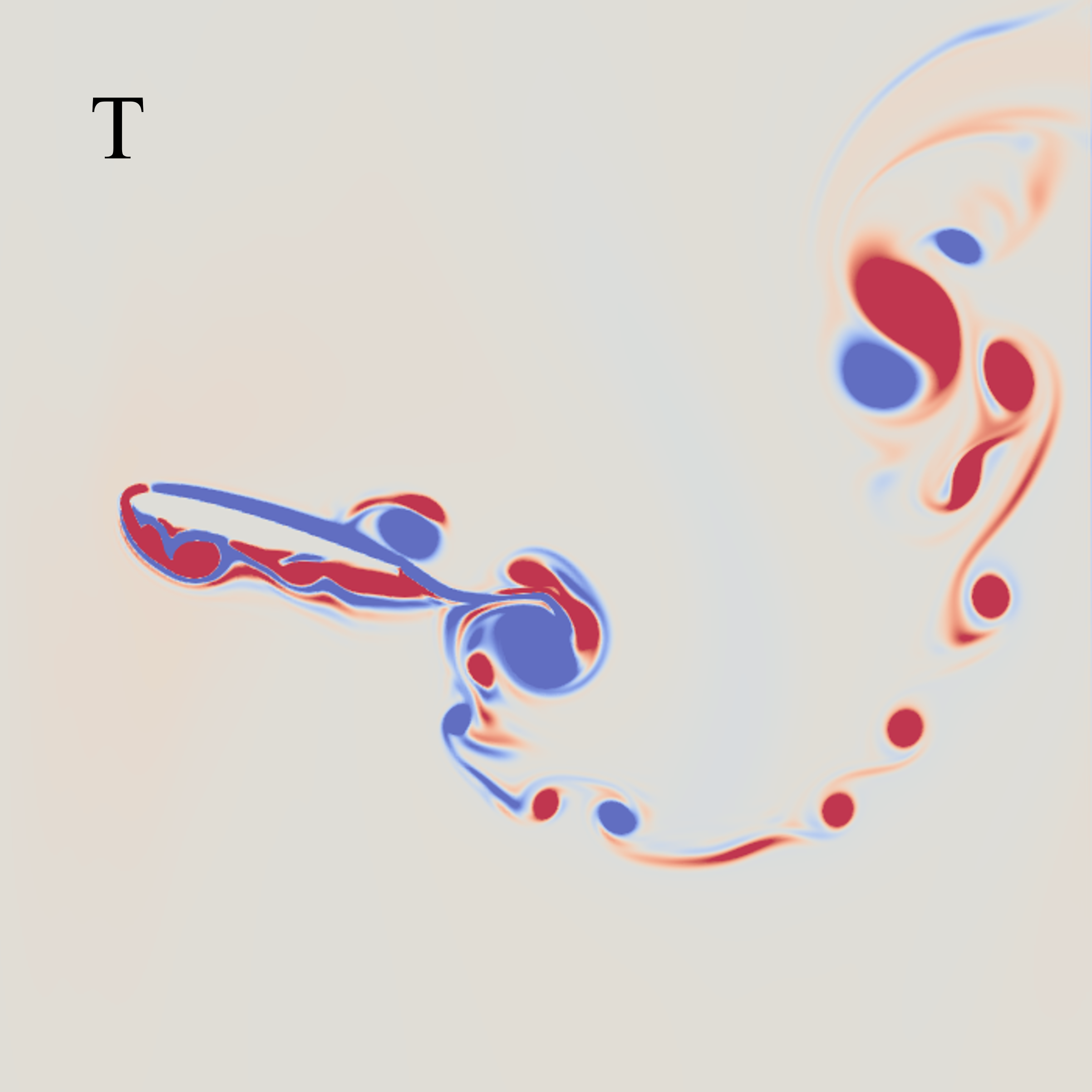} \\
    (a) $t_{0}$ & (b) $t_{0}+\frac{1}{4}T$ & (c) $t_{0}+\frac{1}{2}T$ & (d) $t_{0}+\frac{3}{4}T$ & (e) $t_{0}+T$
    \end{tabular} 
    \caption{Snapshot series of the motion of active hydrofoil and its wake in one cycle ($St_A$=0.08($1^{st}$ row), 0.12($2^{nd}$ row),0.16($3^{rd}$ row), 0.24($4^{th}$ row), 0.44($5^{th}$ row), $\Delta x =  \frac{1}{1280}$ m. The force status is marked in the top left corner, where D is for drag and T is for thrust.).}
    \label{fig:case2_vortex}
\end{figure}

Since $St_A$, e.g. Eq.\ref{eq:st-2}, can be perceived as the ratio between the hydrofoil trailing edge velocity and the velocity of the flow $U_{\infty}$, then for the same uniform flow the results of Fig.\ref{fig:case2_ct} indicate that an intensification in the foil motion (higher $A$ and thus increased $St_A$) leads to greater motion-induced thrust (higher $C_{t}$). This trend is supported by the analysis of various stages of the wake development for different $St_A$s of Fig.\ref{fig:case2_ct}, presented in Fig.\ref{fig:case2_vortex}. The instantaneous forces on the foil are indicated for the cases shown in all the snapshots.

As a whole, during one cycle of the foil's motion, it generates forces that transition between thrust and drag several times. At $t_{0}$, $t_{0}+T$and $t_{0}+1/2T$, (see Fig.\ref{fig:case2_vortex} (a), (c) and (e) columns) the pivot position of the foil overlaps with its initial position, i.e. the velocity in the heave direction is zero and the value of the AoA in the pitch direction is maximum (=15$^\circ$), and thrust is always generated, except for the case when $St_A$=0.08 and time=$t_{0}+T$. At $t_{0}$, $t_{0}+1/4T$and $t_{0}+3/4T$, (see Fig.\ref{fig:case2_vortex} (b) and (d) columns) the pivot position of the foil deviates the most from the initial position, i.e. the acceleration in the heave direction is zero and the value of AoA in the pitch direction is minimum (=0$^\circ$), and drag is always generated.

Specifically, at time $t_{0}$ for $St_A$=0.08 ($1^{st}$ row) the formation of a neutral wake is observed ($1^{st}$ and $4^{th}$ columns), where vortices are shed in-line occurs and the foil experiences a small thrust force (about 0.2N). Then, from $t_{0}+1/4T$ (second column) and onward a transition to a vK vortex street occurs and drag is generated. At $t_{0}+1/2T$, the drag reaches the highest, about 2N. This wake evolution pattern is consistent with boarder-line negative $C_{t}$ (=-0.0716) calculated from the computations for one oscillation time period $T$. For $St_A$=0.12 ($2^{nd}$ row), neutral states are observed throughout the whole period, while neither significant drag nor thrust is generated. This also explains the simulation results for $C_{t}$=0.006. For $St_A$=0.16 ($3^{rd}$ row), a neutral street develops soon after the initial observation of a RvK street at $t_{0}$ and the transition repeats each quarter period until next $t_{0}$. However, the drag generated in the neutral state is only one third of the thrust generated in the RvK state, resulting in the generation of thrust and a positive value for $C_{t}$=0.104. This transition from the vK, the neutral to the RvK is in agreement the literature, e.g. Lagopoulos et al. \cite{lagopoulos2019universal} and Andersen et al. \cite{andersen2017wake}. Further increase in $St_A$ leads to a similar drag-to-thrust transition but with more complicated wake patterns, as depicted in the forth and fifth row of Fig.\ref{fig:case2_ct}. For the higher $St_A$, increased heave amplitudes are known to lead to the generation of leading edge vortices the size of which is comparable to (for $St_A$=0.16) and bigger then (for $St_A$=0.24 and 0.44) the trailing edge vortices; see for example the $3^{rd}$, $4^{th}$ and $5^{th}$ contour plot at the $2^{nd}$ column. For all these cases, the leading edge vortices travel along the hydrofoil and blend with the wake. Overall, the more intense the flapping motion becomes, the stronger the wake becomes and increasing values of thrust are produced. Although not in full quantitative agreement with the experimentally calculated thrust coefficient, the computations present a satisfactory qualitative agreement not only with the experimental results but with the literature as well. As such the computations are considered representative of the flapping hydrofoil's behavioural trends for a uniform flow. Having established this, the case of a hydrofoil exposed in the action of waves is now considered.
        
\subsection{Hydrofoil in waves}
\label{Hydrofoil in waves}
Computations of the wave-hydrofoil interaction are evaluated against the physical model tests of Isshiki et al. \cite{isshiki1983theory, isshiki1984theory}. Using a 25 m$\times$1 m$\times$0.71 m wave flume, the authors conducted self-propulsion tests with a hydrofoil exposed to the action of regular waves. For these experiments, the foil was mounted on a towing carriage and the wave induced, time-averaged, thrust was estimated from the difference between the foil-free and the foil-equipped carriage resistance. The experiments verified the generation of thrust through absorption of wave energy by a passive-type hydrofoil and measurements of the time-averaged advance foil speed were also reported for regular wave lengths ranging from approx. 1.5 m to 9 m.
The simulations were conducted in a Numerical Wave Tank (NWT), the general arrangement of which is shown in Fig.\ref{fig:sch_wavefoil}. However, it is noted that the NWT length varied with respect to the length of the waves generated. Specifically, for the wave conditions summarised in Table \ref{tab:wave_con}, the NWT dimensions was set at 10 m$\times$1.3 m for the smaller wave lengths of group I and to 20 m$\times$1.3 m for the longer waves of group II; by varying the length of the numerical domain improved the time efficiency of the computations. All the domains considered a total of three phases, air ($\Omega_a$), water ($\Omega_w$) and the hydrofoil ($\Omega_r$). As in the experiments, a NACA0015 foil with $c$=0.4 m was placed at distance of $h_1$=-0.077 m below the air-water interface, and the restoring force and moment of Eq.\ref{eq:hp_spring} were introduced as the numerical equivalent of the springs used in the experiments for restricting the heave and pitch motion, see also Fig.\ref{fig:sch_wavefoil}. In the simulations,  $k_h$=10388 N/m and $k_p$=129.24 N$\cdot$m/rad were selected to be the same as the heave and pitch spring stiffness in the physical tests.
\begin{table}[!ht]
\centering
\begin{tabular}{ccccc}
\hline
Group              & No. & Wave length $\lambda$ (m) & Wave height $2a$ (m) & Wave steepness \\ \hline
\multirow{4}{*}{I}  & 1   & 2.00 & 0.096 & 0.048 \\ \cline{2-5} 
                    & 2   & 2.37 & 0.108 & 0.045 \\ \cline{2-5} 
                    & 3   & 2.92 & 0.123 & 0.042 \\ \cline{2-5} 
                    & 4   & 3.75 & 0.143 & 0.038 \\ \hline
\multirow{4}{*}{II} & 5   & 4.75 & 0.159 & 0.034 \\ \cline{2-5} 
                    & 6   & 5.92 & 0.170 & 0.029 \\ \cline{2-5} 
                    & 7   & 6.88 & 0.174 & 0.025 \\ \cline{2-5} 
                    & 8   & 7.70 & 0.175 & 0.023 \\ \hline
\end{tabular}
\caption{Wave conditions in the semi-active flapping hydrofoil simulations.}
\label{tab:wave_con}
\end{table}

\begin{figure}[ht!]
    \centering
    \includegraphics[width=0.8\textwidth]{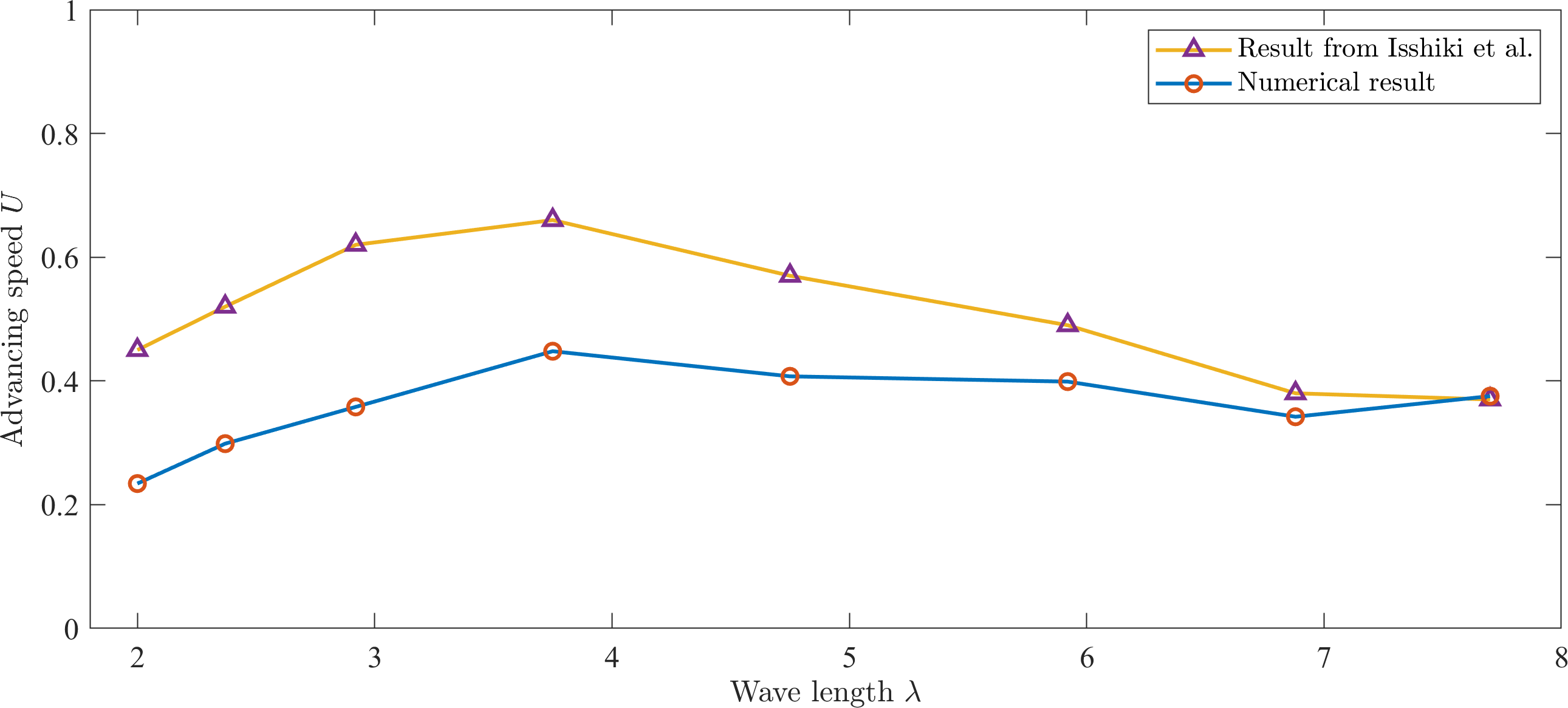}
    \caption{The comparison between the experiment (the upper line) and passive hydrofoil simulation ($\Delta x =  \frac{1}{160}$ m.)}
    \label{fig:com_wavelength}
\end{figure}

The computed and measured time averaged advance velocities of foil for different wave lengths are compared in Fig.\ref{fig:com_wavelength}. The latter comparison is preferred over the comparison of $C_{t}$ over $St_{A}$ presented in Section \ref{Hydrofoil in uniform flow} in order to maintain consistency with the work of Isshiki et al. \cite{isshiki1983theory, isshiki1984theory}, who did not provide values for $C_{t}$ and $St_{A}$; the apparent thrust reported in these works did not take account of the foil's resistance. Similarly to the results presented in Section \ref{Hydrofoil in uniform flow}, Fig.\ref{fig:com_wavelength} confirms a reasonable qualitative agreement between the simulations and the measurements albeit with non-negligible quantitative discrepancies. In particular, the predicted trends in the hydrofoil's response agree with those reported by the physical tests but the amount of thrust generated is notably underpredicted by the numerical simulations. As before, these discrepancies can be likely attributed to the 2D nature of the simulations and/or to insufficient mesh refinement. With regards to the latter, a further refinement in the mesh will likely result in the reduction of $C_{d}$, see Fig.\ref{fig:con_clcd}, and thus in higher velocities for the hydrofoil. In addition, as the wave length increases, the numerical and the experimental solution is seen to converge. It may be that the interaction of the foil's surface with longer waves, results in the evolution of a larger boundary layer, which can be better resolved with the current mesh resolution. However, a further refinement of the mesh would considerably increase the computational cost, thereby reducing the capacity to consider a large range of testing conditions. Nonetheless, and as previously mentioned, the simulations are proven to predict the hydrofoil's response to different flow conditions to a level suitable for the scope of the present work. To this end, the computational results are now used to discuss and justify the generation of thrust and the self-propulsion of the foil against the waves, for all the cases considered.

\begin{figure}[!ht]
 \begin{tabular}{cc}
    \includegraphics[width=0.8\textwidth]{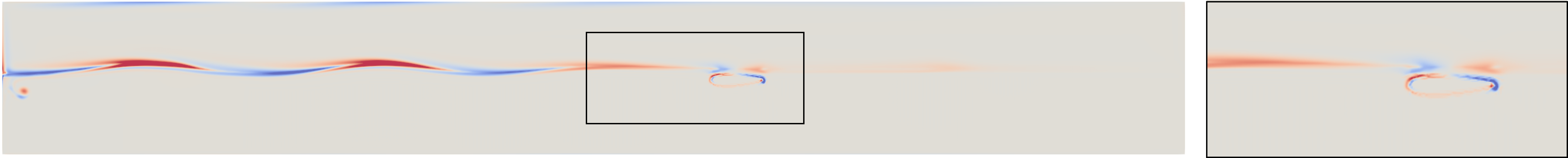} & \multirow{7}{*}{\includegraphics[width=0.12\textwidth]{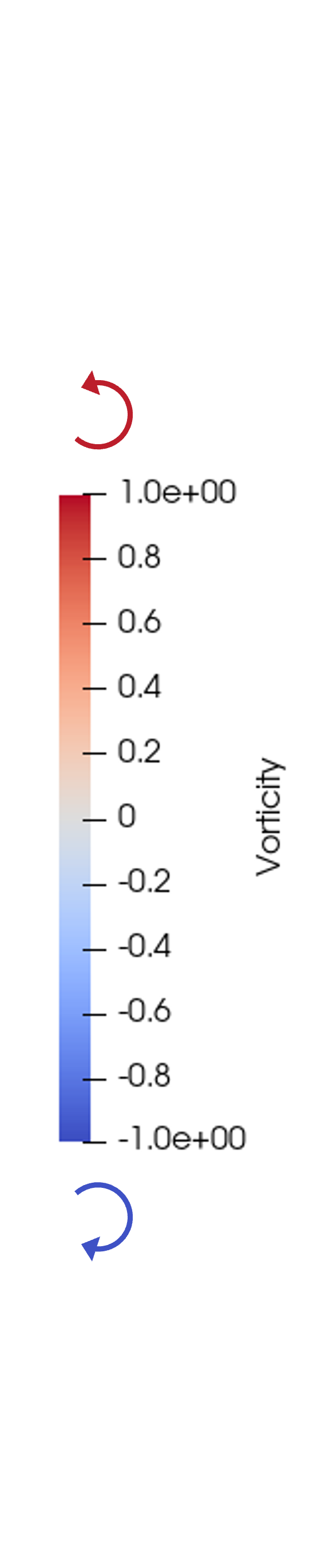}}\\
    (a) $t_{0}=4.10 $ s & \\
    \includegraphics[width=0.8\textwidth]{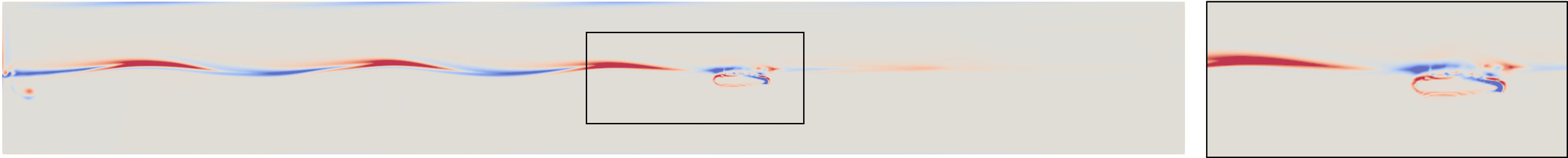} & \\
    (b) $t_{1}=t_{0}+\Delta t$ & \\
    \includegraphics[width=0.8\textwidth]{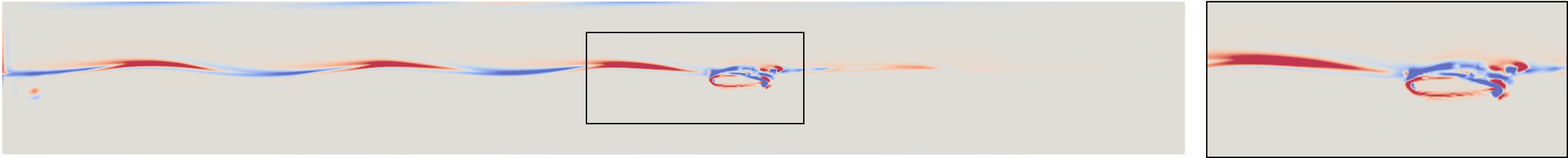} & \\
    (c) $t_{2}=t_{0}+2\Delta t$ & \\
    \includegraphics[width=0.8\textwidth]{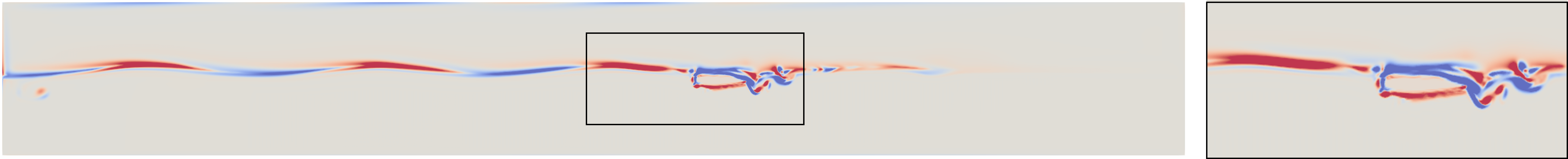} & \\
    (d) $t_{3}=t_{0}+3\Delta t$ & \\
    \includegraphics[width=0.8\textwidth]{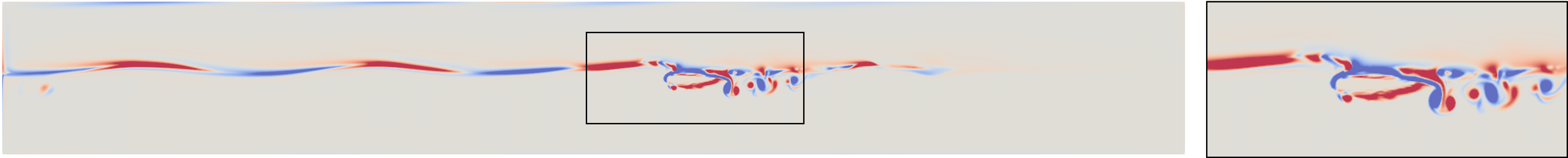} & \\
    (e) $t_{4}=t_{0}+4\Delta t$ &  \\
    \includegraphics[width=0.8\textwidth]{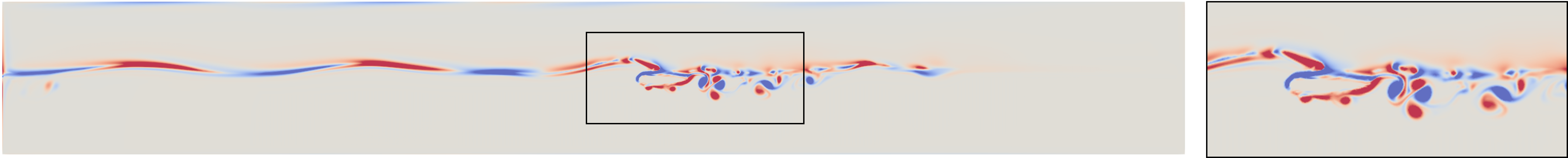} & \\
    (f) $t_{5}=t_{0}+5\Delta t$ & \\
    \includegraphics[width=0.8\textwidth]{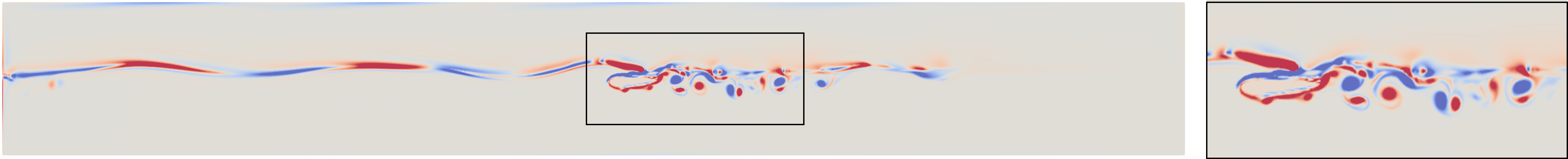} & \\
    (g) $t_{6}=t_{0}+6\Delta t$ & \\ 
\end{tabular} 
    \caption{The position of the passive hydrofoil in wave (wavelength=2.00 m, time interval $\Delta t=1.15$ sec is the wave period). The series of snapshots on the right are the zoomed-in parts.}
    \label{fig:wavefoil_t}
\end{figure}

A characteristic set of results is presented in Fig.\ref{fig:wavefoil_t} in the form of snapshots illustrating the interaction of the hydrofoil with 2.00 m long regular waves at different time instances. Prior to the arrival of the first wave (before $t_{0}$), the foil is located at its original position; it is reminded that the density of the foil was set to be equal to the wave density thereby nullifying buoyancy. The interaction with the first waves, from $t_{0}$ to $t_{3}$,  triggers the movement of the hydrofoil and initiates the wake development at its trailing edge. 4 wave periods later, at $t_{4}$, a neutral street is observed and a small propulsion is noticed. After $t_{5}$ the evolution of a RvK vortex street is evident, leading to a bigger self-propulsion of the foil against the incoming waves. Within 10 wave periods, the hydrofoil is predicted to have covered a horizontal distance approximately equal to the wave length. Leading edge vortices are also generated, they travel downstream as well and blend with the trailing edge vortices, thereby increasing the complexity of the wake. Similar wake patterns and thrust generation is predicted by the computations of all the cases of Table \ref{tab:wave_con} and in agreement with the findings of Isshiki et al. \cite{isshiki1984theory}.

\begin{figure}[!ht]
 \begin{tabular}{cc }
    \includegraphics[width=0.8\textwidth]{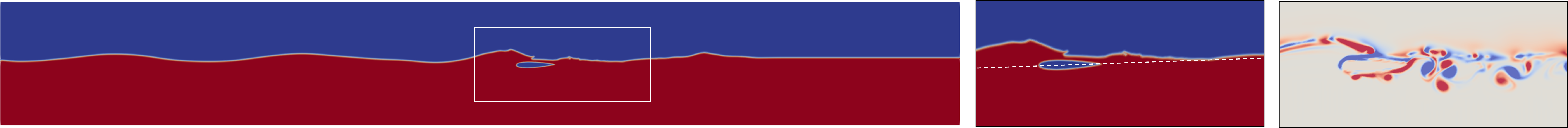} & \multirow{2}{*}{\includegraphics[width=0.2\textwidth]{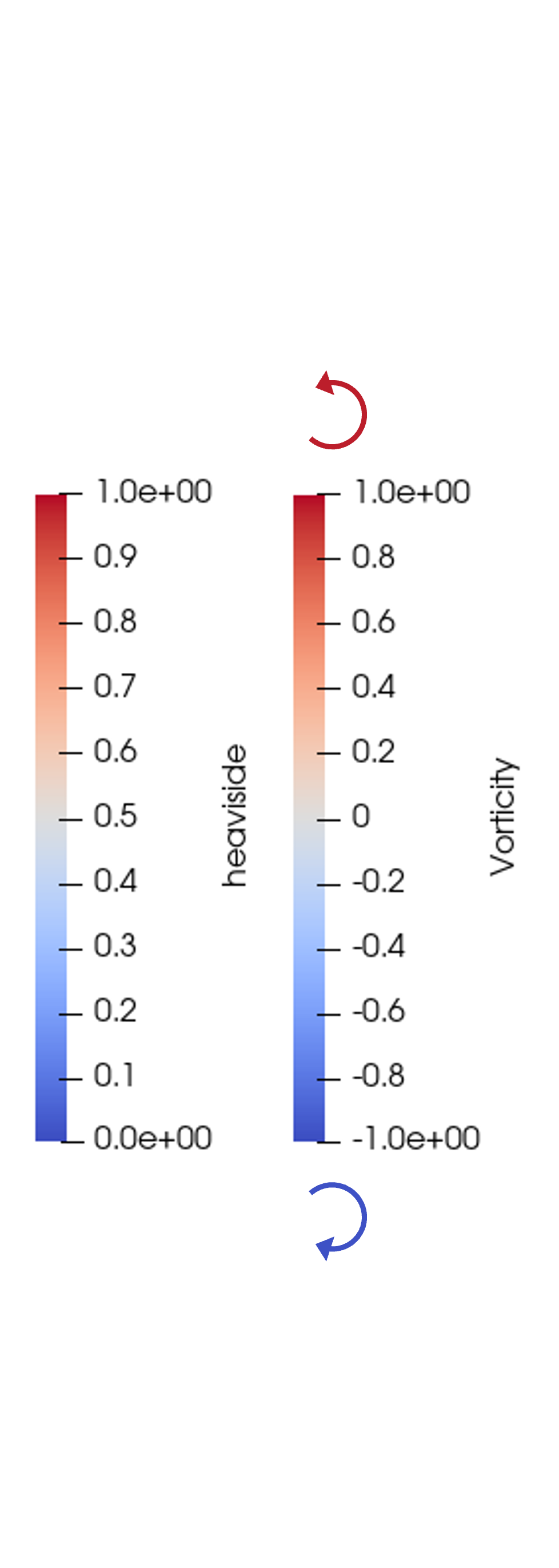}} \\
    (a) $t_{5}=t_{0}+5\Delta t$ \\
    \includegraphics[width=0.8\textwidth]{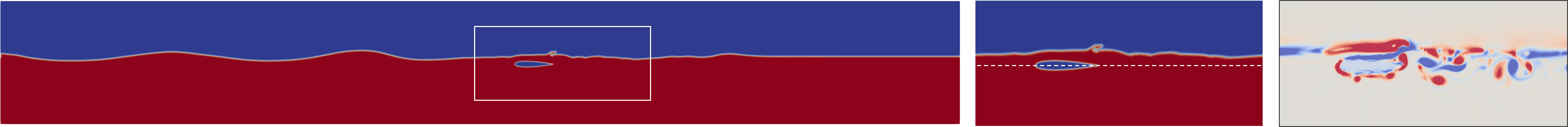} & \\
    (b) $t_{5}+\frac{1}{4}\Delta t$ \\
    \includegraphics[width=0.8\textwidth]{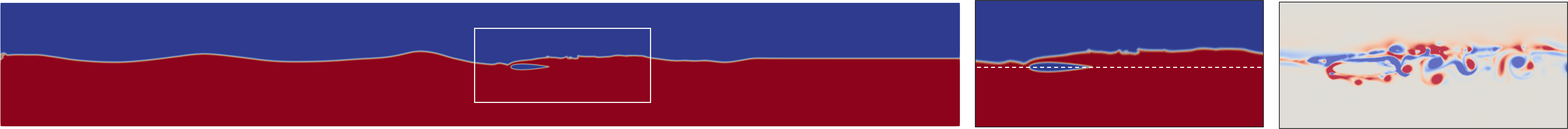} & \\
    (c) $t_{5}+\frac{1}{2}\Delta t$ \\
    \includegraphics[width=0.8\textwidth]{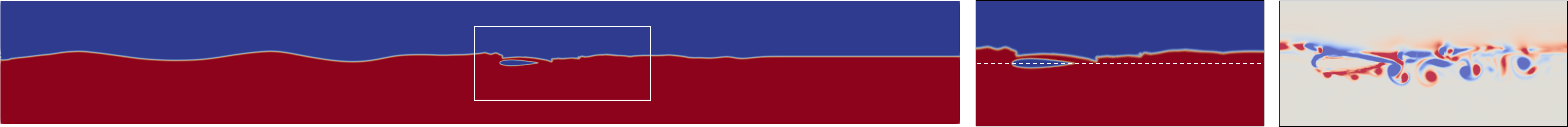} & \\
    (d) $t_{5}+\frac{3}{4}\Delta t$ \\
    \includegraphics[width=0.8\textwidth]{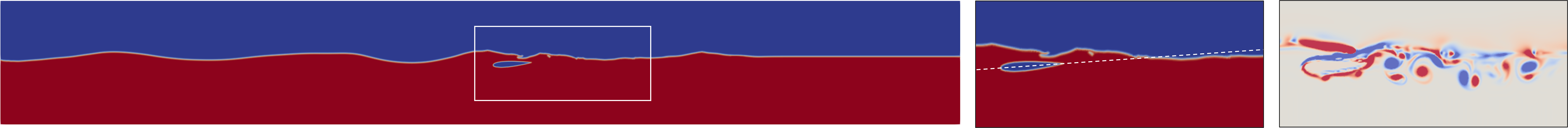} & \\
    (e) $t_{6}=t_{0}+6\Delta t$ \\ 
\end{tabular} 
    \caption{The position of the passive hydrofoil in one wave period (wavelength=2.00 m, $t_{0}=4.10$ sec, time interval $\Delta t=1.15$ sec is the wave period). The series of snapshots on the left shows the Heaviside fields, while those on the right show the zoomed-in vortex fields. The dash lines are the chord lines.}
    \label{fig:wavefoil_t_p}
\end{figure}

The motion of the hydrofoil over one wave period is detailed in Fig.\ref{fig:wavefoil_t_p}, where the orientation of the chord line (dashed line in the second column Heaviside plots) and the wake (contour plots at the third column) at different time instances are shown. The Heaviside plots at the top, show the location of the foil and the wake developed five (5) wave periods after the initiation of the simulation. Under the wave crest, the chord line points downwards, the AoA is negative and a relatively complex RvK vortex street is observed. The thrust generated propels the foil against the incoming wave as shown in the plots of the second row, where the interaction with the following trough strokes the foil upwards and, as seen in the contour plot, leads to the development of a vK street and thus to a short-lived thrust-to-drag transition. A quarter of the wave period later, the foil is still on the upstroke but the vortex street is starting to reverse, initiating a drag-to-thrust transition. The vortex street is fully reversed at $t_{5}+3/4\Delta t$ and the hydrofoil is further propelled against the wave. From wave crest to trough (and vice versa), the thrust-to-drag-to-thrust transition happens all the time after the RvK vortex street is generated. Correspondingly, the force switches from thrust to drag to thrust with an overall net positive thrust force, so that the propulsion can be observed. This wake and force transition is similar with and confirmed by the wake observation of the active flapping foil shown above, see Fig.\ref{fig:case2_vortex}. It is noteworthy that the foil's heave in conjunction with the proximity to the free surface results in the generation of not only LE vortices which travel and mix with the TE wake but also in increased near surface vorticity. The latter observation is in agreement with Isshiki et al. \cite{isshiki1983theory, isshiki1984theory} who commented that the free surface effects are non-negligible but the contribution of these effects to the hydrofoil's self-propulsion remains unclear. Finally, at $t_{6}$ the hydrofoil meets the next wave crest, a vK street is generated on the downstroke and the process repeats periodically. The same pattern in the motion of the hydrofoil and the associated wake was observed to be the same for all the cases considered. However, increasing the wave length was found to reduce the flapping and thus the vortex shading frequency, see Fig.\ref{fig:wavefoil3.75_t} and \ref{fig:wavefoil3.75_t_p} in Appendix (Section \ref{Appendix}).

\begin{figure}[ht!]
    \centering
    \includegraphics[width=0.8\textwidth]{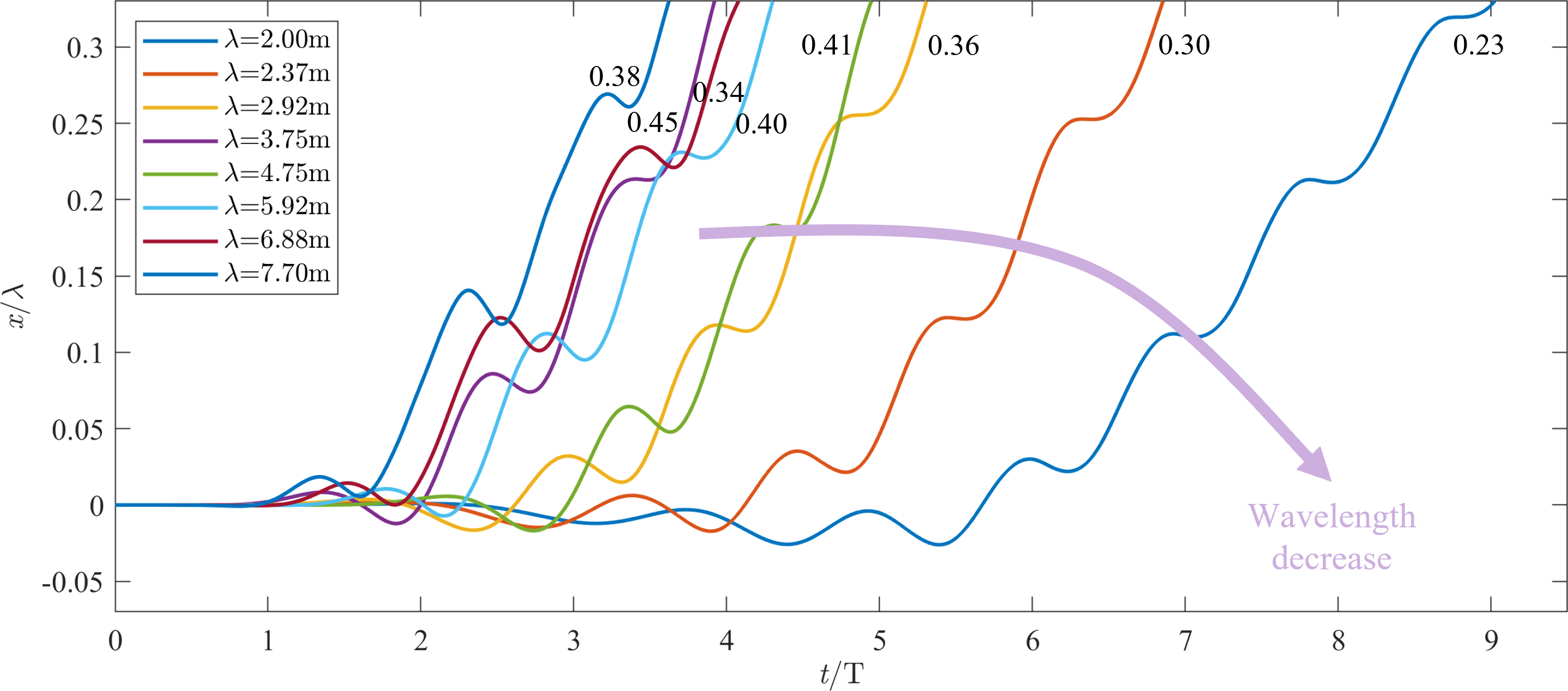}
    \caption{The non-dimensional displacement $x/\lambda$ of the passive hydrofoil with time over wave period $t/$T facing different waves($\Delta x =  \frac{1}{160}$ m).}
    \label{fig:xt_wavelength}
\end{figure}

The effect of the wave length to the hydrofoil non-dimensional displacement is presented as a function of non-dimensional time in Fig.\ref{fig:xt_wavelength}. For all the cases illustrated, the initially weak displacement is due to the interaction the first, underdeveloped wave. Soon after, in every wave period, a relatively longer lasting increase in the foil's displacement is reported, followed by a sharper and faster reduction in the displacement. As previously discussed the former motion is associated with the generation of the RvK when the wave crest/trough is present, while the transition to a vK vortex street under forces between crest and trough the hydrofoil to move along with the wave. Overall, the foil experiences a periodic but asymmetric motion and therefore a positive net displacement, which is also seen to become increasingly non-harmonic / non-asymmetric for longer waves. The periodic nature of the hydrofoil's displacement is aligned with the periodicity in the wave induced flow kinematics but the asymmetry is more likely driven by the time asymmetry in the occurrence of a thrust and a drag generating wake pattern. In other words, the wave enforces the flapping motion of the hydrofoil, which in turn results in the evolution of different wake patterns that then determines the (self-) propulsion of the foil against the incoming waves; as an example, in Fig.\ref{fig:wavefoil_t_p}, a RvK vortex street is observed to occur for about three quarters (3/4) of the wave period. For these tests, the shorter waves were steeper then the longer waves and as such the asymmetry in the hydrofoil's displacement can be seen to be mostly influenced (increase with) by (the increase in) the wave length.

The number next to each line in Fig.\ref{fig:xt_wavelength} is the computed the time averaged advance velocity of the hydrofoil for each test case; also presented in Fig.\ref{fig:com_wavelength}. Starting with the shorter waves, the velocity is observed to increase with the wave length, before reaching a maximum for $\lambda$=3.75 m. Characteristically, the steepness of the non-dimensional displacement curves in Fig.\ref{fig:com_wavelength} decreases as the wavelength increases. This entails, that under the action of shorter waves, the time required for the hydrofoil to advance over the same distance is higher than it is for longer waves. Nonetheless, for waves longer than 3.75 m, the foil's advance velocity follows a fluctuating but in principle decreasing trend, indicating the existence of an operational optimum with respect to $\lambda$.

\subsection{The effect of the pitch stiffness}
\label{The effect of the pitch stiffness}
In the previous section the wave induced flapping motion of the hydrofoil and the resulting thrust-to-drag-to-thrust transition have been discussed. Here, the effect of the pitch stiffness to the generation of thrust is explored. 

The latter effect is consider of particular importance in real life applications, since the extend to which the pitch motion contributes to the overall flapping of the foil will be largely defined by the connection between, e.g., the hydrofoil and the superstructure, which in turn dictates the pitch stiffness. In Isshiki et al. \cite{isshiki1984theory} experiments for example, the foil is connected to the towing mechanism through a series of springs. Altering the stiffness of the pitch springs altered the rotational deviation (and moment) of the hydrofoil thereby affecting the capacity of the foil to pitch freely and influencing the passive response of the foil to the wave action. In the computations, this is expressed through Eq.\ref{eq:hp_spring:2} and it is illustrated in Fig.\ref{fig:sch_wavefoil}.

\begin{figure}[ht!]
    \centering
    \includegraphics[width=0.9\textwidth]{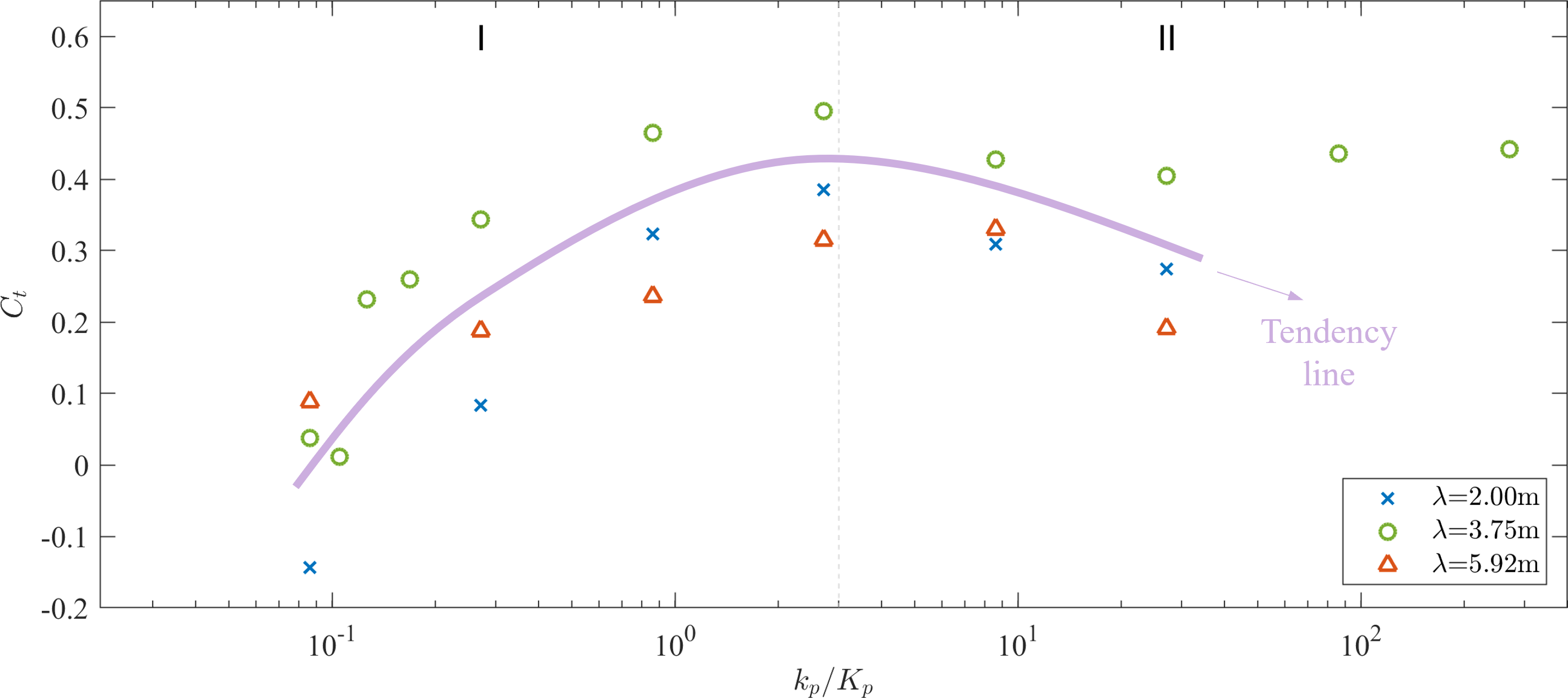}
    \caption{The thrust coefficient $C_{t}$ of the passive hydrofoil with different non-dimensional pitch stiffness $k_p/K_p$($\Delta x =  \frac{1}{160}$ m).}
    \label{fig:u_kp}
\end{figure}

The pitch stiffness effect in the generation of thrust for the case with $\lambda$=3.75 m, which previously yielded the highest speed of foil advance. For the results of Section \ref{Hydrofoil in waves} the pitch spring stiffness $k_p$ was set to 129.24 N$\cdot$m/rad, and additional computations were conducted for $k_p$ ranging between 0 and 12900 N$\cdot$m/rad. The thrust coefficient computed in all simulations is plotted as a function of the non-dimensional $k_p$/$K_p$ ratio in Fig.\ref{fig:u_kp}; $K_p$ is the foil's hydrostatic pitch stiffness calculated to be equal to 47.52 N$\cdot$m/rad (see also Fig.\ref{fig:hp_hydrosp}).

Two distinct areas of the pitch stiffness effect in the generation of thrust can seen in Fig.\ref{fig:u_kp}. I, where no thrust is essentially generated for near zero $k_p$/$K_p$ values but then the thrust coefficient increases sharply $k_p$/$K_p$ until $k_p$=12.9 N$\cdot$m/rad ($k_p$/$K_p$ = 0.27). For higher $k_p$/$K_p$, $C_{t}$ increases further, albeit at reduced rate, and the peak thrust coefficient is achieved for $k_p$/$K_p$=2.71. II, where the further increase $k_p$/$K_p$ reduces $C_{t}$ below its peak value.

\begin{figure}[!ht]
 \begin{tabular}{cc}
    \includegraphics[width=0.8\textwidth]{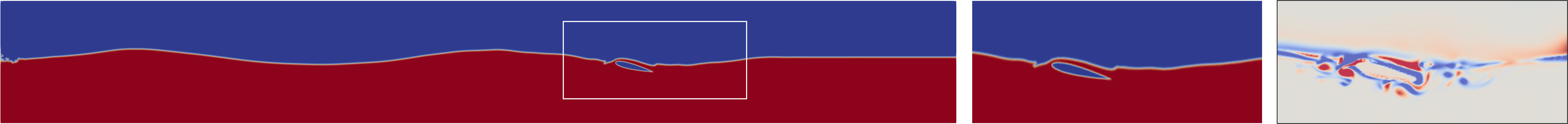} & \multirow{1}{*}{\includegraphics[width=0.2\textwidth]{wl2_hscale_period.png}} \\
    (a) $k_p/K_p$=0 & \\
    \includegraphics[width=0.8\textwidth]{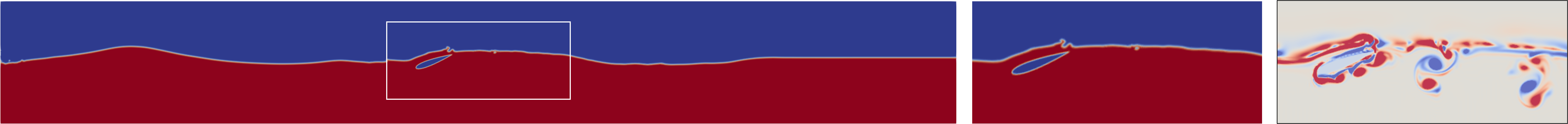} & \\
    (b) $k_p/K_p$=0.27 & \\
    \includegraphics[width=0.8\textwidth]{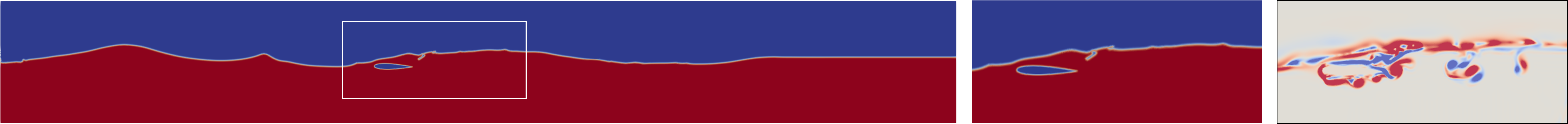} & \\
    (c) $k_p/K_p$=2.71 & \\
    \includegraphics[width=0.8\textwidth]{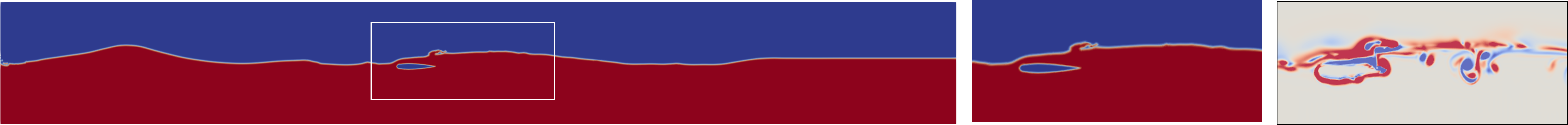} & \\
    (d) $k_p/K_p$=27.15 & \\ 
\end{tabular} 
    \caption{The snapshots of the passive hydrofoil under different non-dimensional pitching spring stiffness $k_p/K_p$  (wavelength=3.75 m, $t_{0}=8.99$ sec). The series of snapshots on the left shows the Heaviside fields, while those on the right show the zoomed-in Heaviside and vortex fields.}
    \label{fig:wavefoil_t_kp}
\end{figure}

\begin{figure}[!ht]
 \begin{tabular}{cc}
    \includegraphics[width=0.8\textwidth]{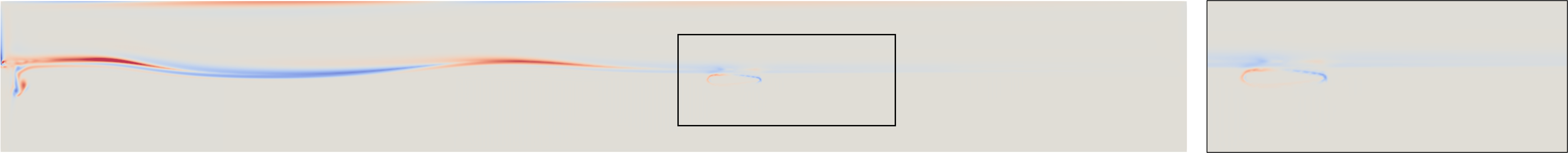} & \multirow{5}{*}{\includegraphics[width=0.12\textwidth]{case3_wl2_sacler.png}}\\
    (a) $t_{0}=1.95 $ s & \\
    \includegraphics[width=0.8\textwidth]{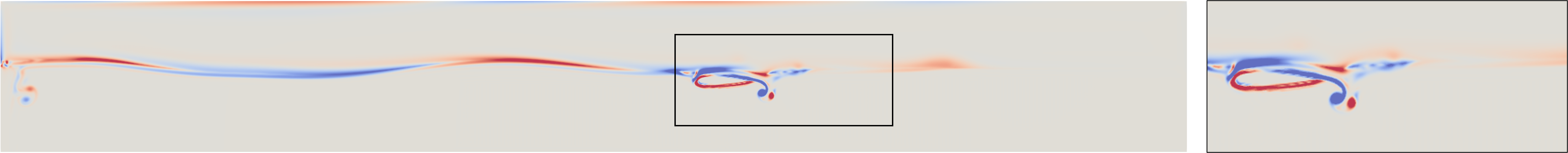} & \\
    (b) $t_{1}=t_{0}+\Delta t$ & \\
    \includegraphics[width=0.8\textwidth]{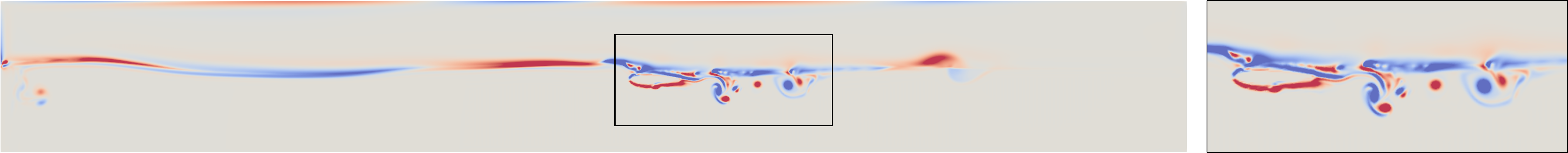} & \\
    (c) $t_{2}=t_{0}+2\Delta t$ & \\
    \includegraphics[width=0.8\textwidth]{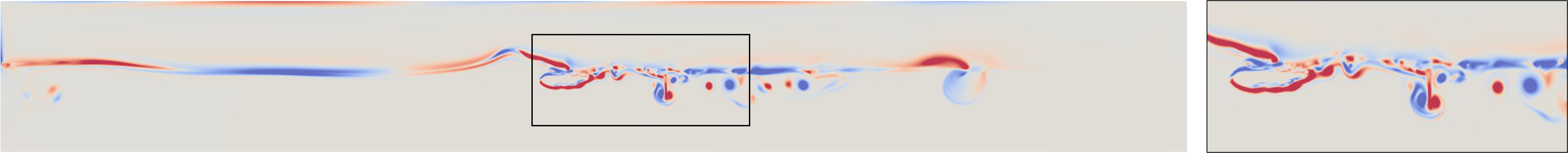} & \\
    (d) $t_{3}=t_{0}+3\Delta t$ & \\
    \includegraphics[width=0.8\textwidth]{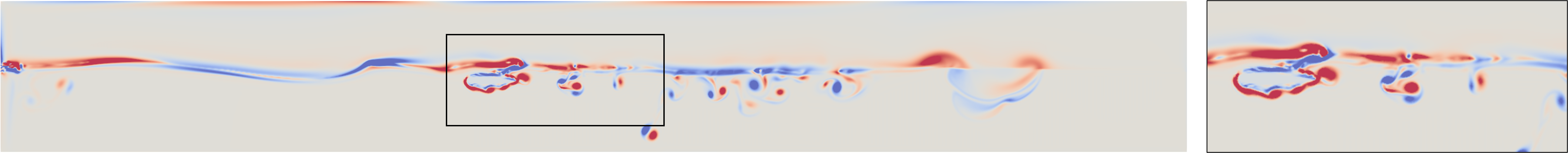} & \\
    (e) $t_{4}=t_{0}+4\Delta t$ &  \\
    \includegraphics[width=0.8\textwidth]{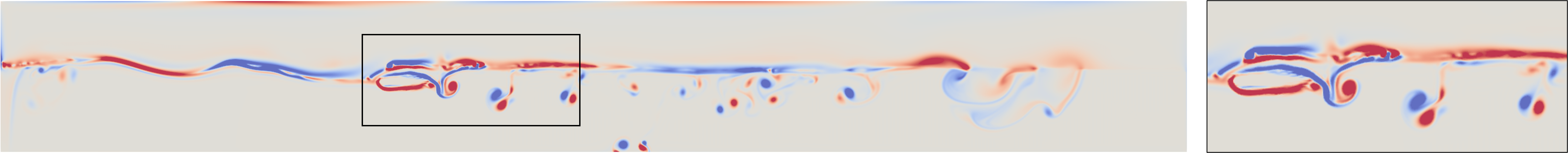} & \\
    (f) $t_{5}=t_{0}+5\Delta t$ & \\
\end{tabular} 
    \caption{The position of the passive hydrofoil in wave (wavelength=3.75 m, time interval $\Delta t=1.70 $ sec is the wave period). The series of snapshots on the right are the zoomed-in parts.}
    \label{fig:wavefoil3.75_t}
\end{figure}

Since for all the cases presented in Fig.\ref{fig:u_kp} the incoming wave is the same, the deviation in the angle of rotation $\Delta \theta$ in Eq.\ref{eq:hp_spring:2} will decrease as $k_p$ increases. In other words, the maximum $\Delta \theta$ will occur for the minimum $k_p$/$K_p$ and vice versa, altering the response of the hydrofoil to the same hydrodynamic stimulation (to the wave). This is depicted in Fig.\ref{fig:wavefoil3.75_t} and best illustrated in the animation attached to this paper. In Fig.\ref{fig:wavefoil_t_kp}, the location of the hydrofoil and the vorticity field is presented for different $k_p$/$K_p$ but at the same time period, $t_{0}$=8.99 sec, after the initiation of each computation. It is noted that $t_{0}$ is considerably higher than 4 wave periods, which according to the simulations is the time required for the full development of a vortex street.

For $k_p$=$k_p$/$K_p$=0, (a) in Fig.\ref{fig:wavefoil_t_kp}, there is no restoring moment applied in the wave induced motion for the hydrofoil, which eventually drifts along with the incoming wave. The absence of the restoring moment and unregulated motion of the foil is seen to prevent the development of a vortex street in the wake of the foil and only a weak vorticity field occurs at the leading and trailing edge. In marked contrast with (a), even a small increase in $k_p$/$K_p$=0.27 is observed to the lead to the generation of a vortex street in (b). The development of a restoring moment regulates the pitch motion, the hydrofoil flap's, and a thrust-to-drag-to-thrust transition occurs and propels the foil against the wave; as described in more details in Section \ref{Hydrofoil in waves}. Once more, the interaction with the free surface and the heave motion lead in the generation of LE vortices which travel towards the TE and cascade into the wake. A similar behavioural pattern is reported for higher $k_p$/$K_p$ ratios, for all of which the hydrofoil is thrust in the opposite to the wave direction. As $k_p$ increases the extend of the pitch motion decreases since $\Delta \theta$ reduces, and smaller in size vortices are shed at a higher frequency from the TE, see e.g. (c) and (d) in Fig.\ref{fig:wavefoil_t_kp}. For $k_p$/$K_p$=2.71, the hydrofoil experiences the higher $C_{t}$, see Fig.\ref{fig:u_kp}, and this is indeed confirmed in (c), where the structure has covered a larger distance in the same time than compared with any other case presented.

According to the results presented in Fig.\ref{fig:u_kp} and Fig.\ref{fig:wavefoil_t_kp}, a non-zero pitch stiffness is necessary for the wave-foil interaction to lead to a controlled flapping motion, leading to the development of a vortex street and to the generation of thrust; it is worth remembering that all the simulations where conducted with a constant heave stiffness. For thrust to occur, a pitch stiffness higher than approx. one third of the hydrofoil's hydrostatic stiffness ($k_p=0.27K_p$) is required, while the highest $C_{t}$ is found at about three (3) times the hydrostatic stiffness, specifically for $k_p=2.71K_p$. Stiffer systems ($k_p$ higher than about 3$K_p$), will experience a reduction in $C_{t}$ bellow its peak value but our results do not provide a full justification for this and thus the reason eludes us. 

For completeness, the effect of $k_p$ in thrust generation has been further explored for a shorter and for a longer wave. The results for waves with $\lambda=2$ m and for $\lambda=5.92$ m are also plotted in Fig.\ref{fig:u_kp}, where the overall effect of $k_p$ to $C_{t}$ appears to be independent of the wave length. Characteristically, for all the cases considered $C_{t}$ increases with $k_p/K_p$ until a maximum is reached, and any further increase in the pitch stiffness of the system affects inversely the thrust coefficient. Interestingly enough, for all the cases considered the optimum $C_{t}$ occurs for $k_p=2.71K_p$, albeit for $\lambda=5.92$ m a high thrust coefficient is observed even for a stiffer case before also reducing for higher $k_p/K_p$. As such, for studies looking to identify the pitch stiffness ($k_p$) resulting in the highest thrust coefficient, a $k_p\approx3K_p$ is recommended as a rational starting point. It is also noteworthy that although optimising the pitch stiffness will lead to a higher thrust coefficient, the maximum threshold for $C_{t}$ is purely defined by the wave length, see Fig.\ref{fig:u_kp}.

\section{Conclusions}
\label{Conclusions}
The present work used an in-house developed, new CFD code to study wave induced thrust in a hydrofoil, placing emphasis on the so-far-not-well-explored effects of the pitch stiffness. To this end, the numerical code was initially validated against a number of previously published experimental results and it was then used for investigating the mechanism of thrust production when a submerged hydrofoil interacts with regular waves. Under the influence of the wave action, the foil followed a flapping motion, which in turn, and in dependence with the wave phase, led to the successive generation of either a RvK or a vK vortex street, the transition from thrust-to-drag and then to thrust again, and finally to the propulsion of the hydrofoil against the incoming wave. Having described the thrust generation mechanism, the stiffness ($k_p$) of the numerical pitch spring was then altered while keeping the heave spring stiffness constant, and the effect of $k_p$ to the thrust produced was examined through a series of simulations with 0 $\leq k_p/K_p \leq$ 271 and for 2 $\leq \lambda \leq$ 5.92 m. 
The main conclusions for this testing range, are: 
\begin{itemize}
  \item A non-zero pitch stiffness is necessary for the hydrofoil to be thrust against the incoming wave.
  \item Two areas of pitch stiffness influence to thrust coefficient have been recognised, regardless to the wave length. Area I, where $C_{t}$ increases with the non-dimensional ration of $k_p$/$K_p$, and a second area (II), where $C_{t}$ reduces bellow its peak values as the system becomes stiffer.  
  \item $k_p\approx3K_p$ is a plausible threshold between the two areas, thus it is recommended as a starting point for studies seeking to achieve the optimum $C_{t}$ in regular waves. 
  \item It is however noted that although an optimum pitch stiffness can be expected to lead to a higher $C_{t}$, the maximum threshold of this $C_{t}$ is defined by the wave length, see Fig.\ref{fig:u_kp}.   
\end{itemize}

The present work is the first in a series of studies focused in improving our understanding of wave induced thrust generation by submerged hydrofoils, the final aim of which is described in \cite{reduce2022} and refers to the implementation of hydrofoils in offshore floating wind platforms.

\section{Appendix}
\label{Appendix}

\begin{figure}[!ht]
 \begin{tabular}{cc }
    \includegraphics[width=0.8\textwidth]{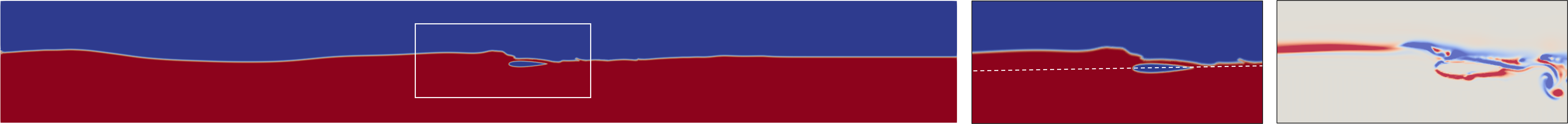} & \multirow{2}{*}{\includegraphics[width=0.2\textwidth]{wl2_hscale_period.png}} \\
    (a) $t_{2}=t_{0}+2\Delta t$ \\
    \includegraphics[width=0.8\textwidth]{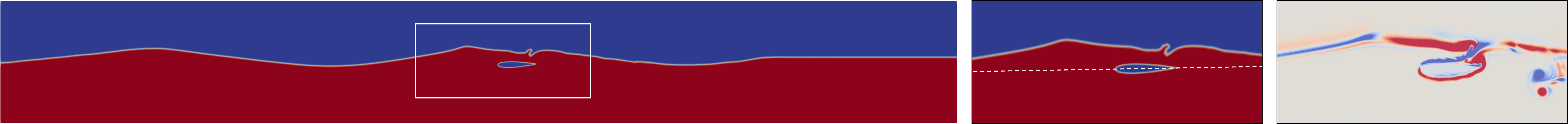} & \\
    (b) $t_{2}+\frac{1}{4}\Delta t$ \\
    \includegraphics[width=0.8\textwidth]{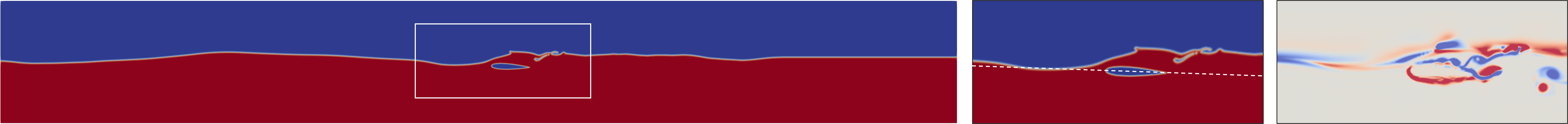} & \\
    (c) $t_{2}+\frac{1}{2}\Delta t$ \\
    \includegraphics[width=0.8\textwidth]{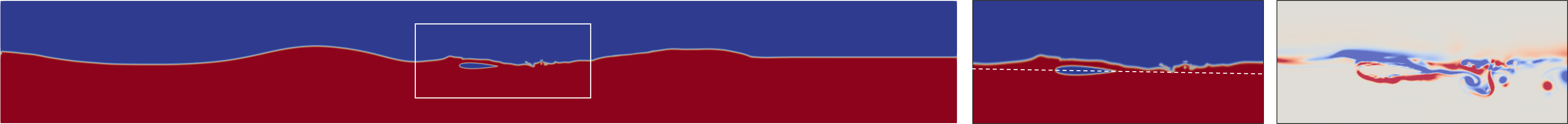} & \\
    (d) $t_{2}+\frac{3}{4}\Delta t$ \\
    \includegraphics[width=0.8\textwidth]{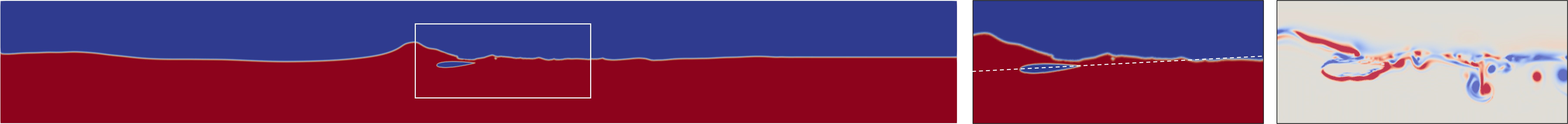} & \\
    (e) $t_{3}=t_{0}+3\Delta t$ \\ 
\end{tabular} 
    \caption{The position of the passive hydrofoil in one wave period (wavelength=3.75 m, $t_{0}=1.95 $ sec, time interval $\Delta t=1.70 $ sec is the wave period). The series of snapshots on the left shows the Heaviside fields, while those on the right show the zoomed-in vortex fields. The dash lines are the chord lines.}
    \label{fig:wavefoil3.75_t_p}
\end{figure}

\section*{Author statements}
J. Xing: Numerical simulations and methodology, formal analysis, writing, editing; D. Stagonas: Original concept development, methodology, formal analysis, writing, supervision, review and editing; C. Zhang: Review and Editing;  J. Yang: Review and editing; P. Hart: Review and Editing; L. Yang: Numerical simulations and methodology, original concept development, formal analysis, reviewing, editing, supervision and project administration. 

\section*{Declaration of competing interest}
The authors declare that they have no known competing financial interests that could have appeared to influence the work reported in this paper.

\section*{Acknowledgements}
J. Xing acknowledges the support by the China Scholarship Council (CSC) from the Ministry of Education of P.R. China. L. Yang acknowledges the support from Supergen ORE ECR Fund and UK-Saudi Challenge Fund `Feasibility study of hybrid propulsion for unmanned surface vehicle for environmental monitoring' from British Council. 

\bibliography{ref}

\end{document}